\documentclass[a4paper,fleqn,longmktitle]{cas-dc}

\pdfoutput=1

\usepackage[authoryear]{natbib}

\usepackage[british]{babel}
\usepackage[english=usenglishmax]{hyphsubst}
\usepackage{microtype}
\usepackage[switch]{lineno}

\def\tsc#1{\csdef{#1}{\textsc{\lowercase{#1}}\xspace}}
\tsc{WGM}
\tsc{QE}
\tsc{EP}
\tsc{PMS}
\tsc{BEC}
\tsc{DE}
\newcommand{\Mearth}{\mbox{$M_{\oplus}$}}            
\newcommand{\Rearth}{\mbox{$R_{\oplus}$}}            
\newcommand{\Rjup}{\mbox{$R_{\mathrm{J}}$}}          
\newcommand{\Mjup}{\mbox{$M_{\mathrm{J}}$}}          
\newcommand{\Mc}{\mbox{$M_{Z}$}}                     
\newcommand{\Me}{\mbox{$M_{XY}$}}                    
\newcommand{\dMc}{\mbox{$\dot{M}_{Z}$}}              
\newcommand{\dMe}{\mbox{$\dot{M}_{XY}$}}             
\newcommand{\dMp}{\dot{M}_{p}}                       
\newcommand{\Mp}{\mbox{$M_{p}$}}                     
\newcommand{\Rp}{\mbox{$R_{p}$}}                     
\newcommand{\ap}{\mbox{$a_{p}$}}                     
\newcommand{\Msun}{\mbox{$M_{\astrosun}$}}           
\newcommand{\Lsun}{\mbox{$L_{\astrosun}$}}           
\newcommand{\Rhill}{\mbox{$R_\mathrm{H}$}}           
\newcommand{\Rbondi}{\mbox{$R_\mathrm{B}$}}          
\newcommand{\sigu}{\mbox{$\mathrm{g\,cm}^{-2}$}}    
\newcommand{\rhou}{\mbox{$\mathrm{g\,cm}^{-3}$}}    
\newcommand{\K}{\mbox{$\mathrm{K}$}}                 
\newcommand{\AU}{\mbox{au}}                          
\newcommand{\Myr}{\mbox{Myr}}                        
\newcommand{\cisec}[1]{Section~\ref{#1}}             
\newcommand{\cifig}[1]{Figure~\ref{#1}}              
\newcommand{\cieq}[1]{Equation~(\ref{#1})}           
\definecolor{mediumaquamarine}{rgb}{0.4, 0.8, 0.67}
\definecolor{lightcoral}{rgb}{0.9412,0.5020,0.5020}
\newcommand{\aqt}{\mbox{$\alpha$43}}
\newcommand{\aqq}{\mbox{$\alpha$44}}
\newcommand{\aqtlz}{\mbox{$\alpha$43LZ}}

\let\oldequation\equation
\let\oldendequation\endequation
\renewenvironment{equation}
  {\linenomathNonumbers\oldequation}
  {\oldendequation\endlinenomath}

\input{my_aas_macros.sty}

\begin{document}
\let\WriteBookmarks\relax
\def\floatpagepagefraction{1}
\def\textpagefraction{.001}
\shorttitle{Jupiter's Formation and Evolution to the Present Epoch}
\shortauthors{\textcolor{mediumaquamarine}{G.~D'Angelo, S.J.~Weidenschilling, J.J.~Lissauer, \& P.~Bodenheimer}}

\title [mode = title]{Growth of Jupiter: Formation in Disks of Gas and Solids and Evolution to the Present Epoch}\tnotemark[1]

\tnotetext[1]{This document is the result of a research project partly funded by NASA.}

\author[1]{Gennaro D'Angelo}[orcid=0000-0002-2064-0801]
\cormark[1]
\ead{gennaro@lanl.gov}
\address[1]{Theoretical Division, Los Alamos National Laboratory, Los Alamos, NM 87545, USA}

\author[2]{Stuart J.\ Weidenschilling}[]
\ead{sjw@psi.edu}
\address[2]{Planetary Science Institute, 1700 East Fort Lowell Road, Suite 106, Tucson, AZ 85719, USA}

\author[3]{Jack J.\ Lissauer}[orcid=0000-0001-6513-1659]
\ead{Jack.J.Lissauer@nasa.gov}
\address[3]{Space Science and Astrobiology Division, NASA-Ames Research Center, Moffett Field, CA 94035, USA}

\author[4]{Peter Bodenheimer}[orcid=0000-0001-6093-3097]
\ead{peter@ucolick.org}
\address[4]{UCO/Lick Observatory, Department of Astronomy and Astrophysics, University of California, Santa Cruz, CA 95064, USA}

\cortext[1]{Corresponding author}

\begin{abstract}
The formation of Jupiter is modeled via core-nucleated accretion,
and the planet's evolution is simulated up to the present epoch. 
Throughout the phases when the planet acquires most of its heavy-element
content, the calculation of solids' accretion accounts for interactions 
with an evolving disk of planetesimals. The phase of growth from 
an embryo of a few hundred kilometers in radius until the time when 
the accretion of gas overtakes solids' accretion was presented by
\citet{gennaro2014}, and the same numerical methods are applied here.
Those calculations followed the formation for about $4\times 10^{5}$
years, until the epoch when the heavy-element and hydrogen/helium
masses were $\Mc\approx 7.3$ and $\Me\approx 0.15$ Earth's masses
($\Mearth$), respectively, and $\dMe\approx\dMc$.
Herein, the calculation is continued through the phase when $\Me$ 
grows to equal $\Mc$, at which age, about $2.4\times 10^{6}$ years, 
the total mass of the planet is $\Mp\approx 20\,\Mearth$. 
About $9\times 10^{5}$ years later, $\Mp$ is approximately $60\,\Mearth$ 
and $\Mc\approx 16\,\Mearth$, three-quarters of which are delivered 
by planetesimals larger than $10\,\mathrm{km}$ in radius.
Around this epoch, the contraction of the envelope 
dictates gas accretion rates a few times $10^{-3}\,\Mearth$ per year,
initiating the regime of disk-limited accretion, whereby the planet
can accrete all the gas provided by the disk, and its evolution is 
therefore tied to disk's evolution.
Growth is continued by constructing simplified models of protosolar 
accretion disks that evolve through viscous diffusion, winds, and 
accretion on the planet. 
Jupiter's formation ends after $\approx 3.4$--$4.2\,\Myr$,
depending on the applied disk viscosity parameter, when nebula gas
disperses. The young Jupiter is $4.5$--$5.5$ times as voluminous as 
it is presently and thousands of times as luminous, 
$\sim 10^{-5}\,L_{\astrosun}$.
The heavy-element mass is $\approx 20\,\Mearth$.
The evolution proceeds through the cooling and contraction phase, in
isolation except for solar irradiation.
After $4570\,\Myr$, the age of the solar system, radius and luminosity 
of the planet are within $10$\% of current values, accounting also
for uncertainties in the power absorbed from the Sun.
During formation, and soon thereafter, the planet exhibits features, 
e.g., luminosity and effective temperature, that may probe aspects 
of the latter stages of formation, if observable.
These possibly distinctive features, however, seem to disappear 
within a few tens of $\Myr$.

\end{abstract}

\begin{keywords}
Accretion\sep
Jovian planets\sep
Jupiter\sep
Planetary formation\sep
Planetesimals

\vspace*{10mm}
Received:~10 Mar 2020\sep
Revised:~22 Aug 2020\sep
Accepted:~25 Aug 2020

\vspace*{10mm}
\textcolor{lightcoral}{
This is an unofficial
preprint prepared by the authors.
}

\end{keywords}

\begin{graphicalabstract}
\resizebox{0.5\linewidth}{!}{%
\includegraphics{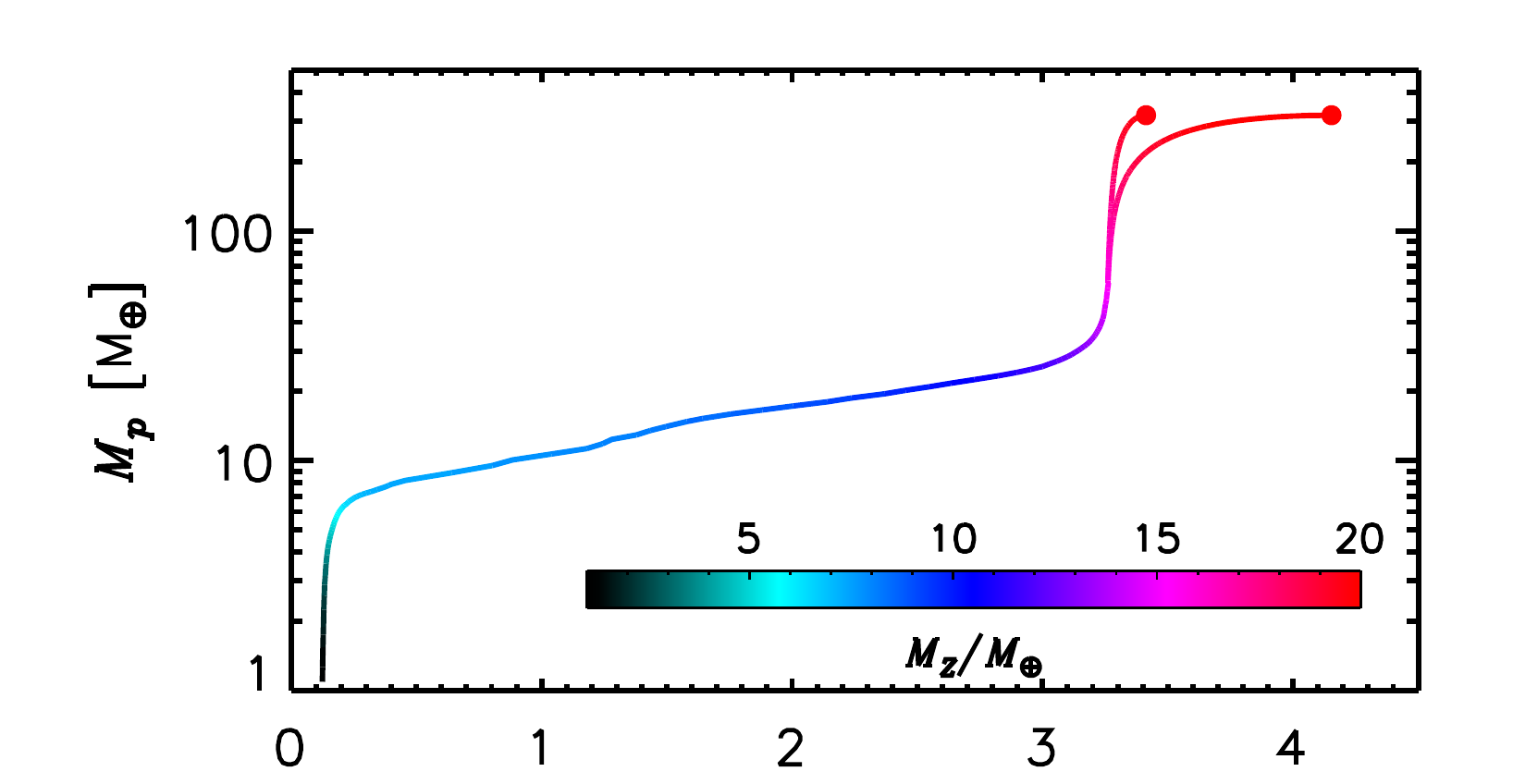}}\\
\resizebox{0.5\linewidth}{!}{%
\includegraphics{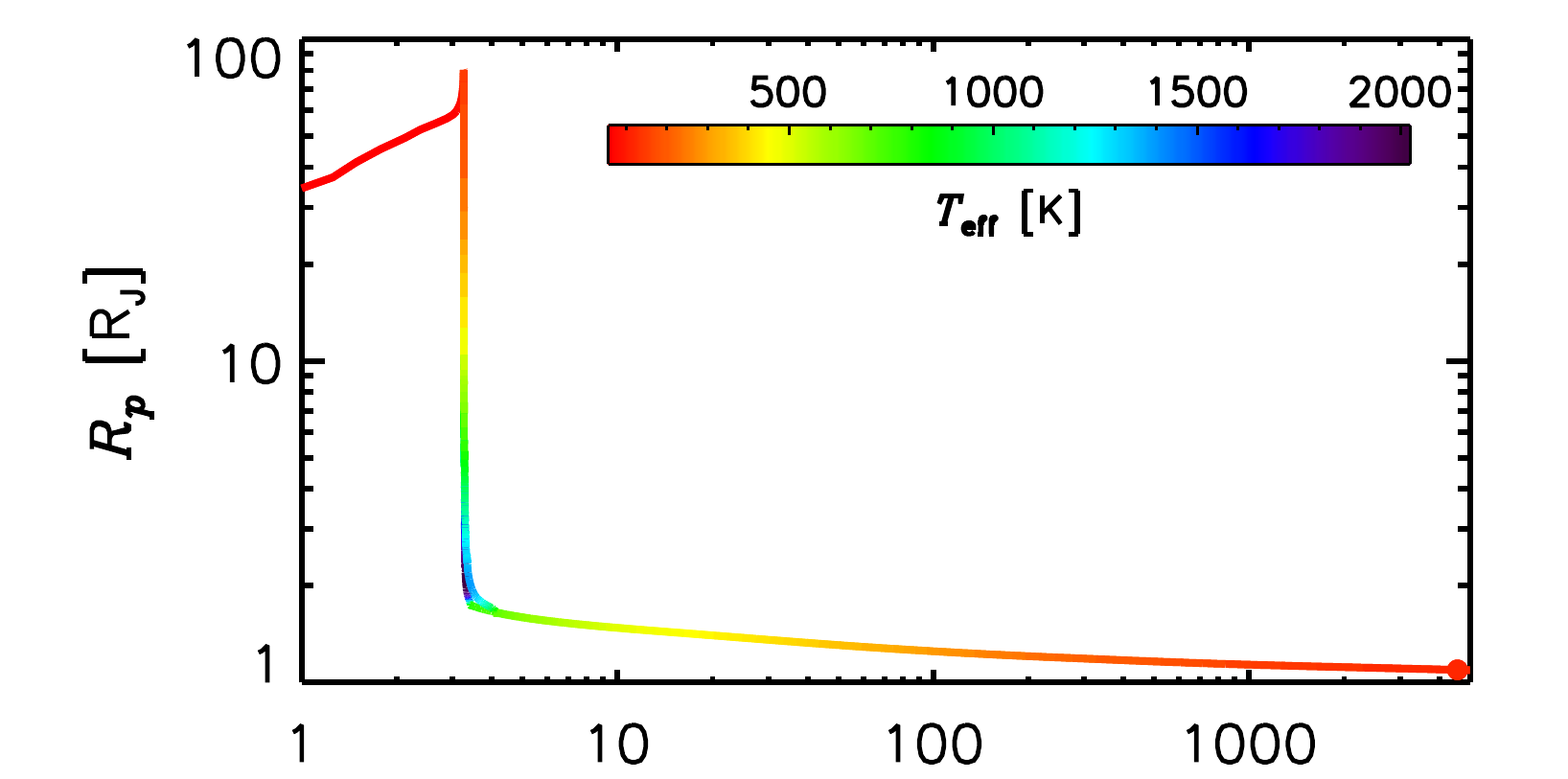}}\\
\resizebox{0.5\linewidth}{!}{%
\includegraphics{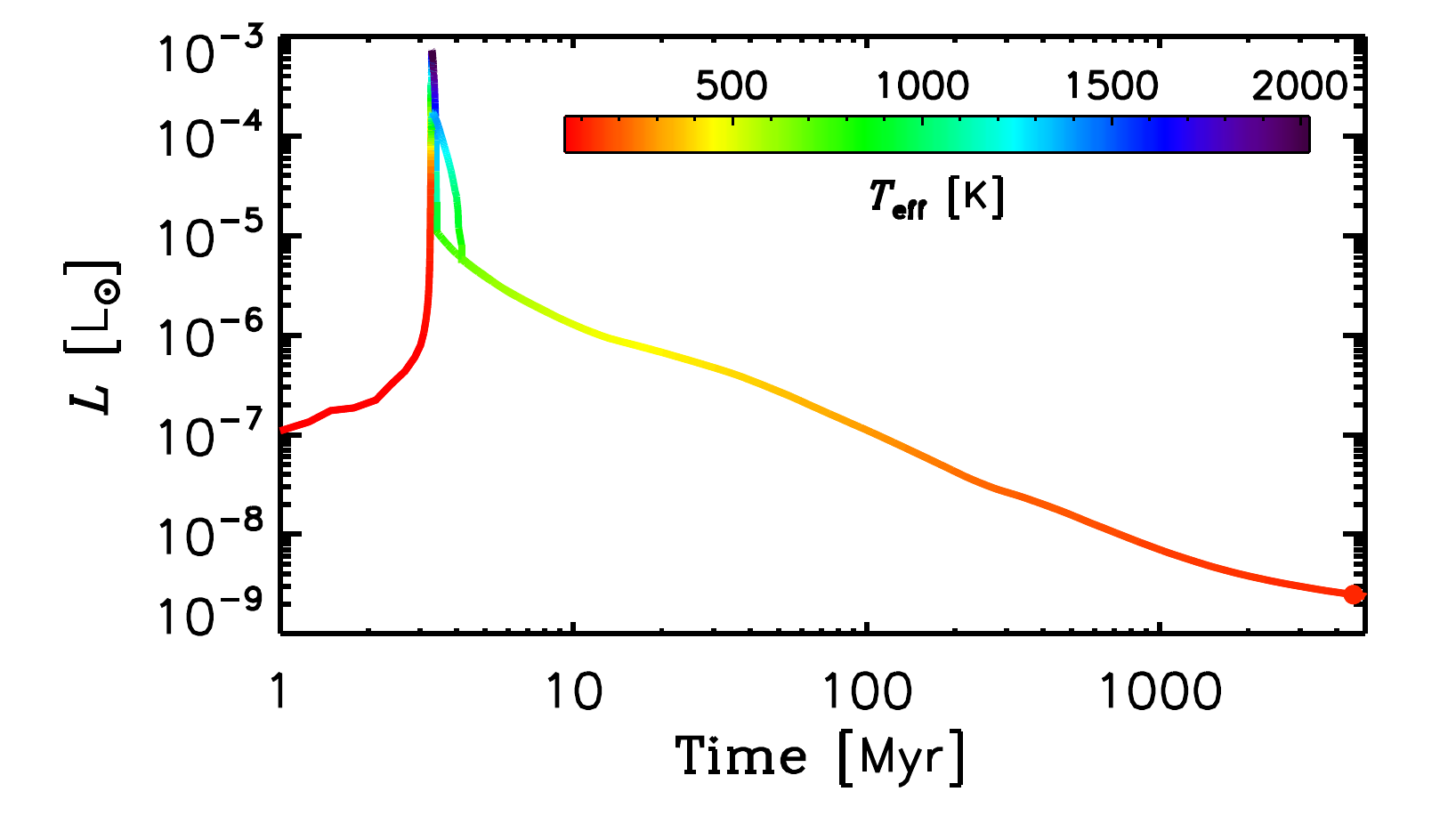}}
\end{graphicalabstract}

\begin{highlights}
\item Jupiter's formation is modeled via core-nucleated accretion
\item Solids' accretion accounts for interactions with an evolving disk of planetesimals
\item Evolution within a dispersing nebula is approximated via simplified disk models
\item Structure calculations are performed throughout the planet's history, for $5$~Gyr
\item Final radii and luminosities agree with those of Jupiter within $10$\%
\end{highlights}

\maketitle

\microtypesetup{activate=true}

\section{Introduction}
\label{sec:intro}
\defcitealias{gennaro2014}{Paper~I}

Jupiter is by far the most massive planet in the Solar System, 
and its large abundances of hydrogen and helium imply that 
it must have formed rapidly, while a significant amount of gas 
still remained within the \emph{solar nebula}, the protosolar
accretion disk. 
Although it is not known at what point of the nebula evolution 
the young Jupiter began to accrete significant amounts of gas,
the planet likely acquired most of its H/He content toward the
latter stages of the nebula life, before the gas around Jupiter's 
orbit was ultimately dispersed by thermal and magnetic winds 
\citep[e.g.,][]{armitage2020}.

Formation according to core-nucleated accretion \citep{pollack1996}
requires the assembly of a planet of at least several Earth masses
($\Mearth$) from the solids originally present in the solar nebula.
In fact, only then can the planet start collecting significant
amounts of hydrogen and helium.
In a scenario in which the core's building blocks are mostly 
planetesimals of tens of kilometers (or larger) in size, the reservoir 
of solids is nearly local, radially extending a few to several Hill 
radii from the planet's orbit \citep{lissauer1987}.
One requirement to initiate formation is then that the swarm of 
planetesimals be massive enough to allow for the rapid ``monarchical'' 
growth of a seed body \citep{weidenschilling2005}, the planetary embryo.
In this case, the growth of the embryo, which becomes the planet's core, 
can proceed until the swarm around its orbit is severely depleted.

Modeling the initial evolution of the embryo of a gas giant, and 
the subsequent accretion of solids, is challenging in several 
aspects \citep[e.g.,][and references therein]{chambers2010}. 
Previous models of Jupiter's formation used simplified methods 
for computing the accretion of heavy elements, based primarily 
on semi-analytical approximations
\citep[e.g.,][and references therein]{lissauer2009,mordasini2012a}.
In \citet[henceforth referred to as Paper I]{gennaro2014}, 
we modeled the early stages of Jupiter's growth applying a numerical/statistical
approach to calculate the accretion of solids, starting from a $350\,\mathrm{km}$
radius planetesimal evolving in an active swarm of solid bodies tens of meters 
to hundreds of kilometers in size. A number of effects were included to mimic
the evolution and interactions of the bodies within the swarm, although 
we neglected any possible radial migration of the planet due to
interactions with the swarm and the surrounding gas
\citep[e.g.,][and references therein]{rnelson2018}.

\citetalias{gennaro2014} modeled the rapid growth of the planet's core, 
up to 
$\approx 7.3\,\Mearth$, until the accretion rates of gas and solids 
were approximately equal. At that point of the evolution, only $2$ 
percent of the planet's mass was in hydrogen and helium. 
Results indicated that planetesimals larger than $10\,\mathrm{km}$ 
in radius still provided the bulk of the planet's heavy-element 
content and that the voluminous -- but initially very low-density 
gaseous envelope -- substantially increased the rate of accretion 
of solids for planet masses as small as $\approx 1\,\Mearth$.

We extend herein the simulation of the growth of Jupiter presented 
in \citetalias{gennaro2014}, from about $4\times 10^{5}$ years 
until the planet reaches one Jupiter mass,
and then evolve it forward in the presence of solar radiation until 
the present epoch. The numerical methods are described in 
\cisec{sec:methods}. \cisec{sec:results} presents the results of our 
simulations. 
We summarize the results and discuss their implications in \cisec{sec:dac}.
Our results are compared with those from some previous formation 
studies
\citep[][]{naor2010,alibert2018,shibata2019}
in \cisec{sec:dac} and with post-formation calculations 
\citep[][]{burrows1997,baraffe2003,baraffe2008}
and measurements of Jupiter's atmosphere \citep[][]{seiff1998}
in Appendix~\ref{sec:CSC},
where the effects of some model assumptions are also discussed.

\section{Numerical Methods}
\label{sec:methods}

The structure and evolution of the planet is calculated according to the methods 
described in \citetalias{gennaro2014}. 
In that study, we investigated the initial assembly of Jupiter at $5.2\,\AU$
from the Sun, in a nebula with a local surface density of solids of initial
value $\sigma^{0}_{Z}=10\,\sigu$ (see \cieq{eq:sigma0}), starting from a planetary 
embryo of $350\,\mathrm{km}$ in radius ($\approx 10^{-4}\,\Mearth$). 
The evolution was calculated for $\approx 4\times 10^{5}$ years, at which point 
the accretion rate of gas had surpassed that of solids. At the epoch when 
the two accretion rates were approximately equal, 
$\approx 3\times 10^{-6}\,\mathrm{\Mearth\,yr}^{-1}$, the mass in heavy elements 
was $7.3\,\Mearth$ and that of hydrogen and helium gas was $\approx 0.15\,\Mearth$.

In the remainder of this section we briefly describe the main components of 
the model used to simulate the formation up to Jupiter's mass, and calculate 
the subsequent evolution of the planet up to an age of 
$\approx 5\times 10^{9}\,\mathrm{years}$.
We indicate with $\Mc$ the mass of heavy elements and with $\Me$ 
the hydrogen/helium 
mass of the planet. Heavy elements are supplied by accretion of planetesimals, 
here assumed to be solid bodies larger than $30\,\mathrm{m}$ in 
diameter.
The total mass of the planet is $\Mp=\Mc+\Me$.
In these calculations, for compositional purposes, there is no dissolution 
of heavy elements, which are assumed to rain out onto a central (condensed) core.

\subsection{Structure Calculations}
\label{sec:sc}

The interior structure of the planet consists of an inner condensed heavy-element 
($Z>2$) core and an exterior envelope of hydrogen and helium.
The condensed core is assumed to be incompressible, i.e.,
\begin{equation}
    \frac{dR_{Z}}{d\Mc}=\frac{1}{4\pi\rho_{Z}R^{2}_{Z}},
    \label{eq:dRZdMZ}
\end{equation}
where $R_{Z}$ is the core radius and $\rho_{Z}$, the core density, is a constant.
The effects of this approximation are evaluated in Appendix~\ref{sec:CSC} through 
core structure calculations, by applying temperature and pressure at 
the core-envelope boundary as $\Mc$ and $\Me$ grow.

The envelope structure is computed by applying the standard equations of stellar 
structure \citep[e.g.,][]{cox2006,kippenhahn2013}, in which the gravitational
energy released by incoming (i.e., accreted) solids represents a depth-dependent
energy source.
These equations are solved as described in, e.g., \citet{pollack1996},
\citet{bodenheimer2000b}, and \citet{hubickyj2005}, with the modification 
detailed in \citetalias{gennaro2014}. 
The composition of the envelope gas is assumed to have a near-protosolar 
\citep[see][]{asplund2009} mass ratio of hydrogen and helium (mass fractions 
$X=0.74$ and $Y=0.24$, respectively), with a small admixture of heavier elements.

The structure equations are integrated from the inner boundary of the envelope, 
$R_{Z}$ (which varies according to \cieq{eq:dRZdMZ}), out to the planet radius
$\Rp$. 
In these models $\Rp$ is approximated by an effective radius $R_{\mathrm{eff}}$ 
(also referred to as ``accretion'' radius), where
$1/R_{\mathrm{eff}}\ge 1/\Rbondi+k/\Rhill$, in which $\Rbondi$ and $\Rhill$ are 
the Bondi and Hill radius of the planet, respectively. 
When the equality holds, which occurs for most of the planet formation history,
the planet is \emph{in contact} with the surrounding nebula and the density and 
temperature at $\Rp$ are those of the nebula gas. During later stages of formation,
when the planet gains most of its mass, it contracts rapidly and detaches from 
the nebula. The envelope's boundary conditions for these stages are discussed in 
\citet{bodenheimer2000b} and \citet{gennaro2016}.
For $\Mp/\Msun \ll 1$, the quantity $k$ above has a theoretical lower bound of 
$k\approx 1.415$ 
\citep[set by the volumetric mean-radius of the planet's Roche lobe,][]{eggleton1983}, 
which would be applicable in the limit of an inviscid and pressureless nebula gas. 
This quantity is expected to be an increasing function of the nebula kinematic 
viscosity and pressure scale-height. For typical nebula conditions at Jupiter's 
location, it was determined through high-resolution three-dimensional 
hydrodynamic simulations that $k\approx 4$ \citep{lissauer2009}.
For a given gas viscosity, $k$ is expected to decrease at smaller heliocentric
distances \citep{bodenheimer2018}.
Since $\Rbondi\propto \Mp$ and $\Rhill\propto M_{p}^{1/3}$, the planet radius 
is more likely governed by the Bondi radius at low planet masses and by the Hill 
radius at later evolutionary stages \citep{gennaro2013}, before the planet starts 
the phase of rapid contraction (see \cisec{sec:P3}). 
Nebula-planet tidal interactions, quantified on average by the factor $k$, 
can potentially modify this connection so that $R_{\mathrm{eff}}\approx\Rhill/k$ 
even at relatively early stages of evolution. 
Moreover, since $\Rbondi$ is also inversely proportional to the nebula temperature, 
the radius of a planet may become more closely tied to $\Rhill/k$ as 
the nebula ages and cools down, especially for planets orbiting close to their
star.

The calculation of the envelope structure includes the computation of 
\emph{a)} the trajectories and the ablation rates of the solids moving through 
the envelope \citep{podolak1988,pollack1996}; 
\emph{b)} the opacity of dust grains released in the envelope by ablating solids,
taking into account the coagulation and settling of the particles
\citep{naor2010};
and
\emph{c)} the accretion of nebular gas, including limiting accretion rates 
dictated by disk-planet tidal interactions \citep{lissauer2009,bodenheimer2013}.

The calculation of the interactions between incoming solids and envelope gas 
also provides the effective cross-section of the planet for the capture of
planetesimals, as a function of the planetesimal radius 
\citepalias[see details in][]{gennaro2014}.
Once in the envelope, solids are tracked until they break up,
are completely consumed by ablation, or reach the core. 
As solids travel through the envelope, they shed mass and release energy. 
The energy 
released at a given depth contributes to the energy budget of that envelope 
layer. The mass deposited by ablation provides the input for the dust opacity 
calculation. This mass, however, is assumed to sink to the bottom of the envelope 
and adds to the mass of the core.
The effects of height-dependent dissolution of heavy elements in the envelope, 
such as those considered by \citet{bodenheimer2018}, are not included in these
calculations.
Jupiter formation models that account for the atmospheric deposition of heavy
elements (via ablation and break-up) will be presented elsewhere.

\subsection{Accretion of solids}
\label{sec:aos}

\begin{figure*}
    \centering
    \resizebox{0.9\linewidth}{!}{%
    \includegraphics[]{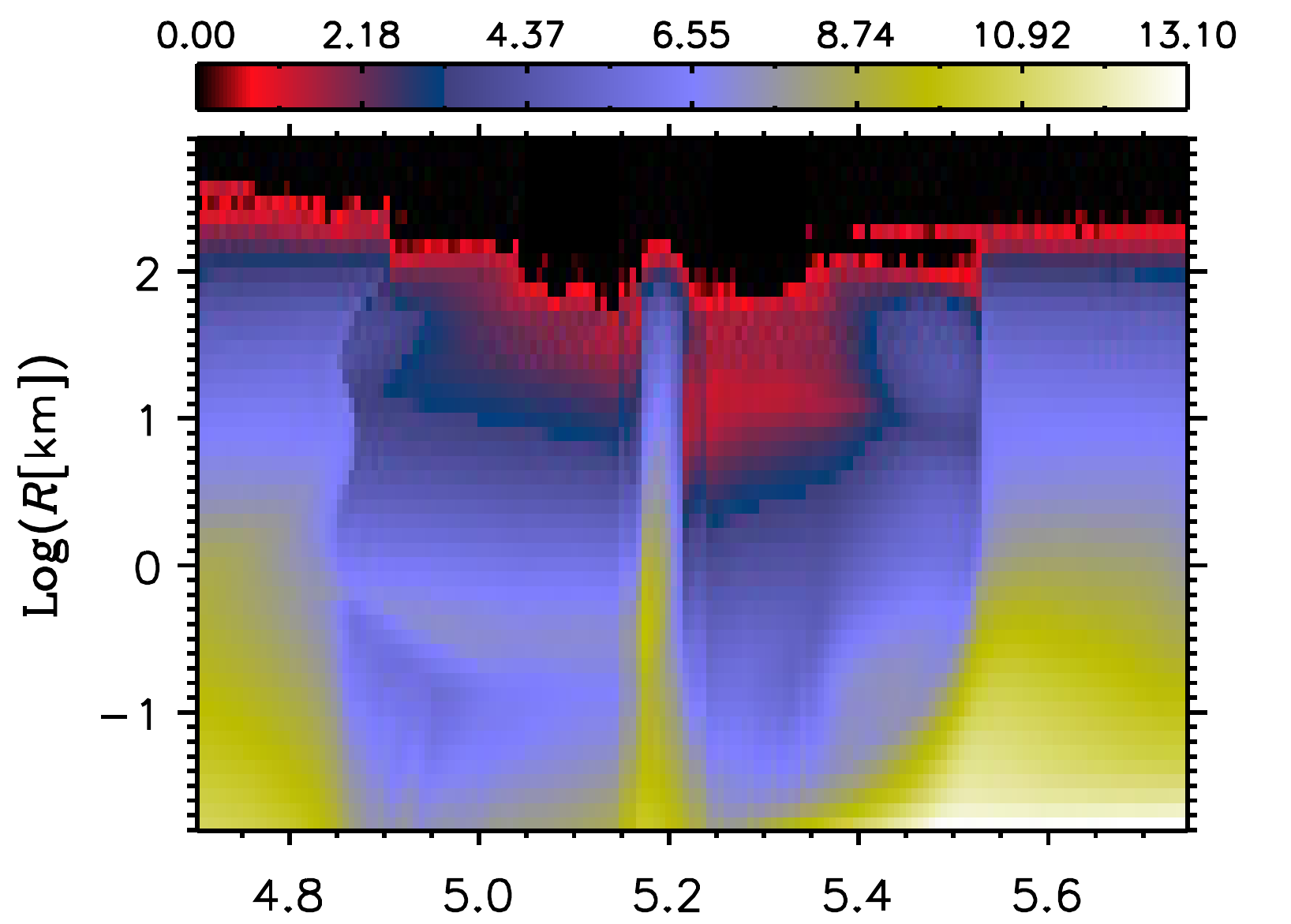}%
    \includegraphics[]{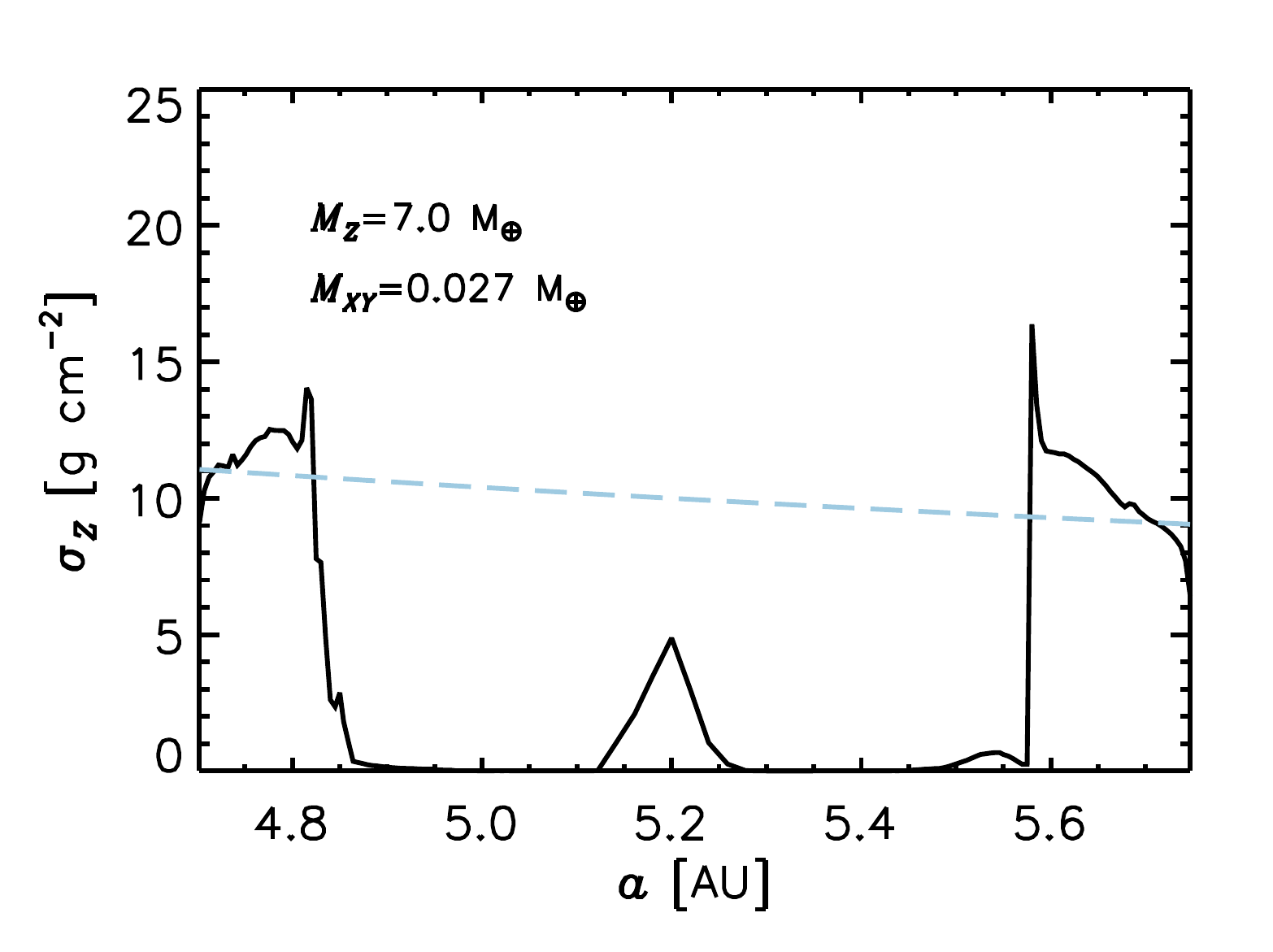}}
    \resizebox{0.9\linewidth}{!}{%
    \includegraphics[]{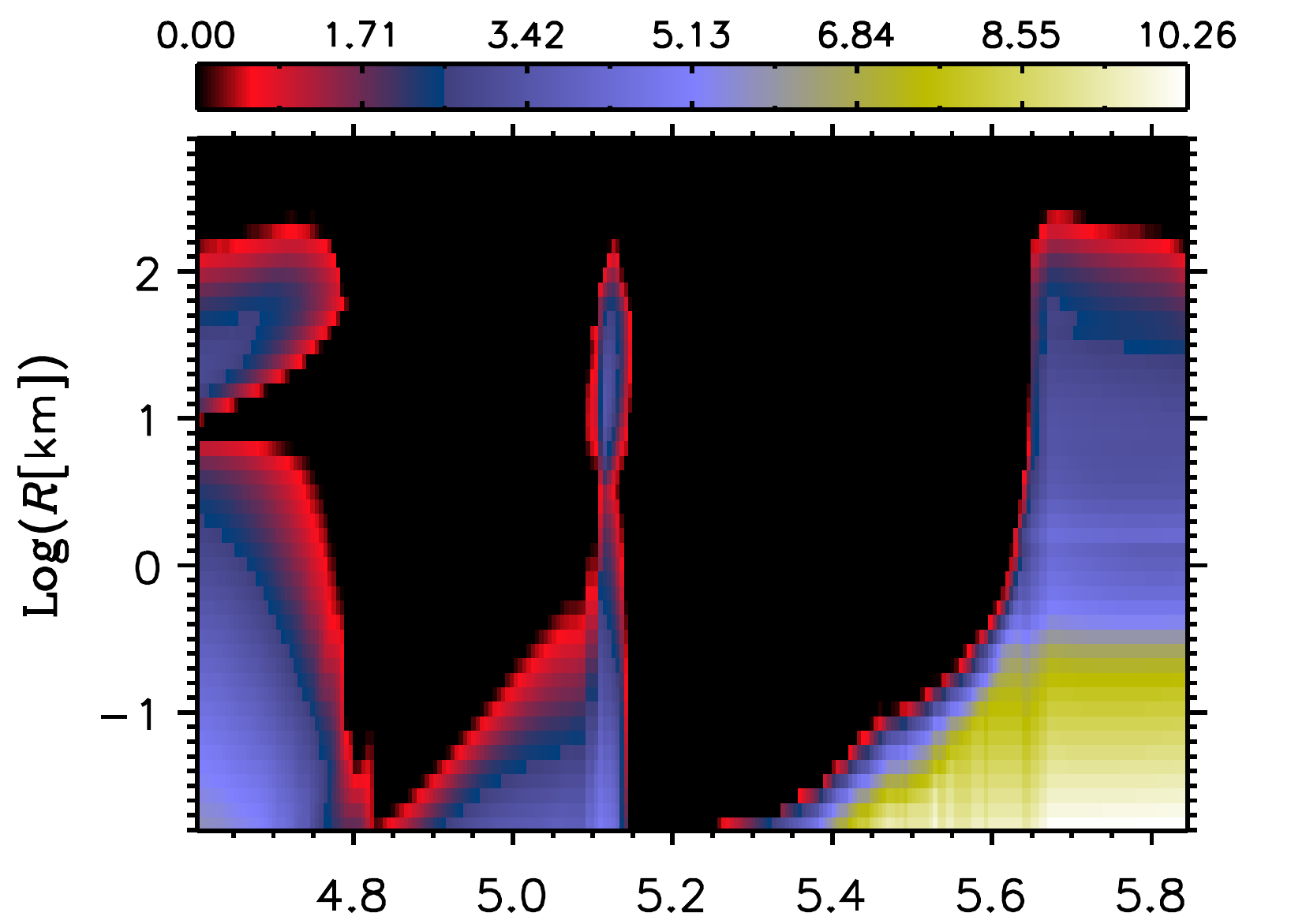}%
    \includegraphics[]{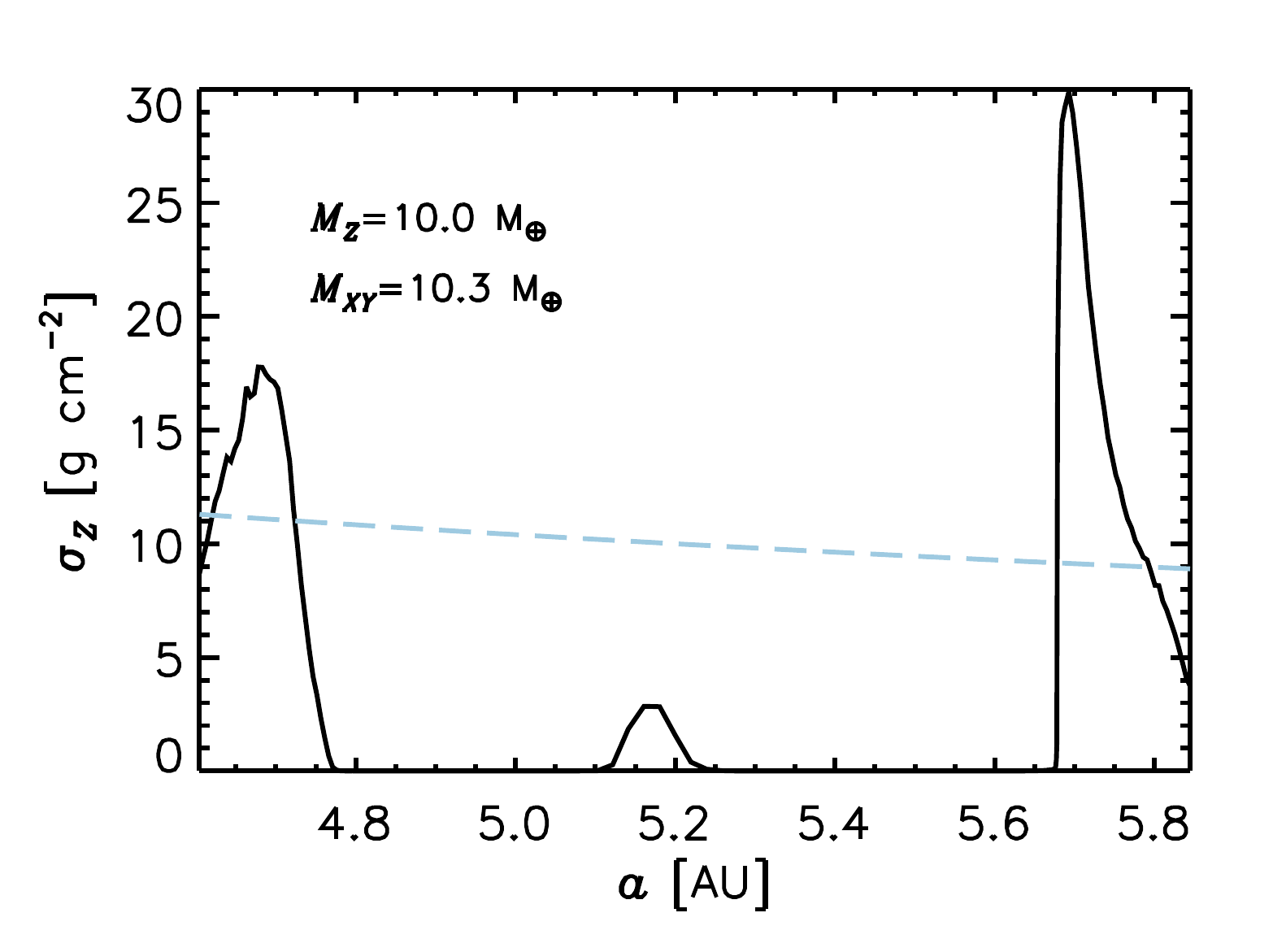}}
    \resizebox{0.9\linewidth}{!}{%
    \includegraphics[]{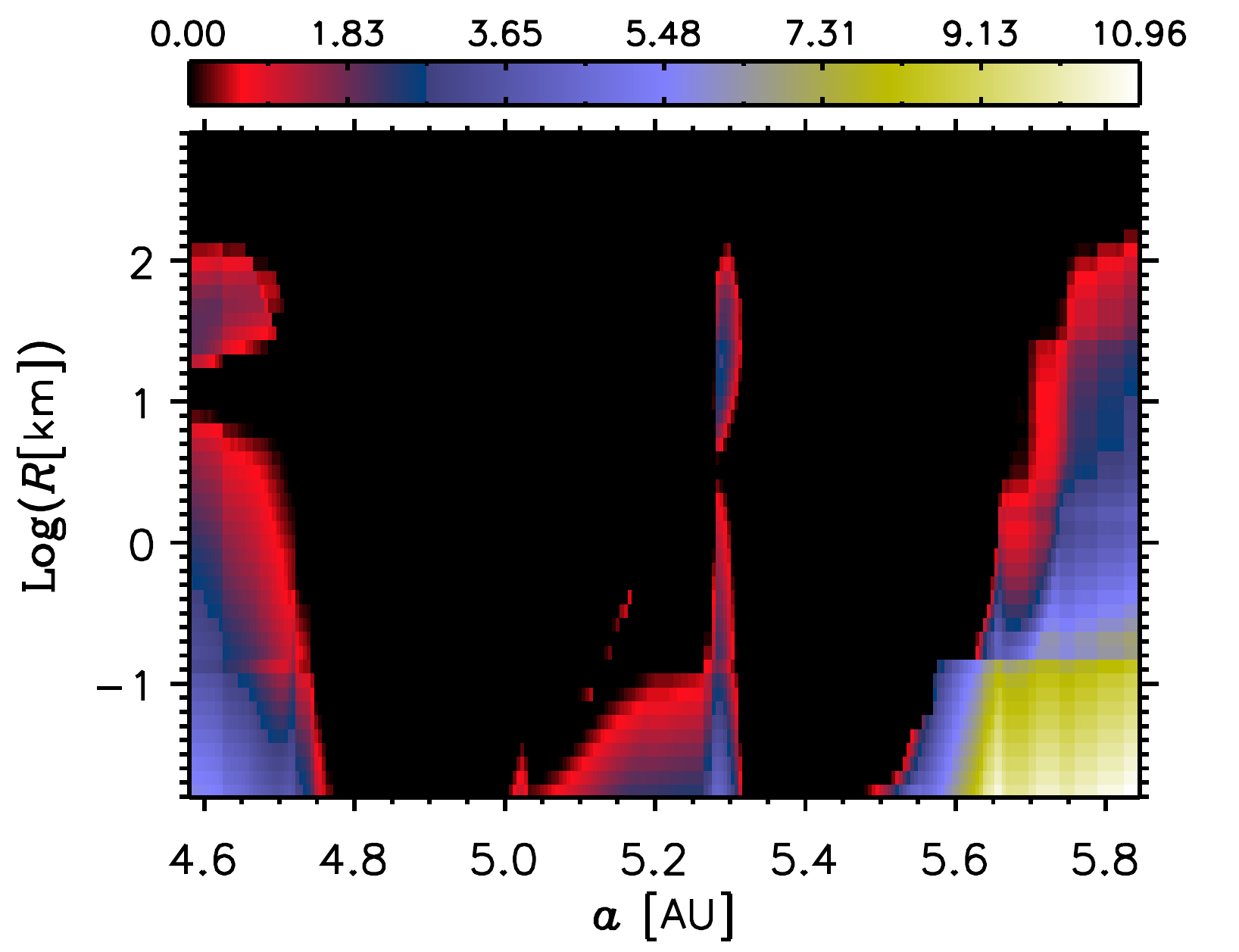}%
    \includegraphics[]{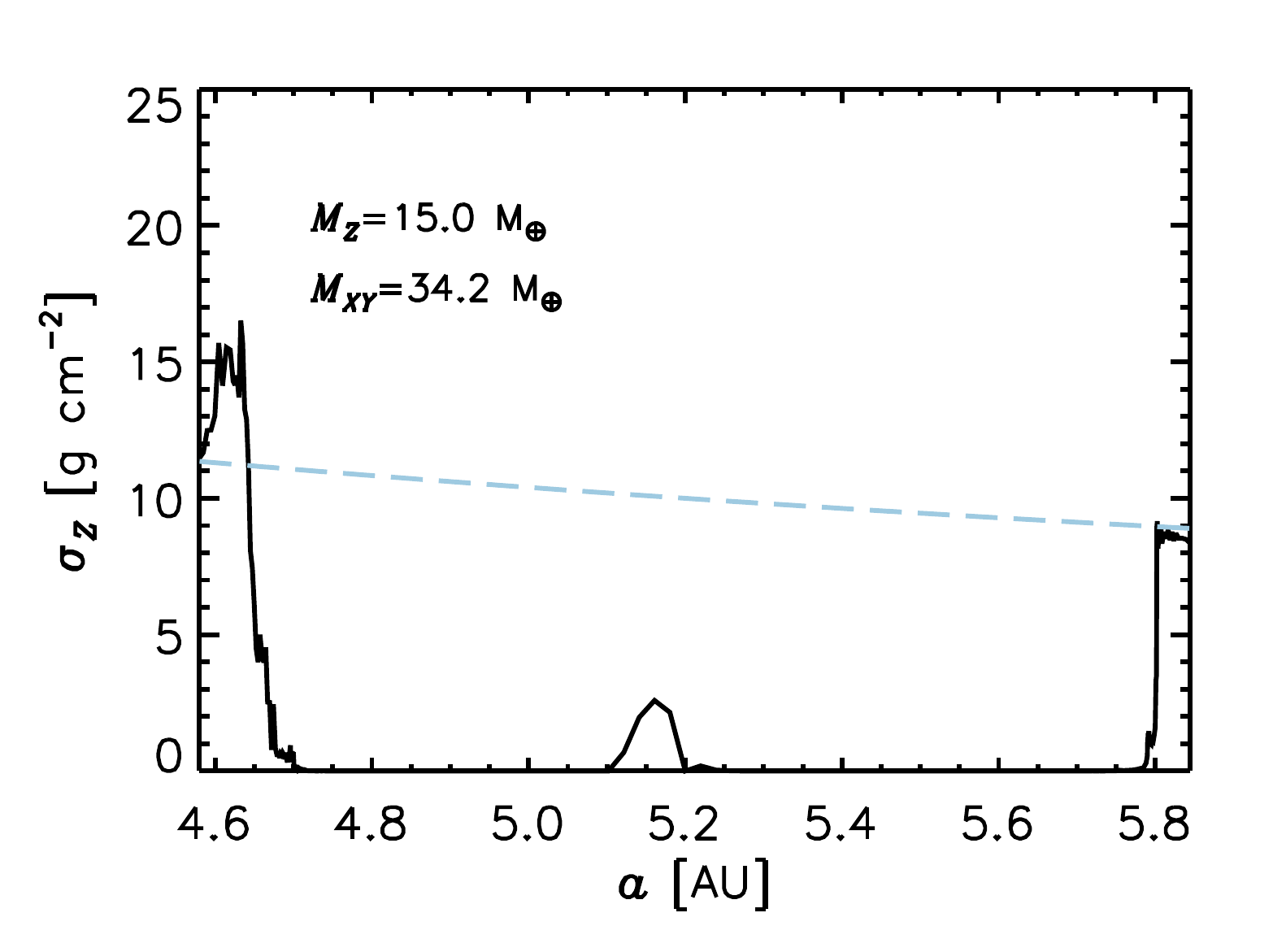}}
    \caption{
    Left: Distributions of planetesimals versus semi-major axis $a$ and radius $R$.
    The logarithm of the number of bodies per radial zone and size bin is
    color-coded at three stages of the planet's growth: $\Mc=7\,\Mearth$
    (top), $10\,\Mearth$, and $15\,\Mearth$ (bottom). The envelope mass is
    indicated in the right panels.
    The width of the radial zones varies between $0.0049$ and $0.0196\,\AU$
    in the top panel (for a total of $170$ zones) and between $0.0003$ and 
    $0.02\,\AU$ in the others (for a total of up to $250$ zones).
    Right: Surface density versus semi-major axis (solid lines) corresponding
    to the distributions of the left panels. The dashed line represents 
    the initial surface density of solids. 
    }
    \label{fig:numsig}
\end{figure*}
\begin{figure}
    \centering
    \resizebox{0.9\linewidth}{!}{%
    \includegraphics[]{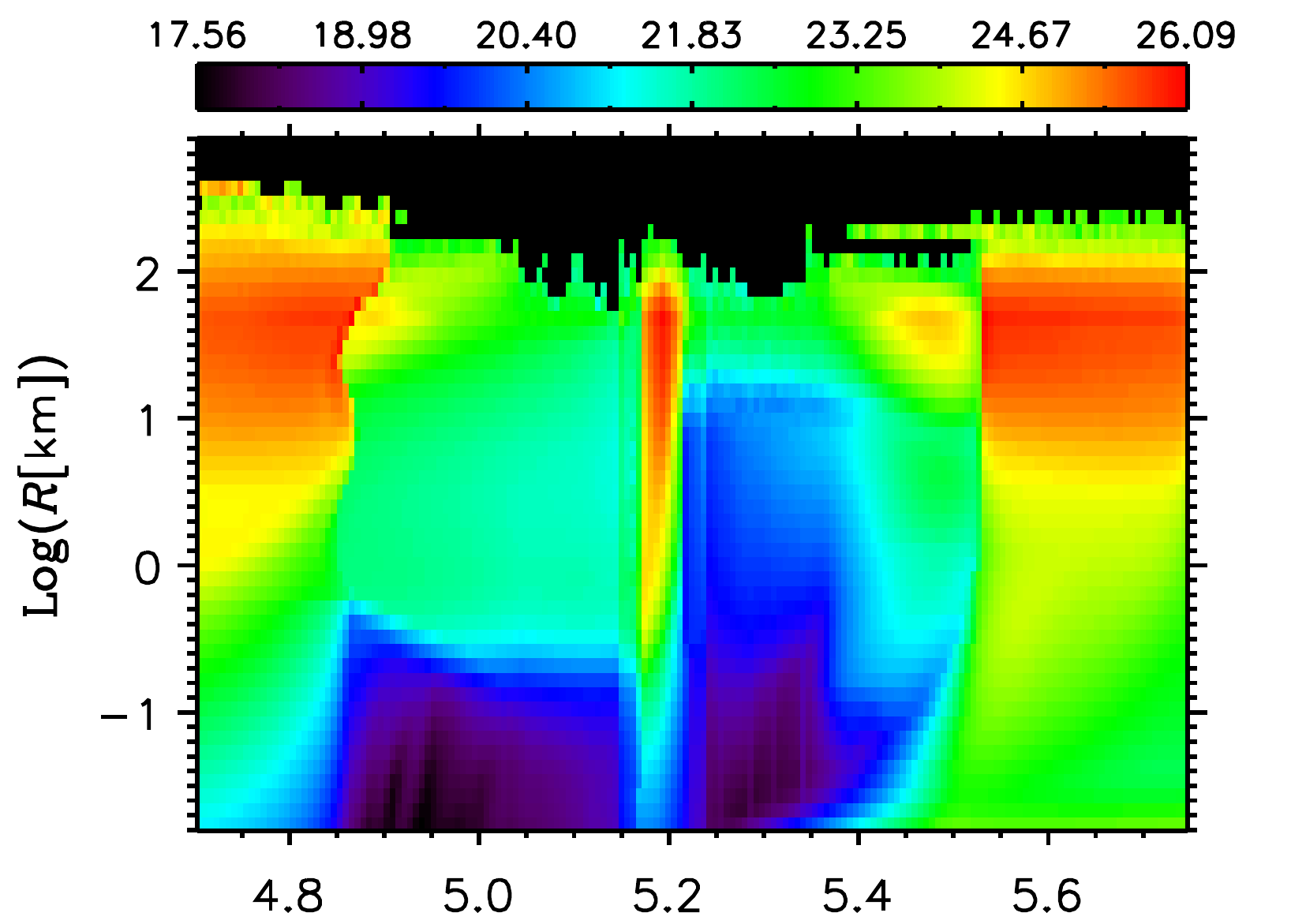}}
    \resizebox{0.9\linewidth}{!}{%
    \includegraphics[]{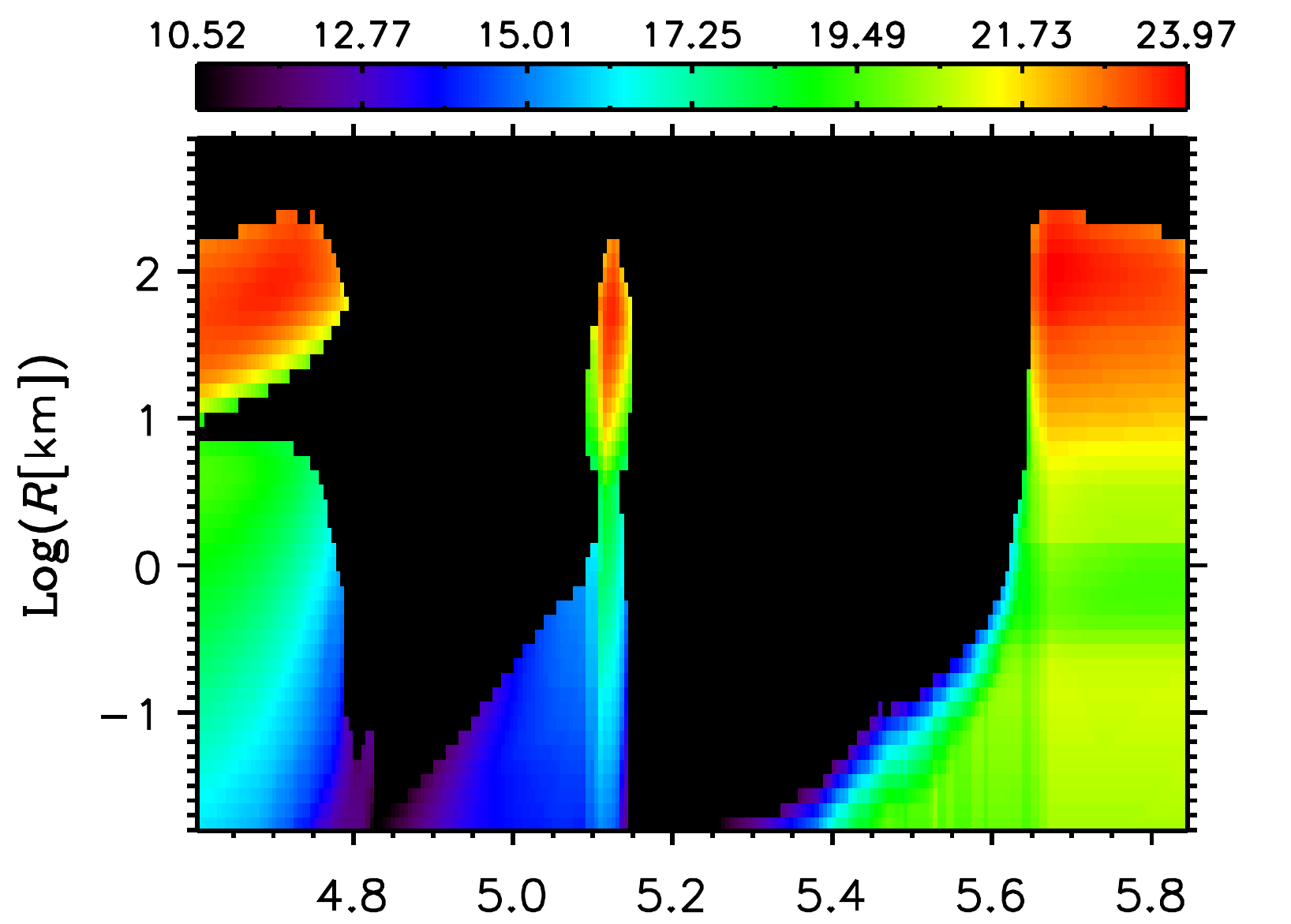}}
    \resizebox{0.9\linewidth}{!}{%
    \includegraphics[]{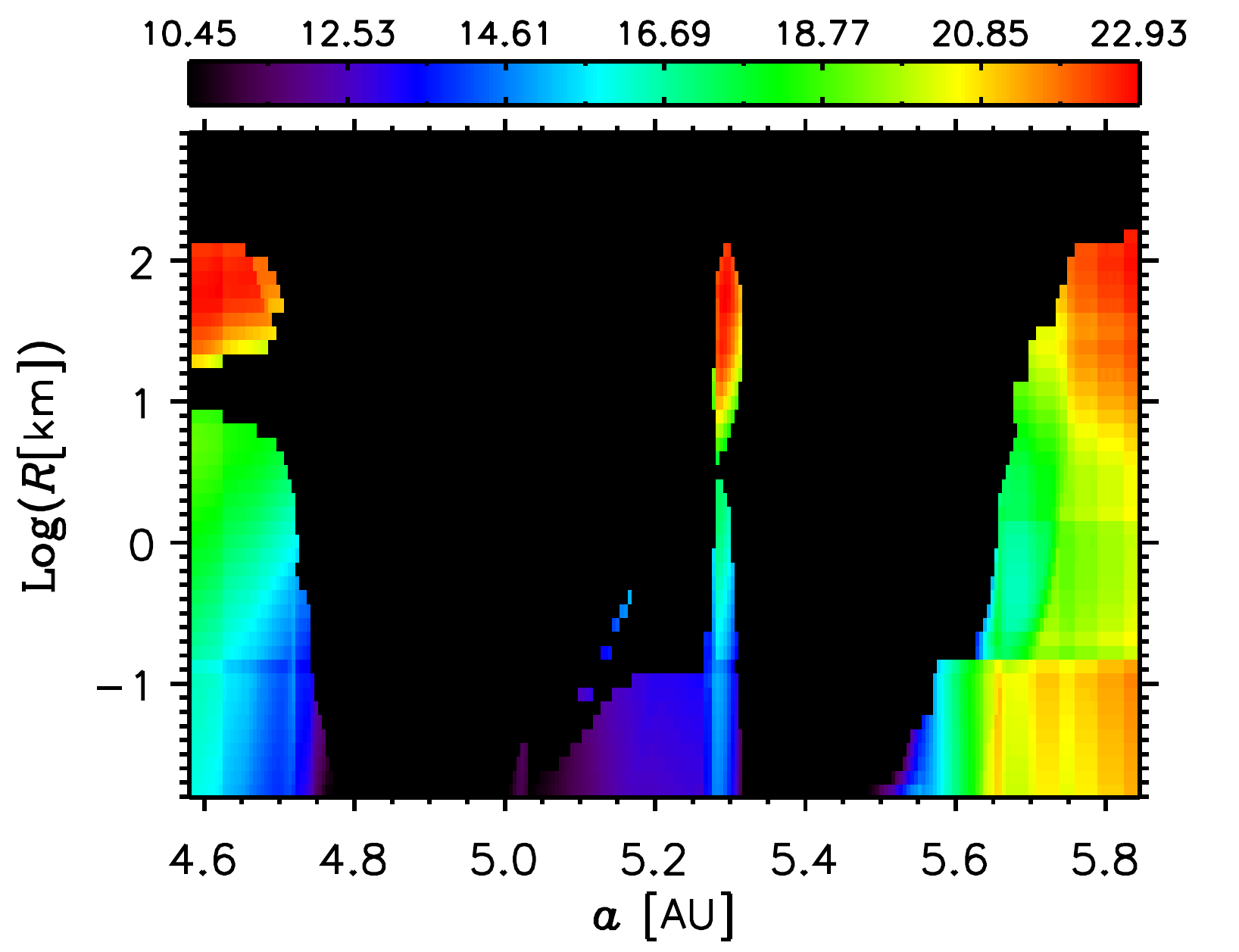}}
    \caption{
    Distributions of the solids' mass versus orbital semi-major axis and 
    planetesimal radius.
    The logarithm of the mass (in grams) is plotted for each radial zone
    and size bin at the same epochs as in \cifig{fig:numsig}, following the
    same order from top to bottom.
    }
    \label{fig:massRa}
\end{figure}
The numerical methods employed to simulate the evolution of a disk of planetesimals 
are described in \citet{weidenschilling1997} and \citet{weidenschilling2011}. 
The code computes collisional and gravitational interactions within a swarm of 
planetesimals that extends over a radial distance of several to many Hill radii 
on either side of the embryo's or planet's orbit. 
The swarm is divided into a number of radial zones. Both the number and width of
the radial zones vary as the core grows (e.g., see details in the caption of
\cifig{fig:numsig}).
In each zone, solids are distributed within a series of size bins
so that the mean planetesimal mass in two adjacent bins differs 
by a factor of two
(i.e., a factor of $\approx 1.26$ difference in radius).
Each bin is also characterized by mean values of orbital eccentricity and 
inclination, which evolve due to viscous stirring, dynamical friction, and 
gas drag.
The effects of gas drag on planetesimal orbits are implemented
according to the formulations of \citet{adachi1976},
\citet{weidenschilling1977b}, and \citet{kary1993}.
The evolution of solids also includes the process of shepherding, 
as discussed in \citetalias[][]{gennaro2014} (and references therein),
and gravitational scattering due to close encounters inside 
$\approx \Rhill$ of the growing planet.

The orbital elements of the bodies are used to detect intersecting orbits and to
compute impact probabilities. Collisions may result in erosion, fragmentation or
merging, which cause the size distribution of the solids within each zone 
to evolve over time and transfer mass between radial zones
\citep[when orbits are sufficiently altered by collisions, see details in][]{weidenschilling1997}.
Since the planetesimals' dynamics is assumed to be Keplerian
and gas drag dominates the orbital evolution of small solids, bodies smaller 
than $\approx 15\,\mathrm{m}$ in radius are removed from the calculation 
(i.e., by assumption, these small bodies do not further contribute 
to the evolution of the swarm and the growth of the planet).
Capture and concentration of bodies caused by drag forces arising from variations 
of the radial gradient of the gas pressure in the nebula are not modeled 
and neither is capture in mean motion resonances 
\citep[for further discussion, see][]{kary1993,kary1995}.
These calculations also neglect the effects of additional embryos
orbiting nearby \citep[see, e.g.,][]{levison2010}.

The planetesimal disk has an initial surface density of solids
\begin{equation}
    \sigma^{0}_{Z}=10\left(\frac{\ap}{a}\right)\,\mathrm{g\,cm^{-2}},
    \label{eq:sigma0}
\end{equation}
where $\ap=5.2\,\AU$ is the orbital radius of the planet and $a$ is 
the heliocentric distance of the solids.
The initial size distribution is a power law of the planetesimals'
radius,
$\propto R^{-11/6}$, with $R$ (initially) ranging from 
$\approx 15\,\mathrm{m}$ to $50\,\mathrm{km}$.
This slope is characteristic of collisional equilibrium, 
and while the size distribution is allowed to evolve, 
the slope remains close to the initial value during 
the simulations.
Accretion tends to deplete the swarm on both sides of the planet's orbit, 
over an average radial distance $b\,\Rhill$.
The depleted region is referred to as the planet's feeding zone for 
the accretion of solids.
In the calculations the value of $b$ is about $3.5$ when $\Mp\approx 7\,\Mearth$,
decreasing to $2.8$ when $\Mp\approx 49\,\Mearth$ (see \cifig{fig:numsig}). 
At earlier epochs, the region is depleted in solids but not entirely empty
\citepalias[see][]{gennaro2014}.
Gravitational shepherding and gas drag produce a mass pileup 
at the edges of this region, as illustrated in \cifig{fig:numsig}. 
In order to provide higher spatial resolution near the edges, 
the zone width is adjusted. 
This is done by subdividing a zone into two, each with half the
width, and the same surface density $\sigma_{Z}$ at the moment of
division, after which they evolve separately. This procedure is
applied sequentially, as required, and allows us to resolve 
radial 
features using zones as narrow as needed ($< 0.001\,\AU$).
As a reference, in the calculations the largest zone width at the boundaries
between depleted and undepleted regions can be a few percent of $\Rhill$, 
but it is typically much smaller.

Assuming that depletion occurs only via accretion on the planet and that there 
is no significant re-supply of solids to the region, the heavy-element content
of the planet (i.e., the core mass in our case) is determined by the amount of 
solids initially contained in the region \citep{lissauer1987}
\begin{equation}
    \Mc\approx 4\pi b a^{2}_{p}\sigma^{0}_{Z}\left(\frac{\Mp}{3\Msun}\right)^{1/3},
    \label{eq:MZiso}
\end{equation}
where $\sigma^{0}_{Z}$ refers to the initial surface density of solids 
at the planet's orbit, $a=\ap$. 
When $\Me\ll \Mc$, the mass of heavy elements is proportional to 
$a^{3}_{p}(b\sigma^{0}_{Z})^{3/2}$ and equal to $\approx 9\,\Mearth$
for $b=3.5$ (see top panels of Figures~\ref{fig:numsig} and \ref{fig:massRa}).
The quantity $b$ is somewhat smaller than the classic value of $4$
\citep{kary1994} for the reasons discussed in \citetalias{gennaro2014}.
Accounting for solids trapped in Trojan-like orbits, the difference in mass
is about $20$\%.
When $\Mp\approx 20\,\Mearth$ ($b\approx 3$), \cieq{eq:MZiso} predicts
$\Mc\approx 10\,\Mearth$, in reasonable accord with the calculation 
(see \cifig{fig:numsig}).
In general, and under the assumption of a sufficient reservoir of heavy elements, 
$\Mc\propto \Rhill$, i.e., $\Mc\propto \Me^{1/3}$ when $\Me\gg \Mc$.
Nonetheless, even if the disk of planetesimals were to extend to infinity,
\cieq{eq:MZiso} predicts that the incremental gain in heavy elements
decreases as the planet mass grows:
\begin{equation}
    \frac{d\Mc}{d\Mp}\approx \frac{4}{3}b\pi%
    \left(\frac{a^{2}_{p}\sigma^{0}_{Z}}{\Mp}\right)%
    \left(\frac{\Mp}{3\Msun}\right)^{1/3}.
    \label{eq:dMZdM}
\end{equation}
Other effects, such as 
gravitational scattering and competing depletion by nearby planets
or planetary embryos, are likely important and can reduce the supply of
solids available for accretion.
It should be noted, however, that in the calculations presented here 
\cieq{eq:MZiso} 
does not strictly apply because of the additive effects of drag forces 
and interactions among planetesimals.

The simulation was initiated with a seed body of about $10^{-4}\,\Mearth$,
large enough compared to neighboring planetesimals in the swarm to allow for 
runaway growth. Within $\approx 2\times 10^{4}$ years, its mass grows by 
over a factor of $10$. The gravitational perturbations of this body soon 
become sufficiently large to stir orbital eccentricities in the surrounding 
region and prevent runaway growth of potential competitors, the largest 
of which reach sizes of a few to several hundred km in radius.

The initial growth of the embryo assumes that solids can directly impact
its surface. In fact, at $\approx 5\,\AU$, the radius of the initial seed 
is much larger than the Bondi radius (assuming a nebula temperature of 
$\approx 120\,\K$), and the presence of a tenuous atmosphere (which can 
increase solids' accretion) is not expected until 
$\Rbondi \gg R_{Z}$, when $\Mc\gg 0.01\,\Mearth$. At that point, however,
the swarm is still largely undepleted and $d\Mc/dt$ is already relatively 
large, a few times $10^{-5}\,\Mearth\,\mathrm{yr}^{-1}$.
The envelope structure is modeled starting from $\Mc\approx 1\,\Mearth$.

Ongoing accretion of solids and shepherding effects 
\citep[see, e.g.,][]{greenberg1983}
generate a gap in the swarm along the orbit of the planet, as can be seen 
in \cifig{fig:numsig}. The figure shows the number of bodies for each zone
(horizontal axis) and size bin (vertical axis) in the left panels and 
the surface density of solids versus semi-major axis in the right panels, 
at different epochs, as indicated by the core and envelope masses in the legends.
Examples of the same plots at lower masses, reported in \citetalias[][]{gennaro2014},
indicate that the depletion of solids in the gap region is already significant 
at $\Mp\approx 1\,\Mearth$. As shown in \cifig{fig:numsig}, this region 
becomes severely depleted or virtually empty for $\Mp\gtrsim 7\,\Mearth$.
Gravitational perturbations by the core and drag forces tend to increase 
the density of solids around the edges of the gap, especially exterior 
to the orbit (see right panels), as can be noticed by comparing with the
initial distribution (dashed line). Also note that the gap is narrower
for smaller-size bodies (left panels).
\cifig{fig:massRa} shows the mass distribution of the planetesimals,
as a function of orbital radius and planetesimal radius, for the same
cases as in \cifig{fig:numsig}. The figure indicates that planetesimals 
of several tens of kilometers in radius typically carry most of the mass
in the swarm.

The middle and bottom panels of \cifig{fig:massRa} also show a small 
reservoir of solids in Trojan-like orbits, most of which are present 
from the beginning of the evolution, trapped there during the initial 
runaway growth of the core. These bodies cannot collide with the core 
and can escape only if the orbital eccentricity increases because of 
collisions or encounters with bodies other than the core. Gas drag may 
cause these solids (particularly the small ones) to drift in semi-major 
axis and allow them to escape. Drag-induced decay of orbits is visible 
in the middle and bottom panels of the figure, especially for 
$R < 1\,\mathrm{km}$. 
It is unclear whether the bodies in Trojan-like orbits, visible in 
the figure, may survive much longer after Jupiter's formation;
the present-day Trojans were possibly trapped much later
(e.g., \citeauthor{morbidelli2005}, \citeyear{morbidelli2005};
\citeauthor{nesvorny2013}, \citeyear{nesvorny2013}; and review
by \citeauthor{slyusarev2014}, \citeyear{slyusarev2014},
and references therein)

\begin{figure*}
    \centering
    \resizebox{0.9\linewidth}{!}{%
    \includegraphics[]{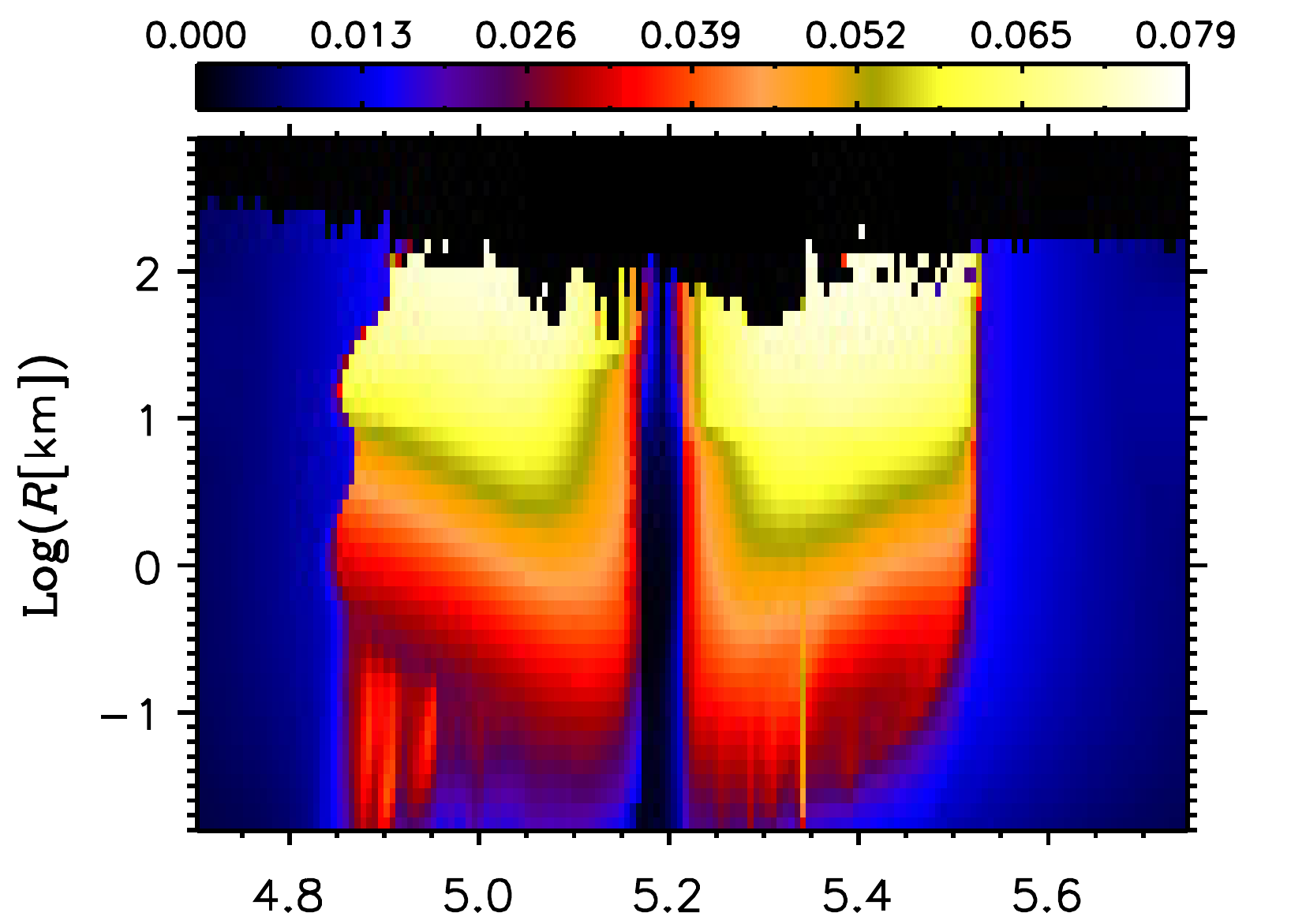}%
    \includegraphics[]{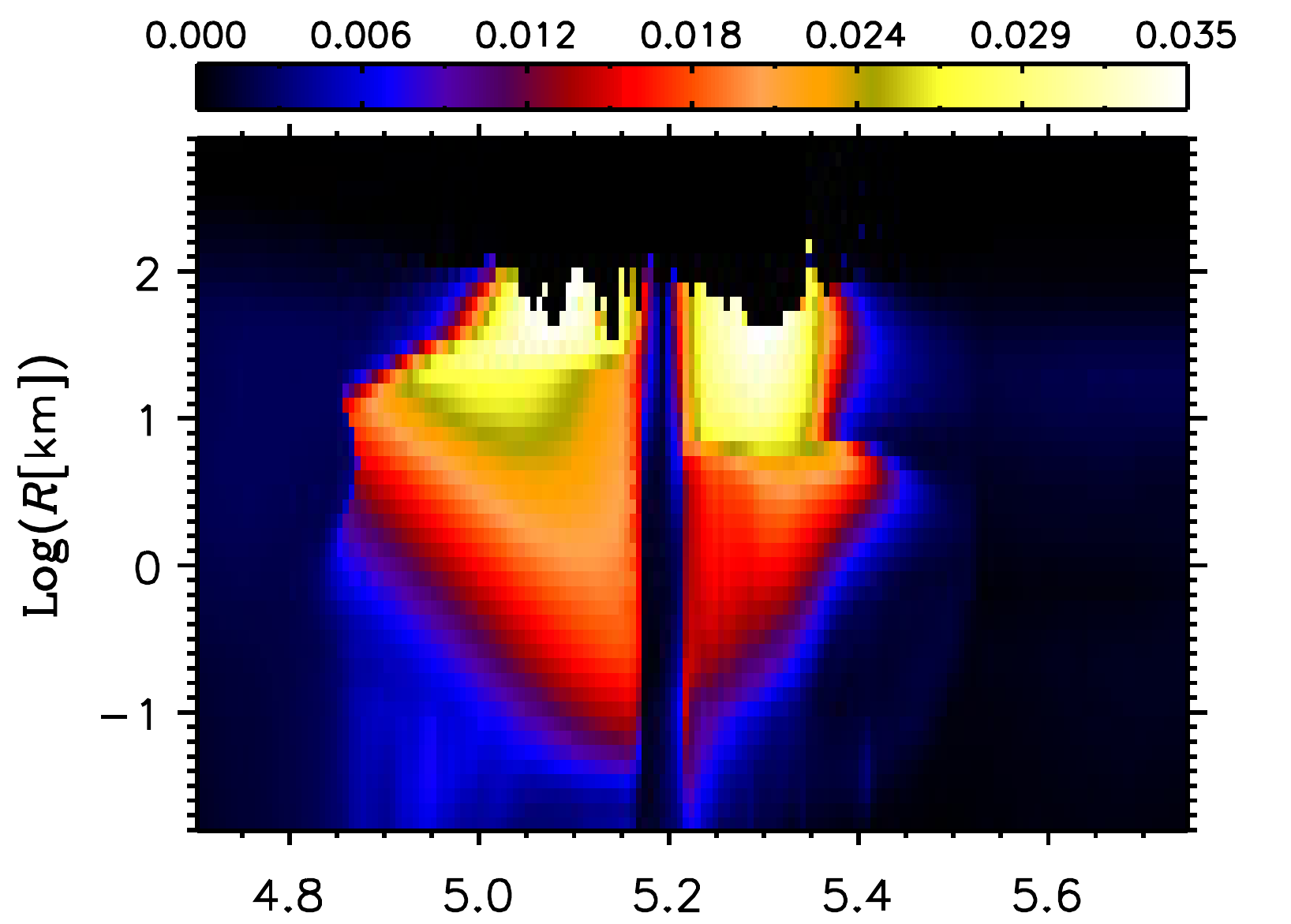}}
    \resizebox{0.9\linewidth}{!}{%
    \includegraphics[]{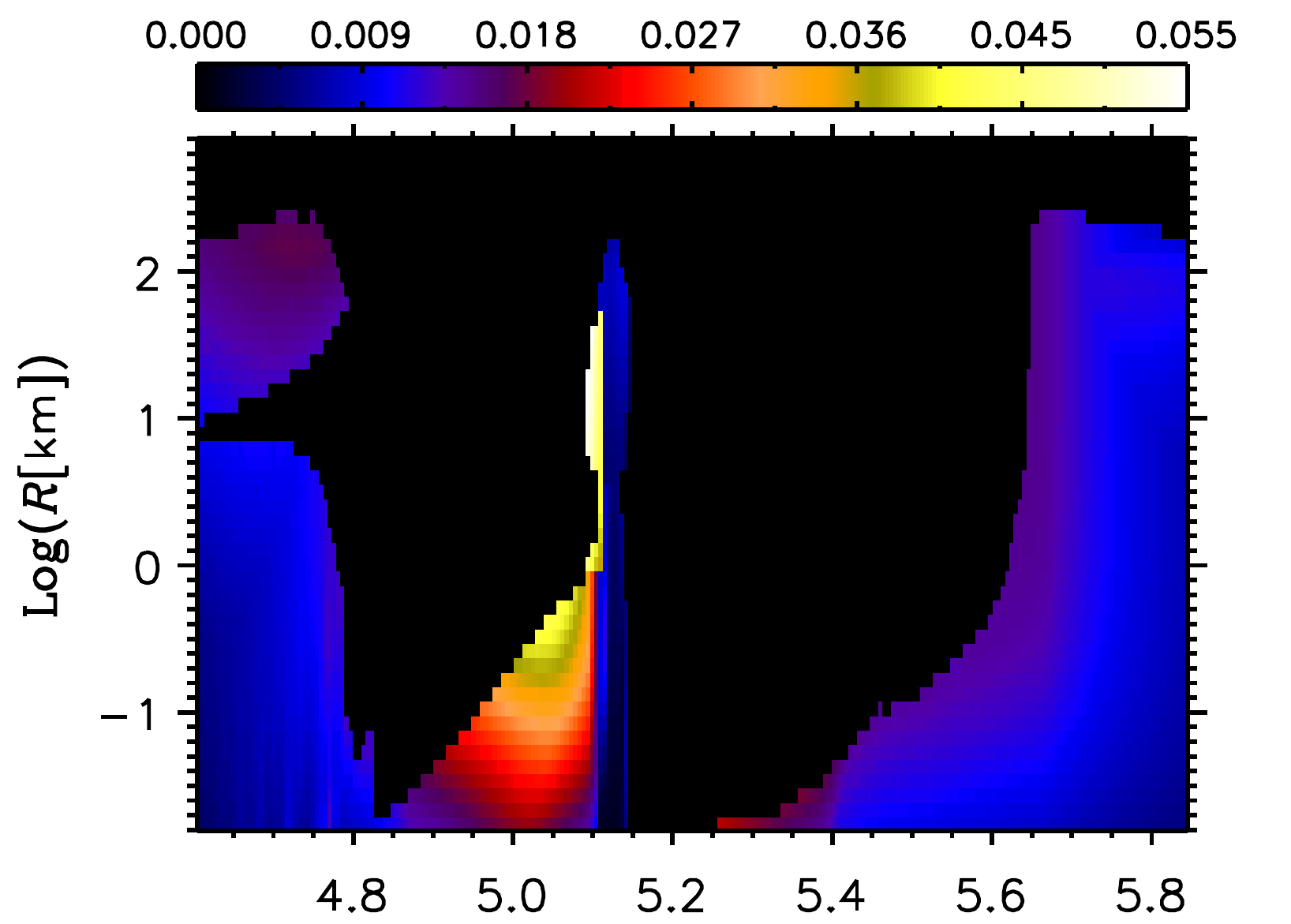}%
    \includegraphics[]{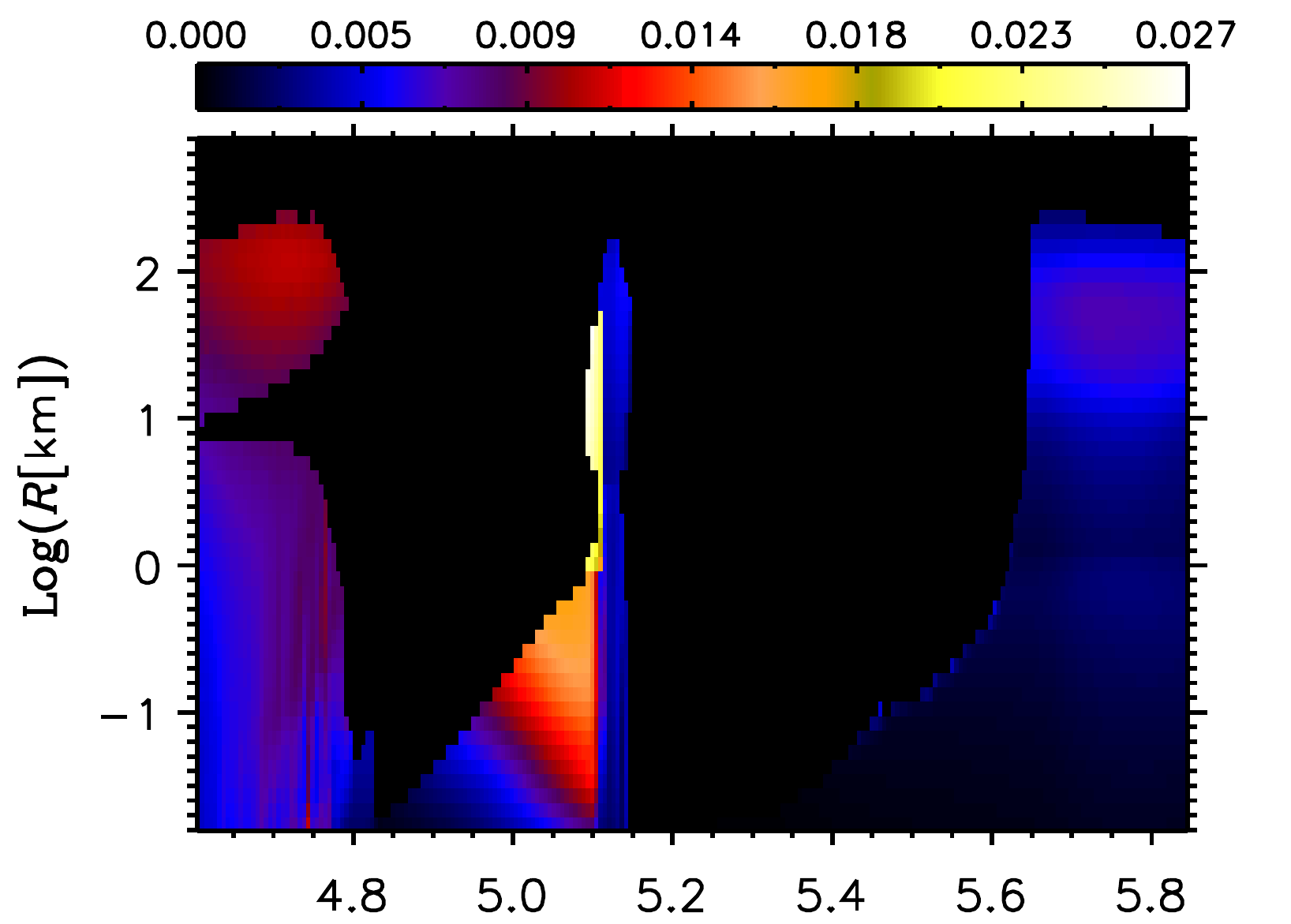}}
    \resizebox{0.9\linewidth}{!}{%
    \includegraphics[]{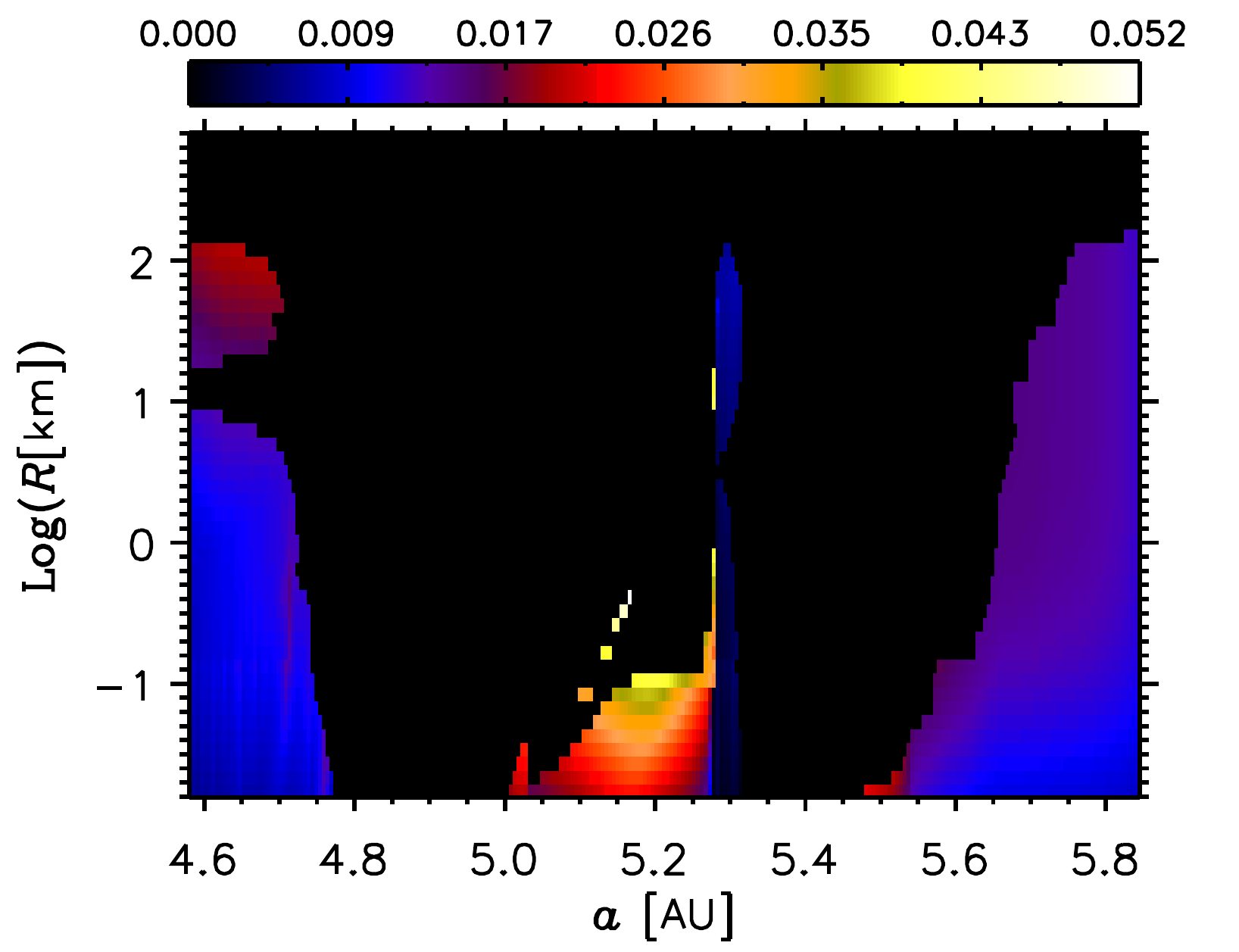}%
    \includegraphics[]{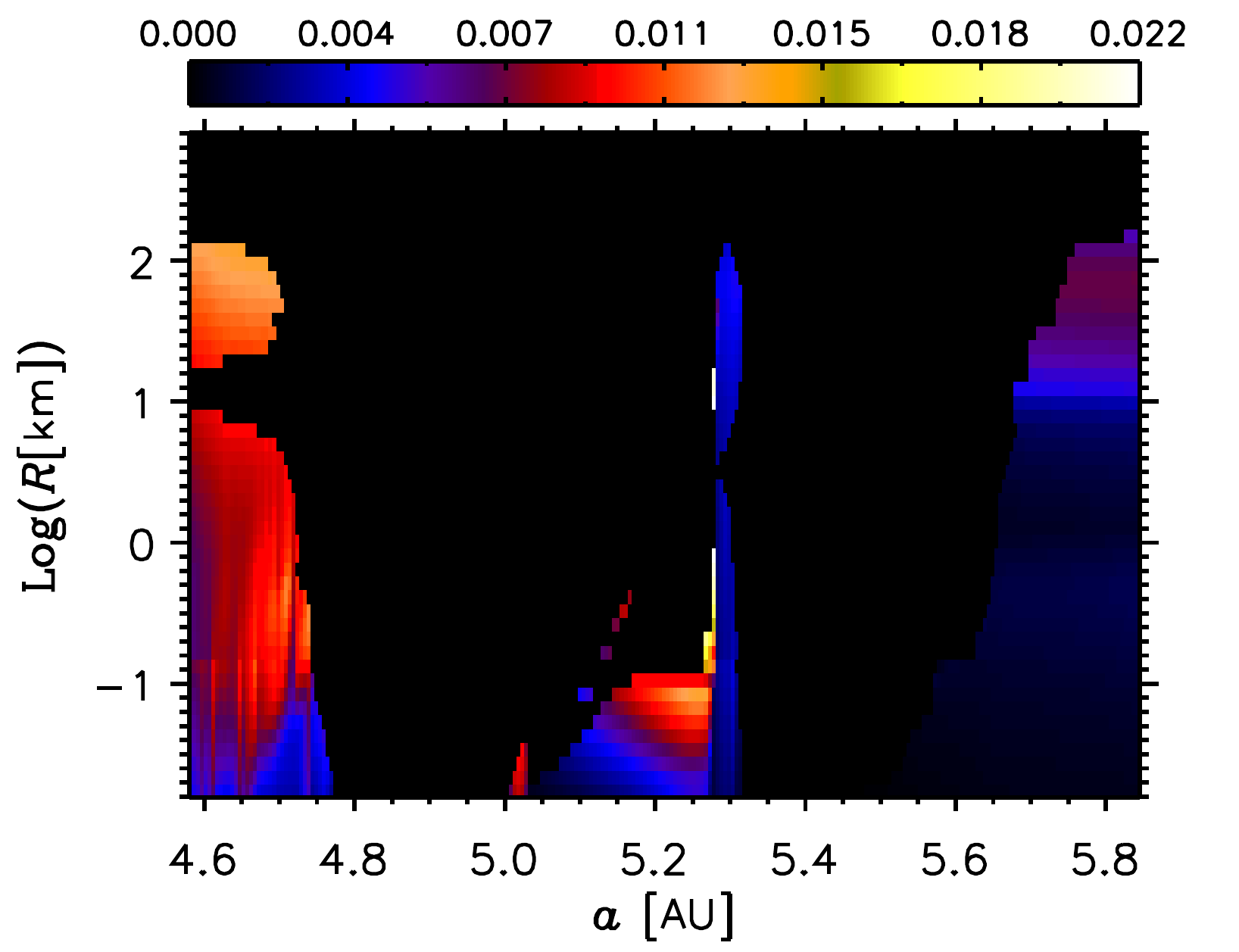}}
    \caption{
    Distributions of the mean orbital eccentricities (left) and of the sine 
    of the mean orbital inclinations (right) of planetesimals as a function 
    of semi-major axis and radius. The distributions are plotted at the same
    times as in \cifig{fig:numsig}, when $\Mc=7\,\Mearth$
    (top), $10\,\Mearth$ ($\Mp\approx 20\,\Mearth$), and $15\,\Mearth$ 
    ($\Mp\approx 49\,\Mearth$, bottom).
    }
    \label{fig:eccicc}
\end{figure*}
At the same epochs as in the two previous figures, \cifig{fig:eccicc} shows
distribution maps of mean eccentricity (left) and inclination (right) of the solids.
The swarm remains relatively flat since stirring mostly occurs in the orbital
plane of the planet, hence exciting orbital eccentricity more than orbital 
inclination (see color bar ranges in \cifig{fig:eccicc}). This is also the
case when the planet has lower mass.

\subsection{Accretion of gas}
\label{sec:AoG}

The accretion of gas is dictated by envelope contraction for most of the 
planet's formation history. During this time, contraction is regulated 
by cooling, which depends on the ability of the outer envelope to radiate 
away the gravitational energy released by contraction and accretion of solids. 
In fact, the accretion of gas is correlated to the accretion of solids 
until the envelope mass exceeds the core mass (see \cisec{sec:results}).
Moreover, the lower the opacity is in the outer envelope gas, the faster
contraction may proceed 
\citep[see, e.g.,][ and references therein]{hubickyj2005,naor2010}.
Here, the accreted gas is assumed to have solar abundances.

Once $\Me\gtrsim\Mc$ and gravitational contraction starts to dominate 
the energy generation in the envelope, the contraction speeds up
and $d\Me/dt$ may grow independently of $d\Mc/dt$. At this point, 
the only limit to growth is set by the available supply of nebula gas.
When tidal interactions between the planet and the disk are negligible, 
the sustained (i.e., non-transient) accretion rate on the planet would be 
limited to the accretion rate through a steady-state accretion disk 
\citep{lynden-bell1974,pringle1981}, which is proportional to the kinematic 
viscosity, $\nu$, and surface density of the gas, $\Sigma$. 
In young, planet-forming nebulae surrounding solar-mass stars, this rate 
is likely $\gtrsim 10^{-2}\,\Mearth\,\mathrm{yr}^{-1}$, although it can be
much smaller in old disks. Therefore, during most of the planet's evolution, 
the rate of supply of gas from the nebula is generally significantly larger 
than the rate of envelope accretion dictated by contraction.
But the situation changes as $d\Me/dt$ grows, the rate of supply by the nebula
declines over time, and disk-planet tidal interactions alter the dynamics
of the gas flux toward the planet.

Gap formation in a gaseous nebula is a consequence of tidal interactions 
and occurs (around a planet's orbit) when the tidal torques exerted by 
the planet's gravity overcome viscous torques exerted by adjacent disk rings 
\citep[e.g.,][ and references therein]{ward1997,lubow2011}.
Applied to a solar-mass star, the condition reads
\begin{equation}
    \left(\frac{\Mp}{\Msun}\right)^{2}\gtrsim 3\pi%
    \left(\frac{\nu}{a^{2}_{p}\Omega}\right)\left(\frac{H}{\ap}\right)^{3},
    \label{eq:gap_con}
\end{equation}
where $H$ is the nebula pressure scale-height and $\Omega$ the local orbital
frequency. In a nebula whose temperature is $\approx 120\,\K$ (which 
determines $H$) and whose viscosity corresponds to an accretion
rate on the star of $\approx 10^{-7}\,\Msun\,\mathrm{yr}^{-1}$ \citep[e.g.,][]{natta2006},
gap formation requires a planet mass $\Mp\gg 10^{-4}\,\Msun$, or several to many
tens of Earth's masses. The flux of gas toward a planet's feeding zone
starts to decline as the gap deepens and widens. The complexity of the accretion
flow, which is three-dimensional in nature even at much smaller masses
\citep{gennaro2013}, does not allow simple estimates.
Disk-limited rates have been evaluated through multi-dimensional hydrodynamic
calculations
\citep[see, e.g.,][ and references therein]{lissauer2009,bodenheimer2013}.
They rapidly decrease as the planet mass increases and, in general, they 
are lower for cooler and less viscous disks \citep[see][]{gennaro2008}.
These rates are applied to our spherically symmetric model; thus,
the radiated luminosity from behind the accretion shock (assumed 
to form where the infalling material hits the surface) is approximately
given by  $G\Mp \dMe/\Rp$, and is close to the total luminosity
of the planet.

Therefore, once Jupiter started its phase of rapid growth, its evolution
likely became tied to that of the surrounding solar nebula. And since 
the supply of gas to a Jupiter-mass planet is non-negligible (under
a broad range of conditions), Jupiter continued to accrete until the
surrounding nebula eventually faded away.

\section{Results}
\label{sec:results}

Adopting a standard terminology (under the approximations applied here),
we denote by ``Phase~1'' the evolution of the planet up to the time when
$d\Me/dt=d\Mc/dt$ and by ``Phase~2'' the evolution from the end of Phase~1 
to the time the crossover mass is attained, $\Me=\Mc$. From then up until 
the time the surrounding disk dictates the accretion of gas, the evolution 
is denoted as ``Phase~3''.

Previously, we divided Jupiter's growth into these three phases, 
based on the results 
presented in \citet{pollack1996}. That paper was primarily concerned with 
the growth of the planet up to the initiation of rapid gas accretion.  
Our prescriptions for $\dMe$ beyond that point were physically-motivated 
in a qualitative sense, but the quantitative implementation was \emph{ad hoc}.
But starting with \citet{lissauer2009}, more physics has been used to model 
the later stages of gas accretion. The transition from cooling-limited 
accretion to disk-limited accretion is now calculated, and indeed it is 
more fundamental than our boundaries between Phases~1, 2, and 3.
Therefore, to recognize disk-limited accretion as a separate stage 
of growth, in this paper and henceforth we will refer to the epoch 
from the onset of disk supply-limited gas accretion until the end 
of gas accretion as ``Phase~4''.

Typically, the growth rate of the planet $\dMp/\Mp$, over its entire 
history, is largest at the end of Phase~3 and is controlled by 
the envelope's contraction rate.
The transition to Phase~4 depends on disk properties and planet mass. 
In this phase, the growth rate becomes tied to the evolution of 
the surrounding disk, and it slows down as the disk evolves;
$\dMe$ is also determined by tidal interactions between the disk 
and the planet.

As anticipated above, since a giant planet may grow as long as there is 
nebula gas near its orbit, the end of Phase~4 typically corresponds with 
the dispersal of the gas from the planet's surroundings. After that time, 
the planet evolves in \emph{isolation}.
The inventory of H/He does not change during the evolution in isolation
(if there is no or negligible loss by solar irradiation), although
the inventory of heavy elements might, due to further collisions with 
solid bodies.

\subsection{Phases~1 and 2}
\label{sec:P12}

\begin{figure}
    \centering
    \resizebox{1.0\linewidth}{!}{%
    \includegraphics[]{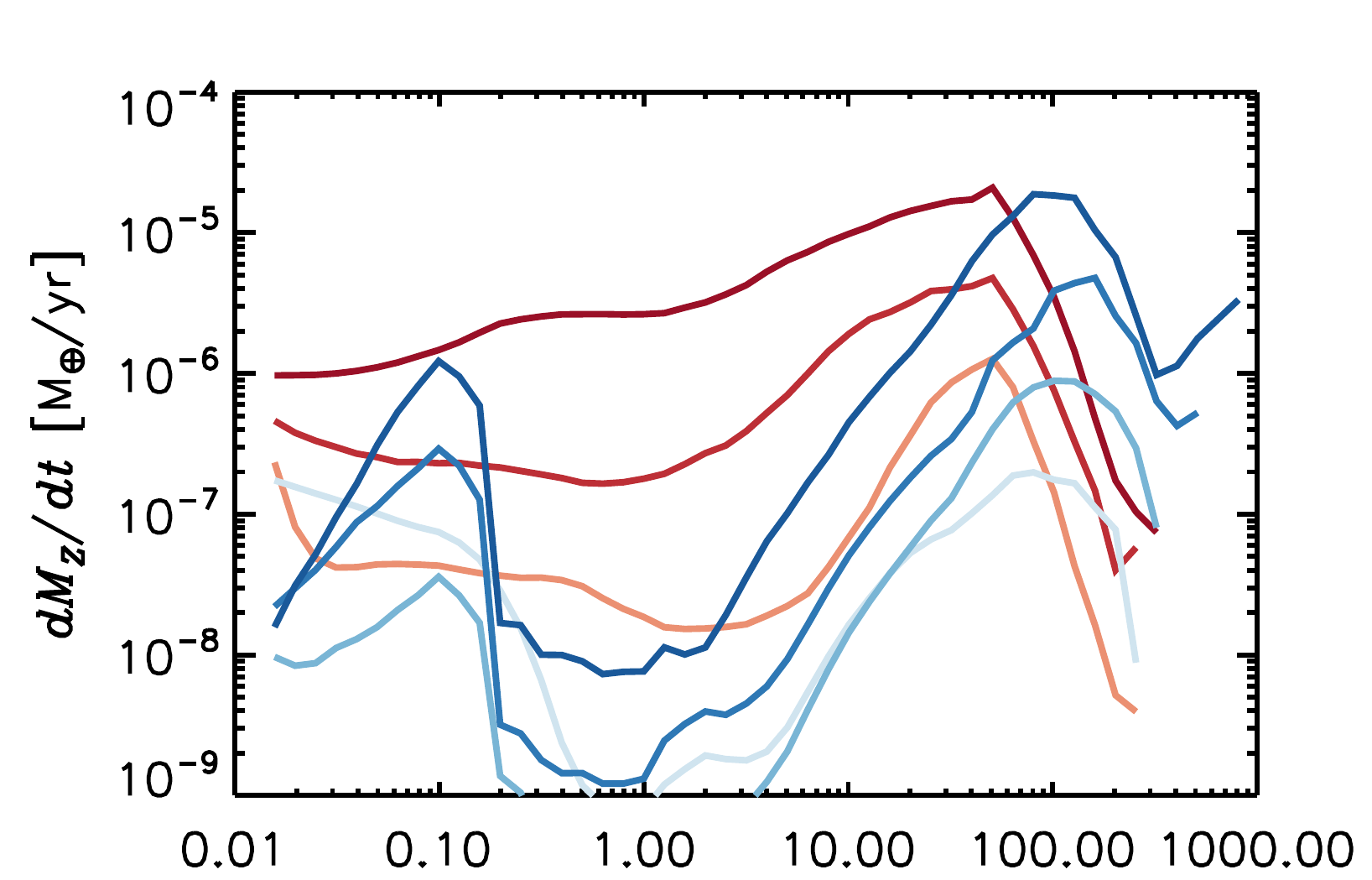}}
    \resizebox{1.0\linewidth}{!}{%
    \includegraphics[]{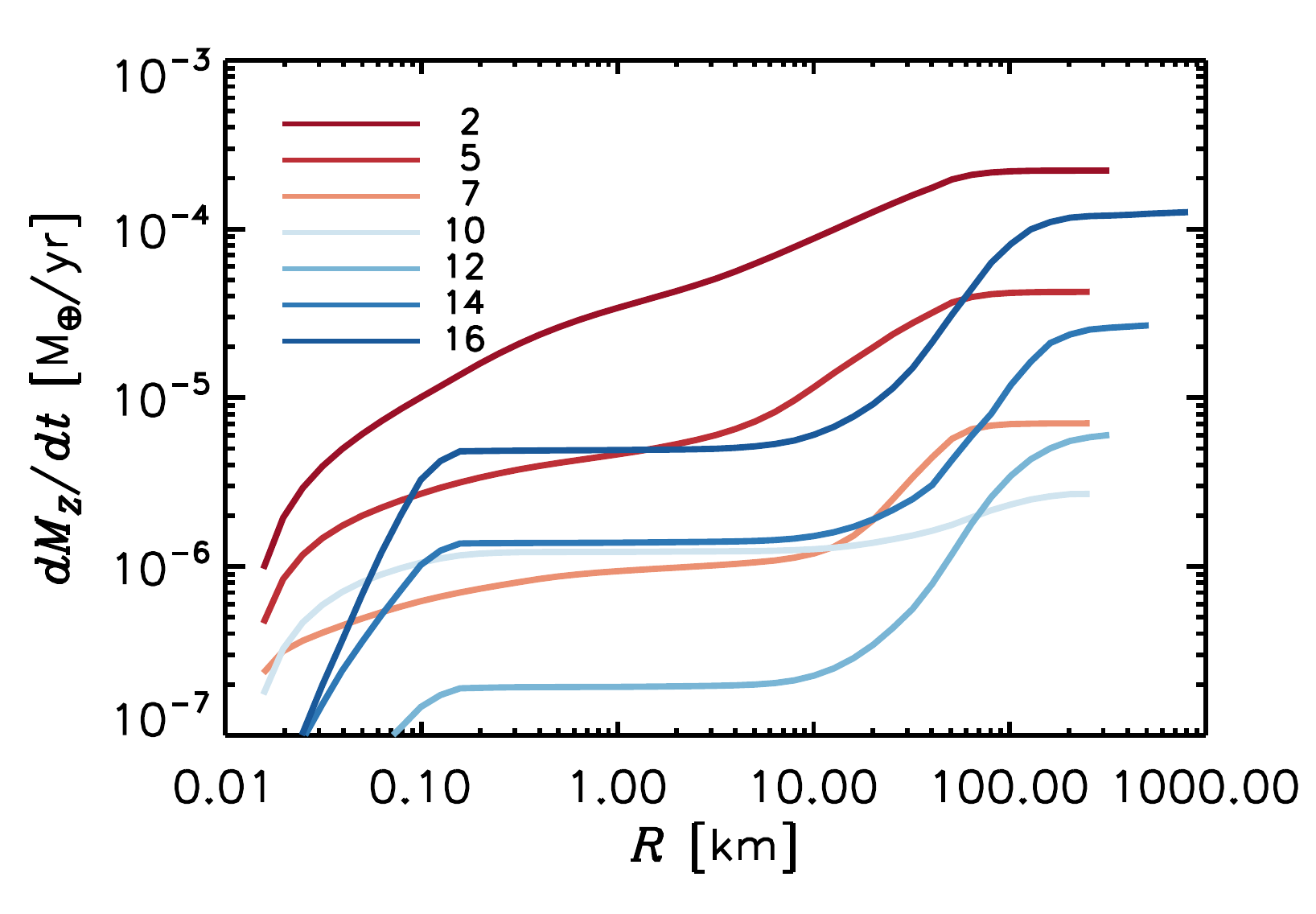}}
    \caption{
    The top panel shows the accretion rate of planetesimals as a function of 
    their
    radius, $R$, for different core masses, \Mc, as indicated in the bottom 
    panel in units of \Mearth. The bottom panel shows the cumulative
    distributions of the accretion rates, obtained by integrating the
    distributions of the top panel.
    The corresponding gas accretion rates are illustrated in \cifig{fig:acc}.
    }
    \label{fig:accvsr}
\end{figure}
The evolution of the planet during Phase~1 is described in detail
in \citetalias{gennaro2014}. The evolution begins at 
$\Mc\approx 10^{-4}\,\Mearth$ although the structure calculation of 
the envelope begins at $\Mc\approx 1\,\Mearth$.
The growth of the core is initially rapid and proceeds in a runaway fashion,
with $d\Mc/dt$ increasing as $\Mc$ increases.
During Phase~1, the solids accretion rate reaches a maximum of 
$2.2\times 10^{-4}\,\Mearth\,\mathrm{yr}^{-1}$
at $\Mc\approx 2.1\,\Mearth$ ($\Me\ll \Mc$ at this point). As the swarm
around the planet's orbit depletes, $d\Mc/dt$ begins to decline. The
gaseous envelope, although small in mass, is large in volume and significantly
increases the capture radius of the planet (which is a function of $R$) for 
the accretion of planetesimals, first of small ones (a few to several 
kilometers in radius) and of all sizes once 
$\Mc\gtrsim 4\,\Mearth$ ($\Me \gtrsim 10^{-3}\,\Mearth$).

The top panel of \cifig{fig:accvsr} shows the distribution of $d\Mc/dt$, 
versus the planetesimal radius $R$, for different values of $\Mc$. 
The cumulative distributions, obtained by integrating over $R$ the curves 
in the top panel, are plotted in the bottom panel for the core masses 
indicated in the legend in units of $\Mearth$.
During Phase~1, $d\Mc/dt$ is maximum for $R\approx 50\,\mathrm{km}$
bodies and accretion of large planetesimals accounts for most of the 
heavy-element content of the planet. In fact, by the time 
$\Mc\approx 7\,\Mearth$, about $70$\% of the heavy element mass
has been delivered by planetesimals $\gtrsim 10\,\mathrm{km}$ in radius 
and only about $13$\% by bodies smaller than $1\,\mathrm{km}$ in radius.

\begin{figure}
    \centering
    \resizebox{1.0\linewidth}{!}{%
    \includegraphics[]{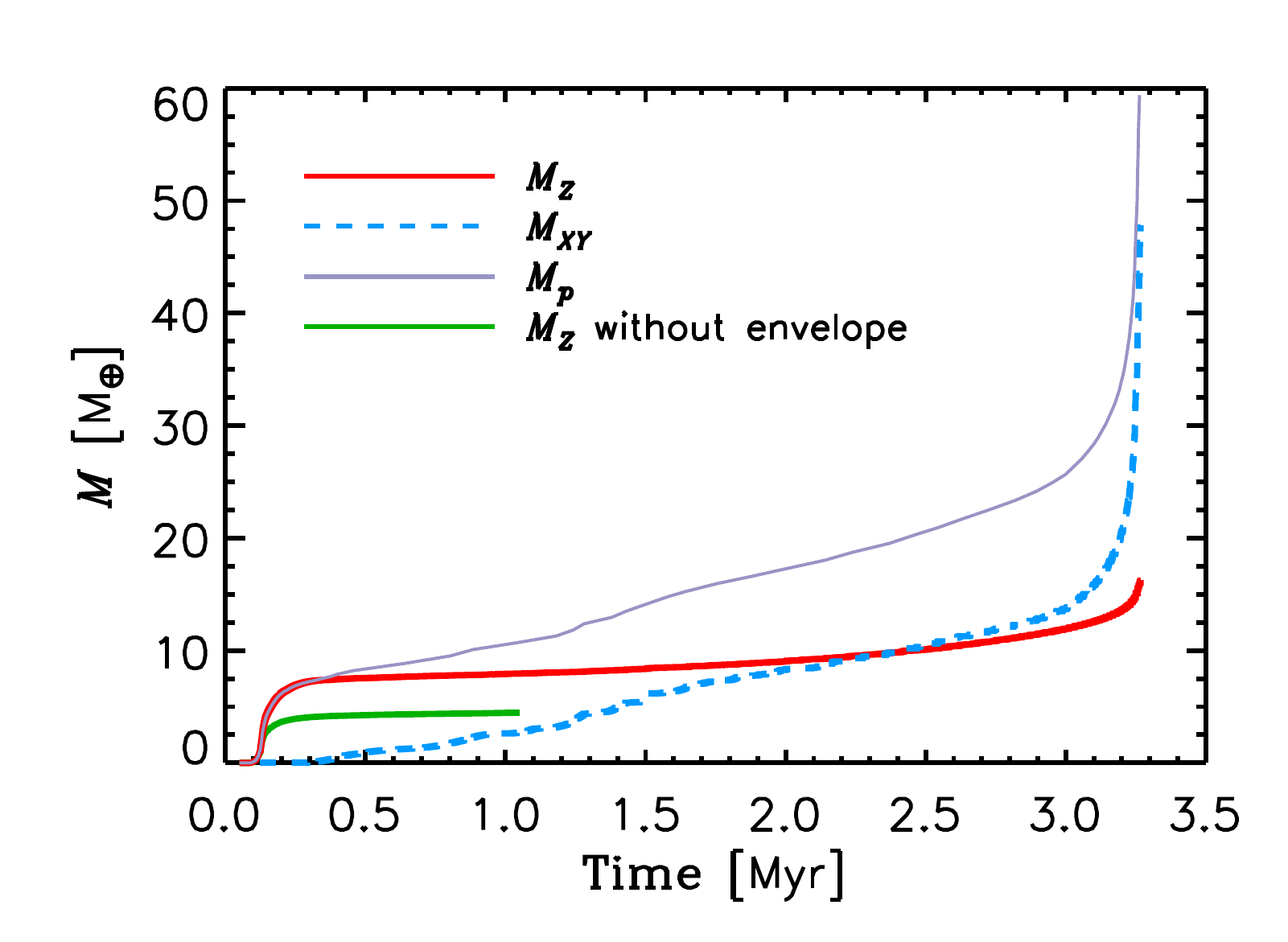}}
    \caption{
    Heavy-element mass, $\Mc$, and envelope mass, $\Me$, versus
    time, as indicated in the legend. The green line represents the
    bare-core model discussed in \citet{gennaro2014}. Core and 
    envelope mass are equal at $t\approx 2.38\,\Myr$,
    $\Mc=\Me= 9.8\,\Mearth$.
    }
    \label{fig:mzmxy}
\end{figure}
The accretion rate of solids decreases by a factor 
of more than $30$ as the planet mass grows from $\approx 2\,\Mearth$ 
to $\approx 7\,\Mearth$. 
Meanwhile, the accretion rate of gas increases by a similar factor.
Although the mass fraction of H/He is still quite small at this time,
one percent when $\Mp\approx 7.3\,\Mearth$ ($\Me\approx 0.08\,\Mearth$, 
see \cifig{fig:mzmxy}), the importance of the envelope can be appreciated 
by comparing
$\Mc$ with the case in which a bare core grows in the same swarm, 
attaining a mass just above $4\,\Mearth$ in $10^{6}$ years, when its 
accretion rate has dropped to a few times $10^{-7}\,\mathrm{\Mearth\,yr}^{-1}$ 
(compare red and green curves in \cifig{fig:mzmxy}; also see discussion
in \citetalias{gennaro2014}).

\begin{figure*}
    \centering
    \resizebox{1.0\linewidth}{!}{%
    \includegraphics[]{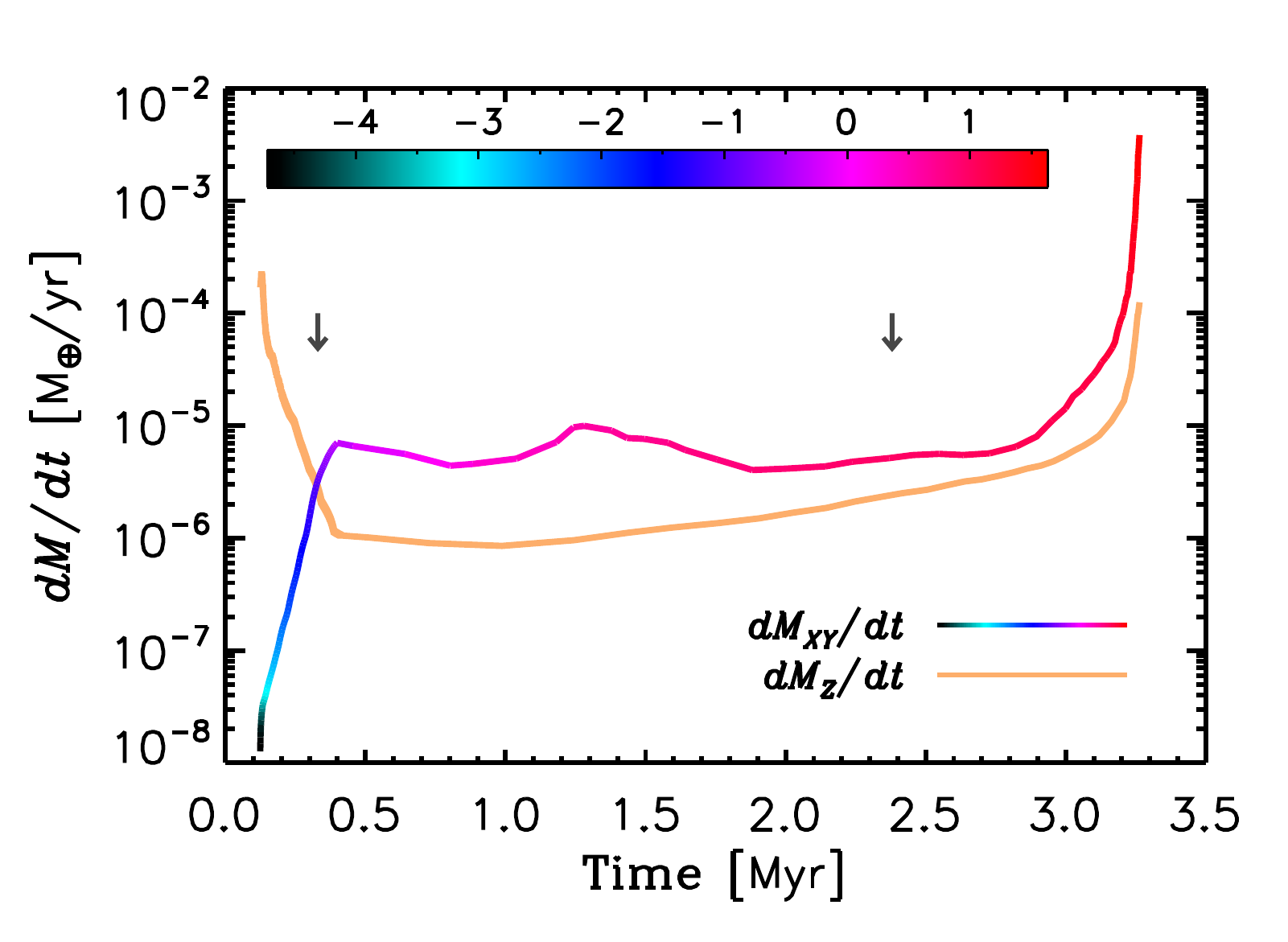}%
    \includegraphics[]{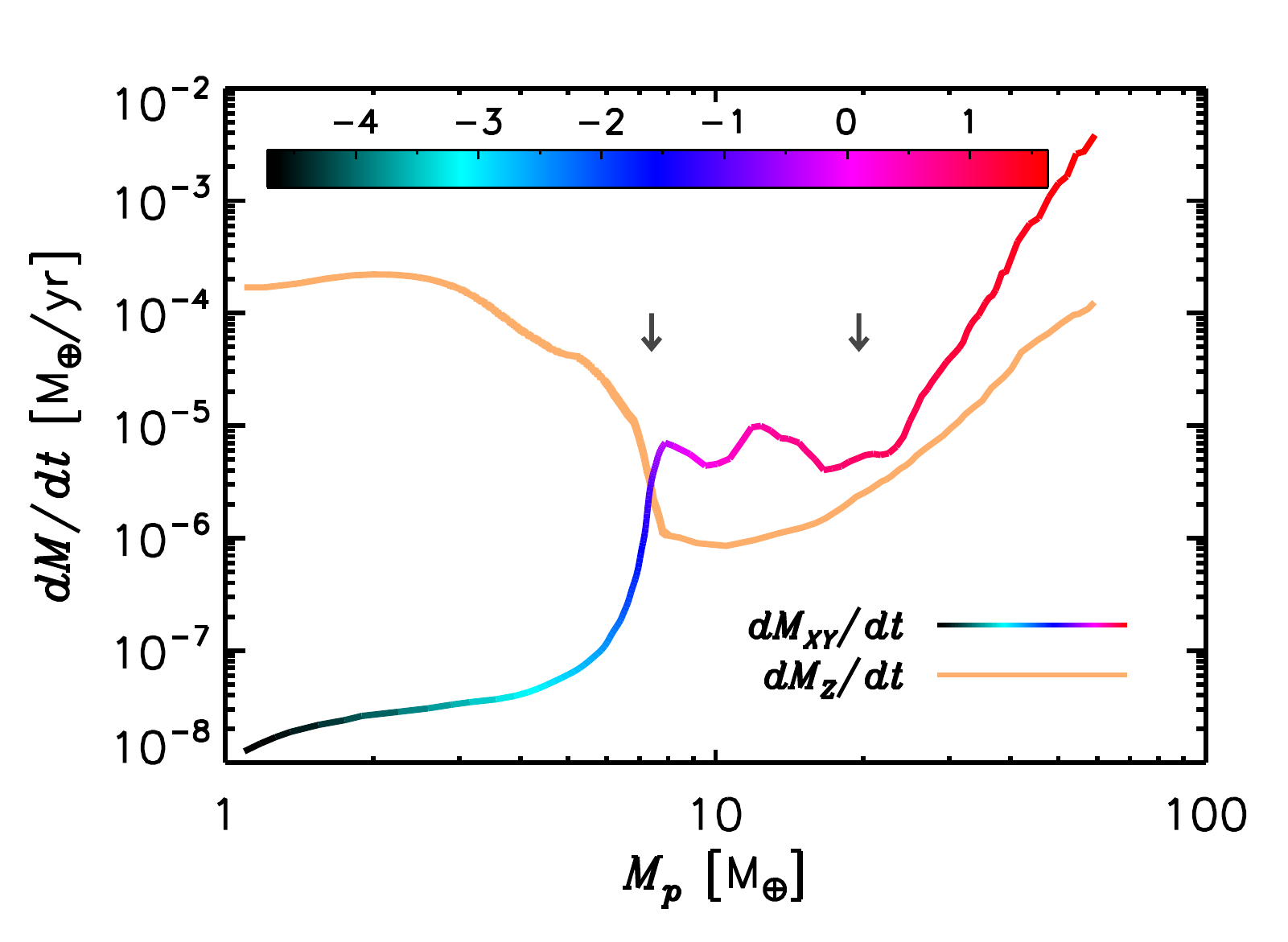}}
    \caption{
    Accretion rate of heavy elements, $d\Mc/dt$, and of gas, $d\Me/dt$,
    as a function of the time (left) and of the total planet mass (right).
    In both panels, the curve representing the gas accretion rate is color-coded 
    by the logarithm of the envelope mass, $\Me$, as indicated.
    Note that the evolution of the planet begins at $\Mc\approx 10^{-4}\,\Mearth$
    but the structure calculation of the envelope begins at 
    $\Mc\approx 1\,\Mearth$.
    Phase~1 ends at $\Mp\approx 7.45\,\Mearth$, when
    $\dMc \approx \dMe \approx 3\times 10^{-6}\,\mathrm{\Mearth\,yr}^{-1}$ (left arrow of pair).
    Phase~2 ends at $t\approx 2.38\,\Myr$, when 
    $\Mp\approx 19.6\,\Mearth$ (right arrow of pair).
    }
    \label{fig:acc}
\end{figure*}

At time when $\Mc\approx 7.3\,\Mearth$ and $\Me \approx 0.15\,\Mearth$, 
the accretion rates of H/He and heavier elements are approximately equal, 
$d\Mc/dt\approx d\Me/dt \approx 3\times 10^{-6}\,\mathrm{\Mearth\,yr}^{-1}$.
This formally signals the end of Phase~1 and the beginning of Phase~2.
The behavior of both quantities can be seen in \cifig{fig:acc}, as a function
of time (left) and planet mass (right, see figure's caption for details).
The arrows mark the end of Phases~1 (left of pair) and 2 (right of pair).
The figure also shows $\log{(\Me/\Mearth)}$ color-coded on the gas accretion
rate curves. The mass fraction of H/He at the end of Phase~1 is $2$\%.
The growth from the initial embryo ($\Mp\approx 10^{-4}\,\Mearth$) to
$\Mp\approx 7.45\,\Mearth$ takes approximately $3\times 10^{5}$ 
years, although the length of Phase~1 should also include the time
needed to assemble the initial embryo.

The trend of $d\Me/dt$ and $d\Mc/dt$ at the end of Phase~1 continues into 
Phase~2 for $\approx 10^{5}$ years, until 
$d\Me/dt \approx 7\times 10^{-6}\,\mathrm{\Mearth\,yr}^{-1}$ and
$d\Mc/dt\approx 10^{-6}\,\mathrm{\Mearth\,yr}^{-1}$, when the total mass
of the planet is $\Mp\approx 8\,\Mearth$ ($\Me\approx 0.6\,\Mearth$).
During the next $\approx 2\times 10^{6}$ years, the accretion rate of gas 
is relatively constant, as can be seen in \cifig{fig:acc} and deduced from 
the near-linear growth of $\Me$ in \cifig{fig:mzmxy}. 
The accretion of solids stays constant over the first $\approx 5\times 10^{5}$ 
years of Phase~2 but it then begins to increase as the planet mass grows. 
During Phase~2, the planet adds nearly $10\,\Mearth$ of H/He 
and about $2.7\,\Mearth$ of heavy elements.
Large bodies continue to deliver most of the heavy elements
to the planet: by the time $\Mc\approx 10\,\Mearth$, about $66$\% of the 
high-$Z$ material has been delivered by planetesimals $\gtrsim 10\,\mathrm{km}$ 
in radius (but only $\approx 1$\% by bodies with $R>160\,\mathrm{km}$)
and about $21$\% by bodies smaller than $1\,\mathrm{km}$ in radius.

As mentioned above, the accretion rates of H/He and heavy elements are
correlated during Phases~1 and 2 (and the first part of Phase~3, see 
\cifig{fig:acc}), when envelope contraction is thermally regulated. 
While the planet is transitioning between Phases~1 and 2, 
$6\,\Mearth\lesssim\Mp\lesssim 8\,\Mearth$, there is a large increase 
in $d\Me/dt$ (by almost two orders of magnitudes) triggered by a large 
drop in $d\Mc/dt$. 
Likewise, during the latter part of Phase~2, $d\Me/dt$ declines by 
a factor of a few as $d\Mc/dt$ increases by a similar factor.
Fluctuations in the relative amounts of small planetesimals ($R\lesssim
0.1\,\mathrm{km}$) delivered to the envelope may also contribute 
to opacity, and hence $d\Me/dt$, variations in Phase~2
(the outer envelope's layers are radiative during this Phase).

Cross-over mass ($\Me=\Mc$) is attained at $t\approx 2.4\,\Myr$, when 
$\Mc$ is around $10\,\Mearth$ (but see details in the caption of \cifig{fig:mzmxy}).
At that time, 
$d\Me/dt \approx 5.5\times 10^{-6}\,\mathrm{\Mearth\,yr}^{-1}$, which is
about twice as large as the accretion rate of heavy elements,
$d\Mc/dt\approx 2.6\times 10^{-6}\,\mathrm{\Mearth\,yr}^{-1}$.

\subsection{Phase~3}
\label{sec:P3}

Past the end of Phase~2, the accretion rate of gas remains roughly constant
until $\Mp\approx 23\,\Mearth$ ($t\approx 2.7\,\Myr$), at which point 
it begins to increase in a runaway fashion, as illustrated in \cifig{fig:acc}.

\begin{figure*}
    \centering
    \resizebox{1.0\linewidth}{!}{%
    \includegraphics[]{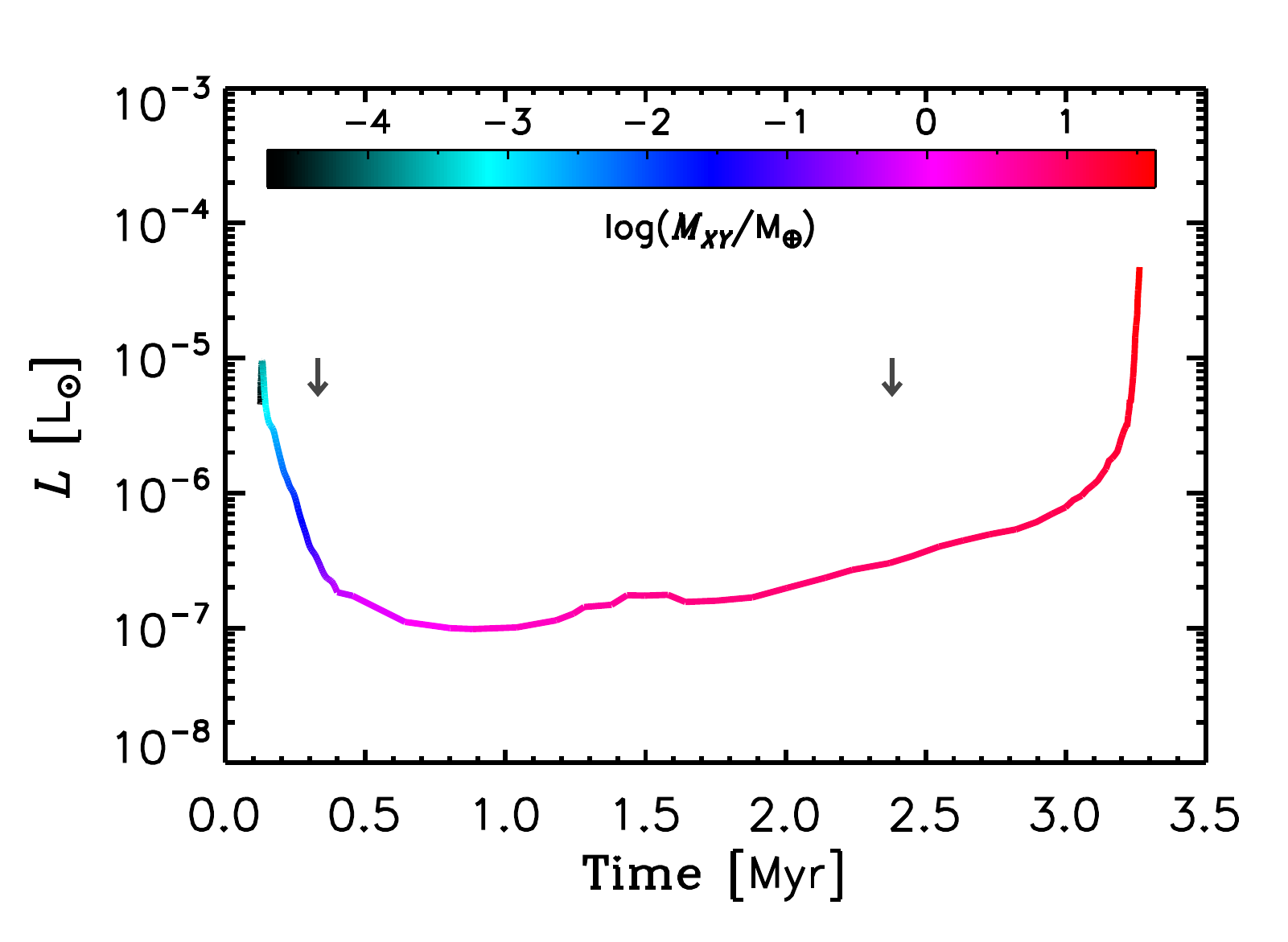}%
    \includegraphics[]{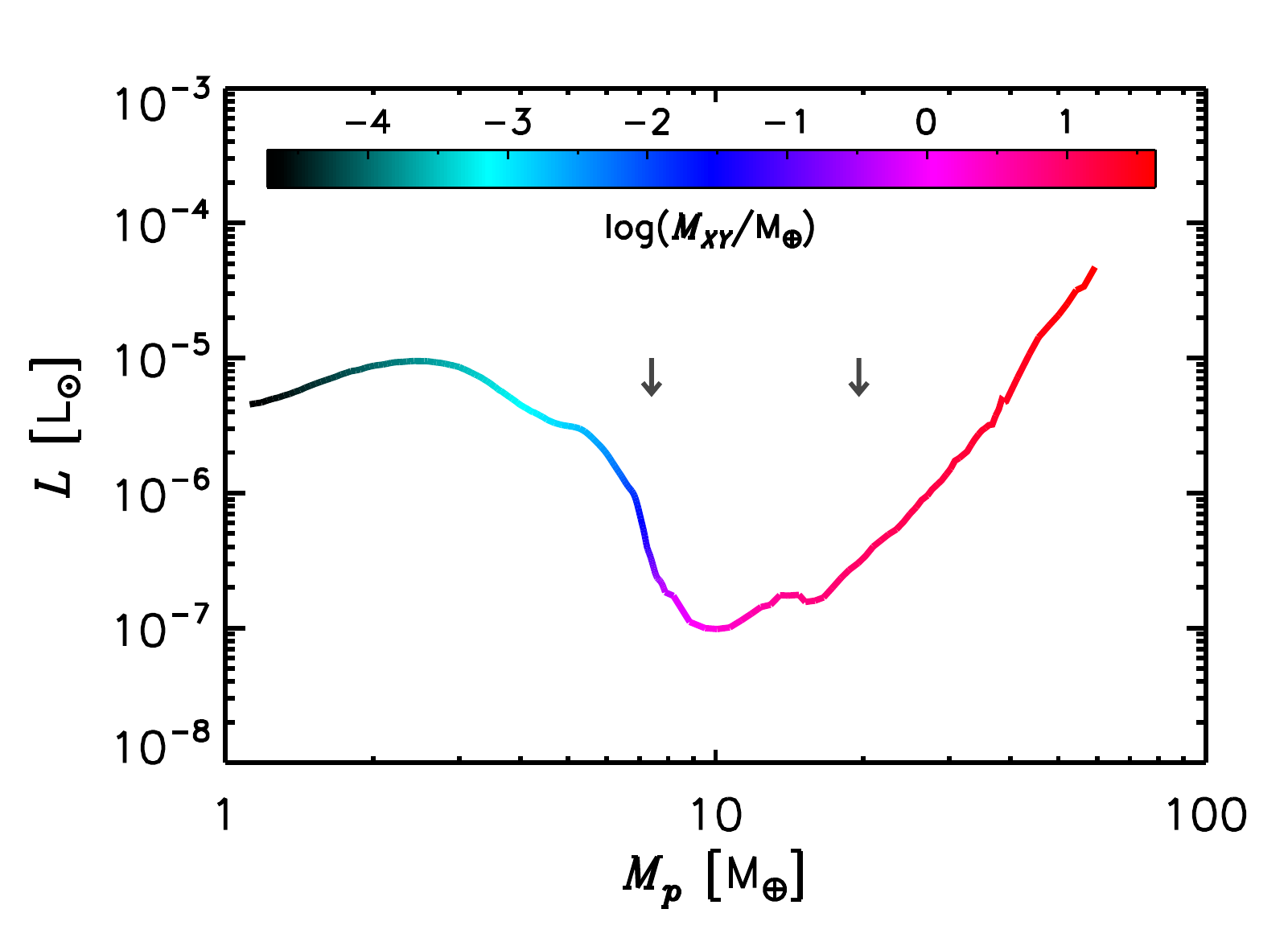}}
    \caption{
    Luminosity of the planet as a function of the time (left) and of the 
    total mass of the planet (right).
    The curves are color-coded by the logarithm of the envelope mass, $\Me$. 
    During Phase~1, the luminosity peak occurs when 
    $\Mp\approx 2.5\,\Mearth$ (see right panel).
    The transitions between Phases~1 and 2 and between Phases~2
    and 3 are marked by the pair of arrows 
    (see also \cifig{fig:acc}).
    }
    \label{fig:lum}
\end{figure*}
\cifig{fig:lum} shows the luminosity of the planet as a function of time (left)
and total planet mass (right). The curves are also color-coded according to 
$\log{(\Me/\Mearth)}$ in both panels.
During Phase~1 the luminosity of the planet is close to the energy per unit time 
released by the accretion of heavy elements
\begin{equation}
    L_{Z}\approx \frac{G\Mc (d\Mc/dt)}{R_{Z}}.
    \label{eq:Lacc}
\end{equation}
But also during Phase~2 and somewhat beyond, in the first part of Phase~3,
the planet's luminosity can be approximated by $L_{Z}$.
\cieq{eq:Lacc} assumes that incoming planetesimals deliver most of their mass
close to the core-envelope boundary, $R_{Z}$. This is not necessarily the case, 
since bodies can be consumed and/or break up farther above in the envelope. 
However, since most mass is delivered by large planetesimals (several tens 
of kilometers in radius), even toward and past the end of Phase~2 
(see \cifig{fig:accvsr}) and large bodies tend to be held together 
by their own gravity, it is expected that accreted planetesimals release 
gravitational energy deep in the envelope.
In the calculation, $L$ is close to $L_{Z}$ during these stages 
also because all accreted solids are assumed to eventually sediment 
onto the core.
 
\begin{figure*}
    \centering
    \resizebox{1.0\linewidth}{!}{%
    \includegraphics[]{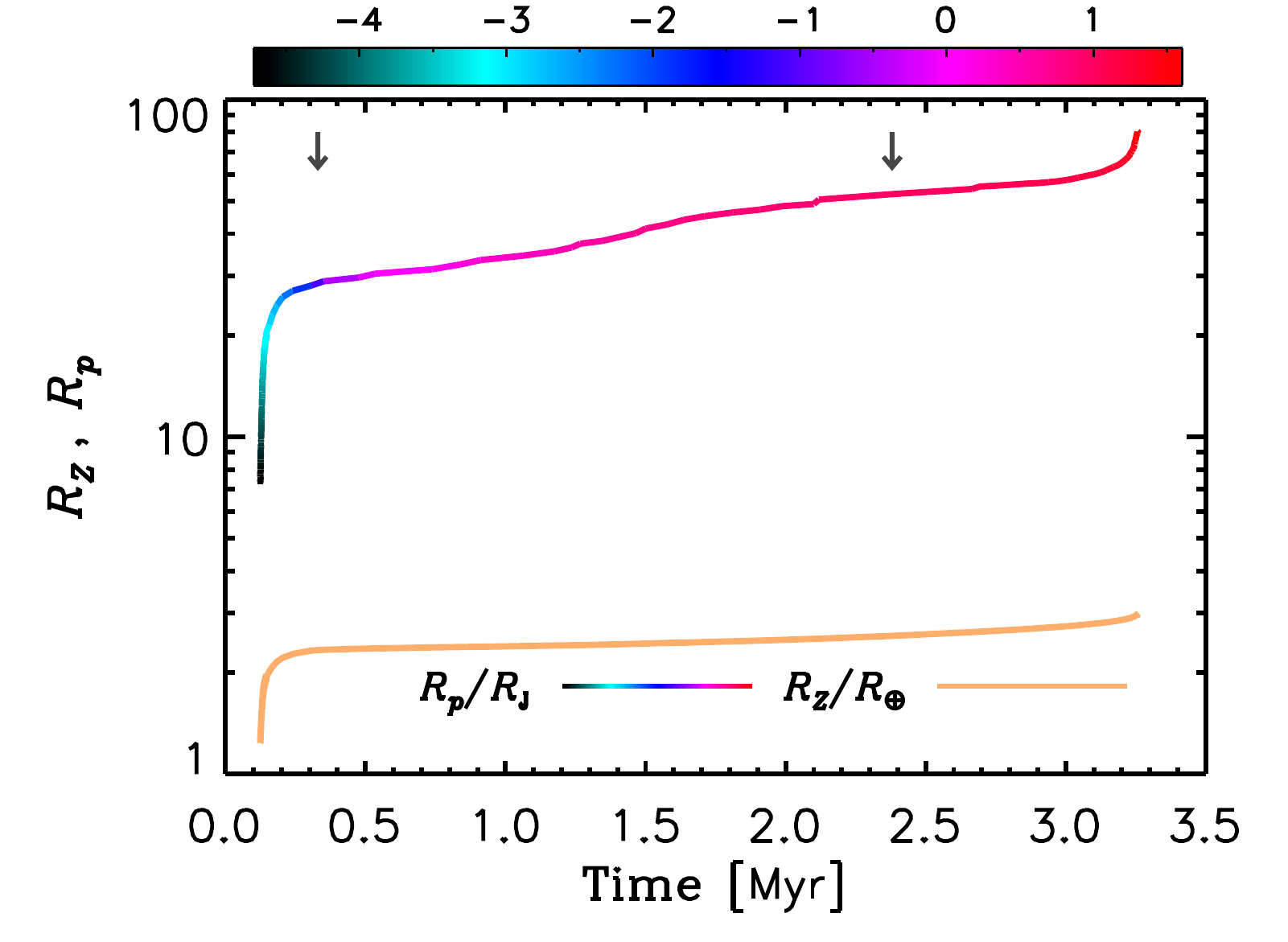}%
    \includegraphics[]{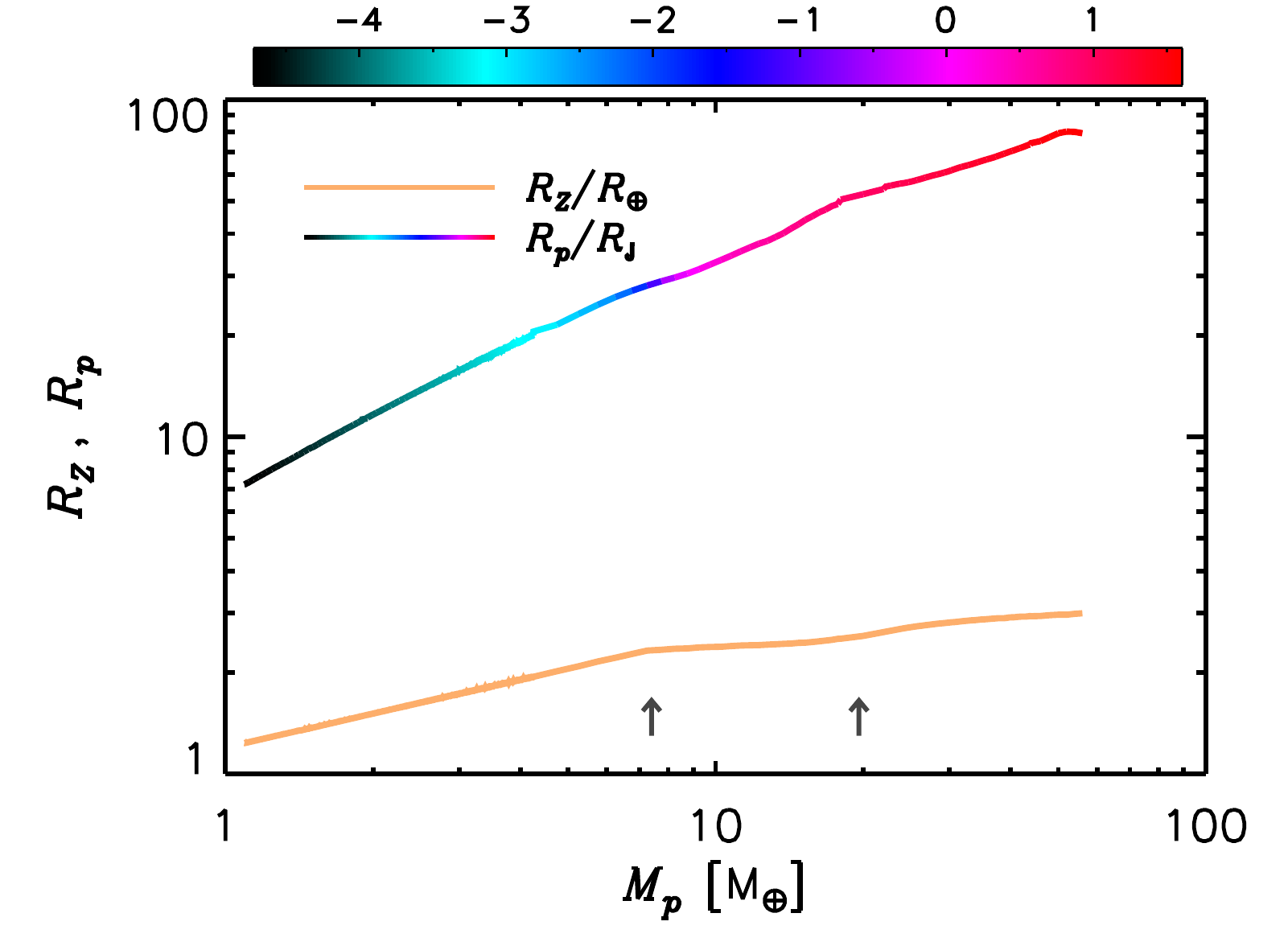}}
    \caption{
    Radius of the planet, $\Rp$, and radius of the heavy-element
    core, $R_{Z}$, respectively normalized to Jupiter's radius, $\Rjup$,
    and to the Earth's radius, $\Rearth$ ($\Rjup\approx 11\,\Rearth$). 
    The radii are plotted as a function 
    of the time (left) and of the total mass of the planet (right).
    The curve representing the planet's radius is color-coded by 
    $\log{(\Me/\Mearth)}$.
    The transitions between Phases~1 and 2 and between Phases~2
    and 3 are marked by the pair of arrows 
    (see also \cifig{fig:acc}).
    Note that, throughout the epoch shown, the ratio $\Rp/R_{Z}$ 
    increases from roughly $65$ to $300$.
    }
    \label{fig:rad}
\end{figure*}
The radius of the planet, i.e., the envelope radius, $\Rp$ is plotted in 
\cifig{fig:rad}, along with the radius of the heavy-element core $R_{Z}$, 
as a function of both time (left) and planet mass (right). 
The radii are normalized, respectively, by Jupiter's mean radius 
(at $1$~bar envelope depth), $\Rjup=69\,911\,\mathrm{km}$
\citep{archinal2011}, and by the Earth's radius, $\Rearth$. 
The planet radius is also color-coded in terms of the envelope mass. 
As mentioned above, since the core is assumed to be incompressible, 
$R_{Z}\propto M_{Z}^{1/3}$ (see \cisec{sec:methods}).
During Phases~1, 2, and 3 the envelope is in contact with the nebula 
and $\Rp$ is basically given by the effective radius $R_{\mathrm{eff}}$, 
defined in \cisec{sec:sc}.
In the model, and throughout most of the evolution shown in the figure, 
$R_{\mathrm{eff}}$ is limited by $\Rhill/4$ (except for the early
stages of Phase~1). 
The envelope starts to shrink when the rate of gas supply dictated 
by contraction exceeds that actually delivered by the nebula, 
which occurs somewhat later, during Phase~4 \citep[see also][]{lissauer2009}.

\begin{figure*}
    \centering
    \resizebox{1.0\linewidth}{!}{%
    \includegraphics[]{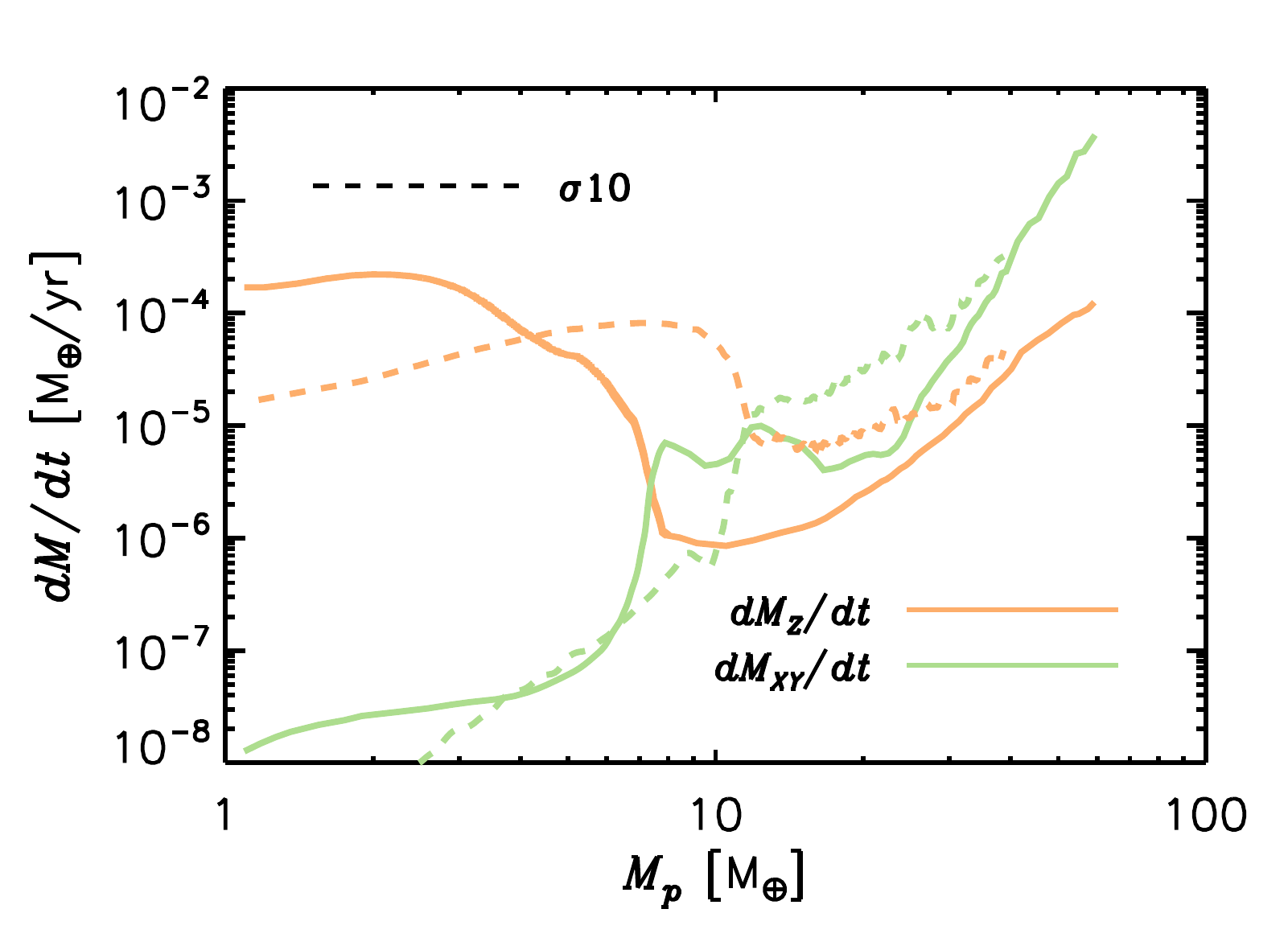}%
    \includegraphics[]{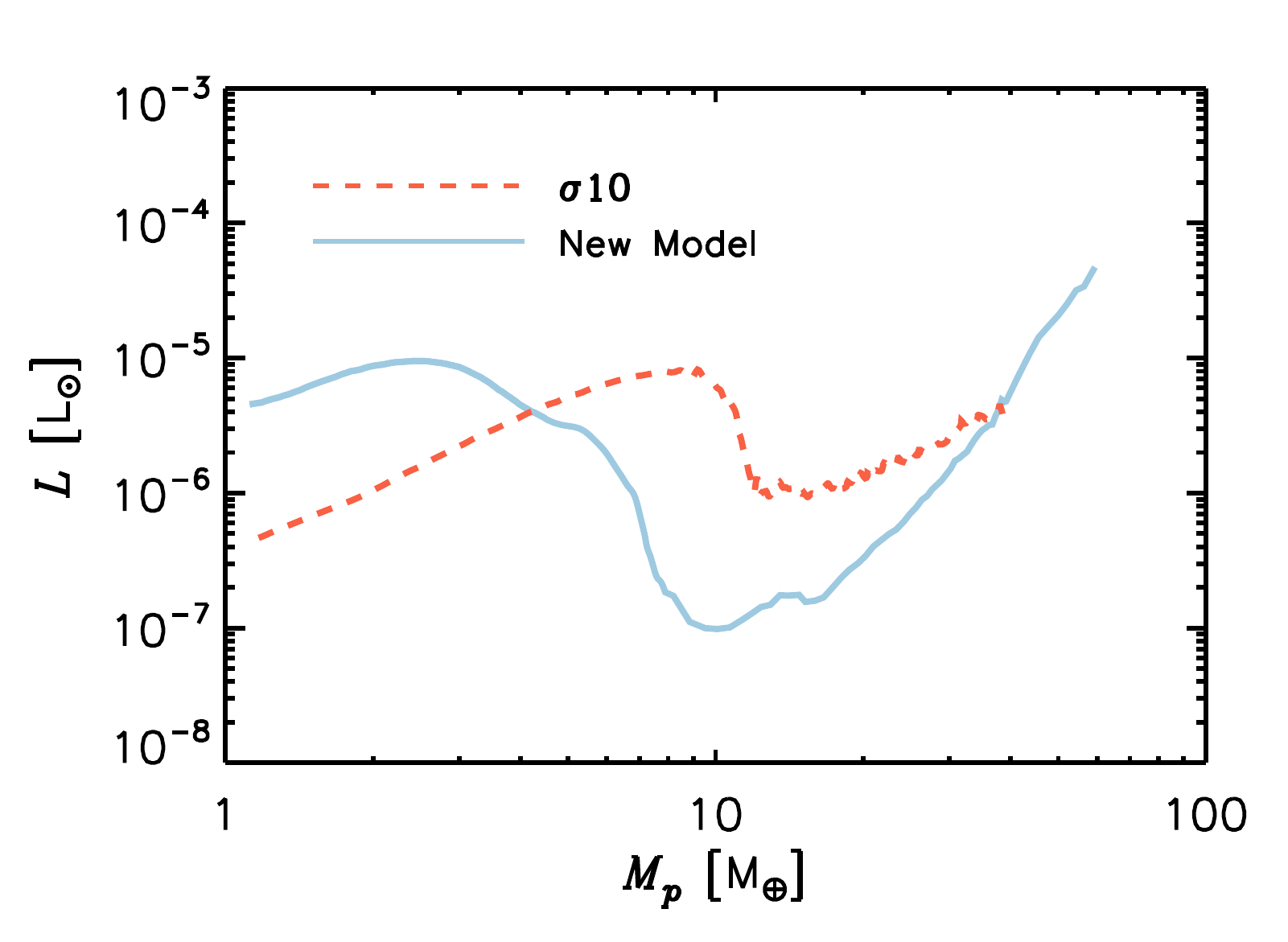}}
    \caption{
    Comparison of the accretion rates of heavy elements and gas (left) 
    and luminosity (right), as a function of the total planet mass, for 
    the new calculation (solid lines) and a model from \citet{naor2010}, 
    labelled as $\sigma 10$ (dashed lines), as indicated in the legends 
    (see text for further details).
    }
    \label{fig:acomp}
\end{figure*}
In \cifig{fig:acomp} we compare accretion rates (left) and luminosity (right) 
of the calculation discussed here (solid curves) with the results from the model 
labelled $\sigma 10$ of \citet{naor2010}. That model also applied a density of
solids $\sigma^{0}_{Z}=10\,\sigu$ at $a=5.2\,\AU$. However, that
calculation did not account for the evolution of the swarm of planetesimals, 
but it rather assumed an accretion rate of solids based on the three-body problem 
results of \citet{greenzweig1992}, and planetesimals were assumed to be all 
$100\,\mathrm{km}$ in radius.
Model $\sigma 10$ achieves the $d\Mc/dt=d\Me/dt$ condition, the end of Phase~1, 
at $\Mp\approx 12.1\,\Mearth$, a mass larger than that in the new model, but 
when $\Mc$ ($11.5\,\Mearth$) is close to the mass predicted by \cieq{eq:MZiso}
for a solid's feeding zone parameter $b=4$.
During Phase~2, the accretion rates of both gas and heavy elements in the 
previous model are on average several times as large, and so is the luminosity
(as also predicted by \cieq{eq:Lacc}). Phase~3 begins at $\Mp\approx 32\,\Mearth$
in model $\sigma 10$, a mass significantly larger than that in the current model. 
However, once into Phase~3, accretion rates and luminosity become comparable
to those of the new model. 

\begin{figure*}
    \centering
    \resizebox{1.0\linewidth}{!}{%
     \includegraphics[]{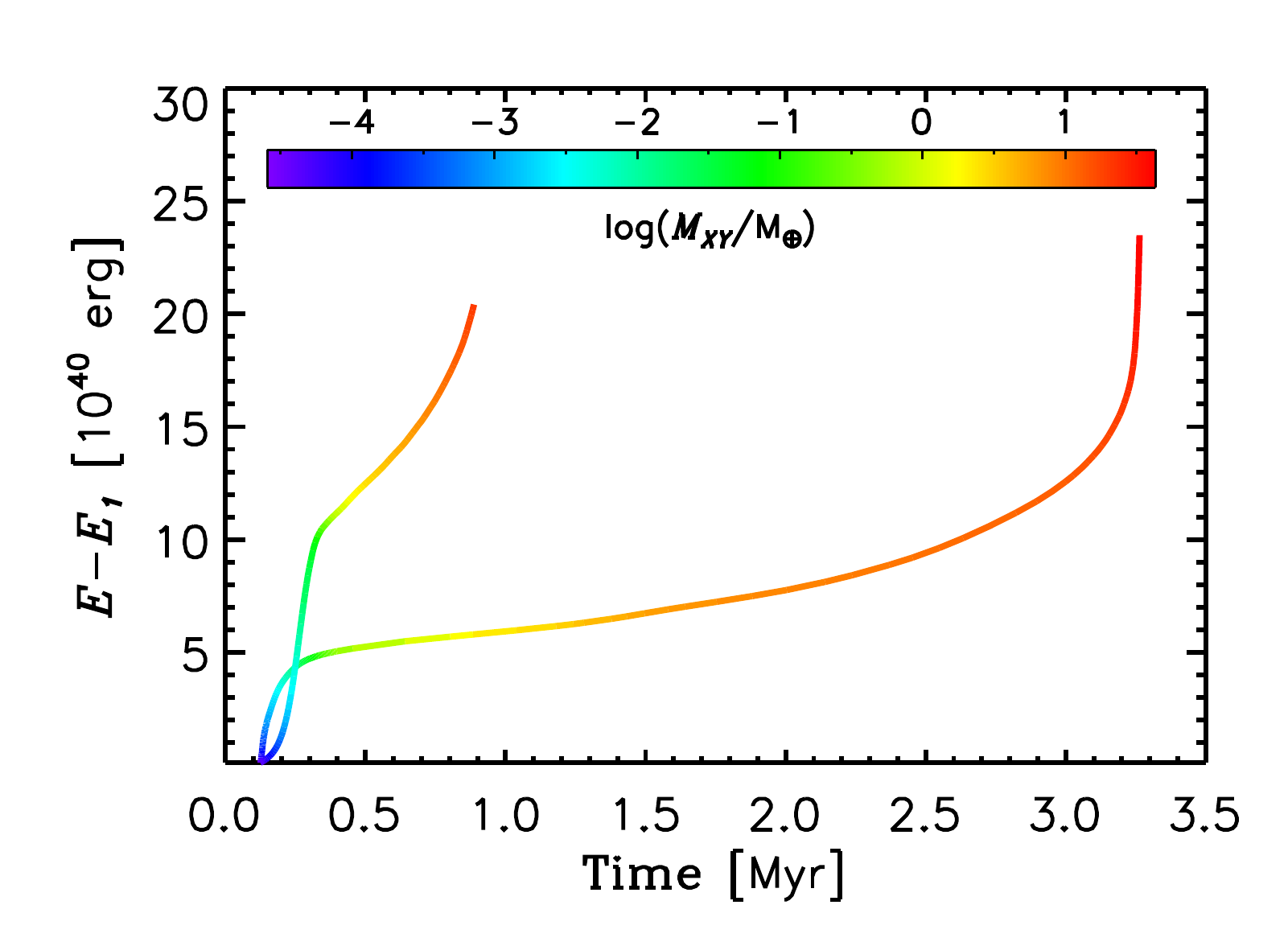}%
     \includegraphics[]{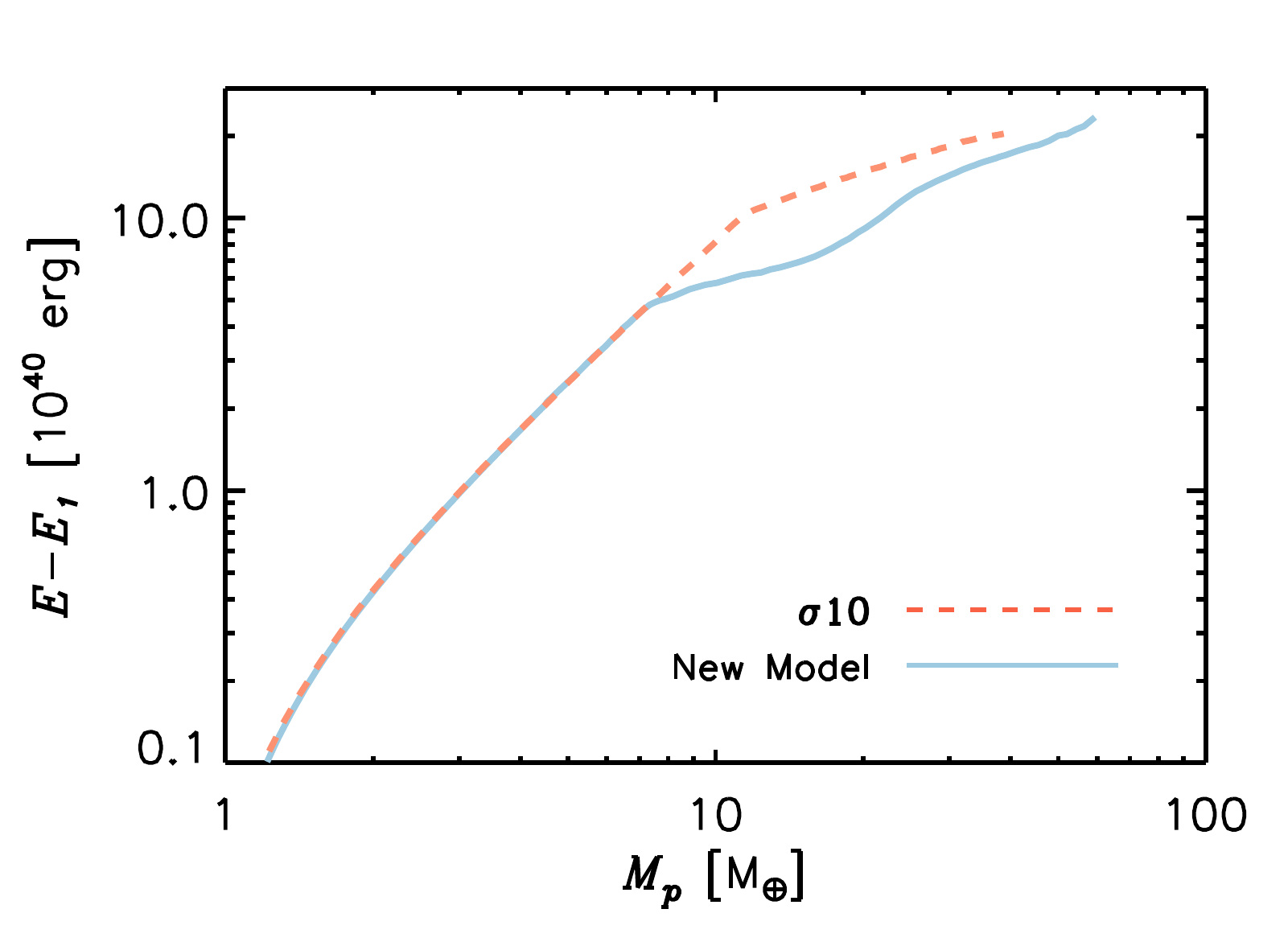}}
    \caption{
    Cumulative energy radiated by the planet, \cieq{eq:Ene}, as a function of
    time (left) and total mass (right). The plots in both panels compare the
    new formation model to model $\sigma 10$ of \citet{naor2010} (see also 
    \cifig{fig:acomp} and text for details), which 
    completes Phase~2 in a little less than $1\,\Myr$. For comparison purposes,
    the evolutionary track of model $\sigma 10$ is shifted in time (see text). 
    The curves of the left panel are color-coded by the value of $\Me$.
    }
    \label{fig:ene}
\end{figure*}
The cumulative energy radiated by the planet during formation
\begin{equation}
    E(t)=\int_{t}L(\Mp) dt',
    \label{eq:Ene}
\end{equation}
where $\Mp$ is a function of time, is illustrated in \cifig{fig:ene} for both 
the new model and model $\sigma 10$ of \citet{naor2010}. In the figure, 
the energy $E_{1}$ represents the cumulative energy radiated by the planet
until its total mass is approximately $1\,\Mearth$. In this calculation, 
the planet growth starts when $\Mp=\Mc\approx 10^{-4}\,\Mearth$, although the 
envelope structure is calculated only for $\Mc> 1.1\,\Mearth$. Nonetheless, 
assuming that for $\Mc\lesssim 1\,\Mearth$ the luminosity $L$ can be approximated 
by the accretion luminosity $L_{Z}$ in \cieq{eq:Lacc}, an approximation that holds 
valid at far larger masses, it is estimated that 
$E_{1}\approx 9\times 10^{38}\,\mathrm{erg}$.
The evolution of model $\sigma 10$ is shifted in time so that it achieves 
$\Mc\approx 1.1\,\Mearth$ at the same time as the new model.
Since $E_{1}$ cannot be computed for model $\sigma 10$ (because that model was 
started at about $1.2\,\Mearth$), we assume the same value for $E_{1}$.
Notice that the evolution of model $\sigma 10$ is faster, completing Phase~2 in
about $0.97\,\Myr$ (compare curves in left panel of \cifig{fig:ene}), when 
$\Mp\approx 32\,\Mearth$.
Nonetheless, the two models predict similar radiated energies up until 
$\Mp\approx 7.5\,\Mearth$ (assuming comparable values for $E_{1}$), which marks 
the end of Phase~1 in the new model. 
The release of energy is larger on average in model $\sigma 10$ 
during Phase~2 and the initial stages of Phase~3, but they again appear 
to become comparable later on.

The structure calculation is continued up to a time 
$t\approx 3.3\,\Myr$, when $\Mc\approx 16.2\,\Mearth$. At that point, 
the total mass of the planet exceeds $60\,\Mearth$.
During this time both $d\Me/dt$ and $d\Mc/dt$ increase, but the rate 
of growth of the former greatly exceeds that of the latter (see
\cifig{fig:acc}).
Accretion of solids is still dominated by large bodies, as can be seen 
in \cifig{fig:accvsr}.
By the time $\Mc\approx 16\,\Mearth$, about $74$\% of the heavy elements
has been delivered by planetesimals larger than $\approx10\,\mathrm{km}$ 
in radius and about $19$\% by bodies $\lesssim 1\,\mathrm{km}$ in radius.

Proper runaway accretion of gas, defined as the mass-doubling timescale 
$\Mp/\dMp$ being a decreasing function of $\Mp$, commences shortly after
the start of Phase~3, when $\Mp\approx 23\,\Mearth$. Even so, the planet 
takes over half a million years to double its mass (see \cifig{fig:mzmxy} 
and \ref{fig:acc}), a reminder that the onset of runaway gas accretion 
does not necessarily imply an immediately short growth timescale.

\subsection{Phase~4}
\label{sec:DLA}

Once $\Mp \gtrsim 65\,\Mearth$, envelope contraction allows
gas accretion rates in excess of $0.01\,\Mearth\,\mathrm{yr}^{-1}$. 
These large values of $d\Me/dt$ may exceed the rate at which a disk
(a few million years old at this point) can supply gas to the planet,
unless the disk is initially very massive and/or very long-lived 
(in excess of several \Myr).
The planet would then have entered Phase~4, the phase of disk-limited
gas accretion \citep{gennaro2008,lissauer2009}, whereby disk evolution
and disk-planet tidal interactions limit the supply of gas that can be
delivered to the vicinity of the planet.

\begin{figure*}
    \centering
    \resizebox{1.0\linewidth}{!}{%
    \includegraphics[]{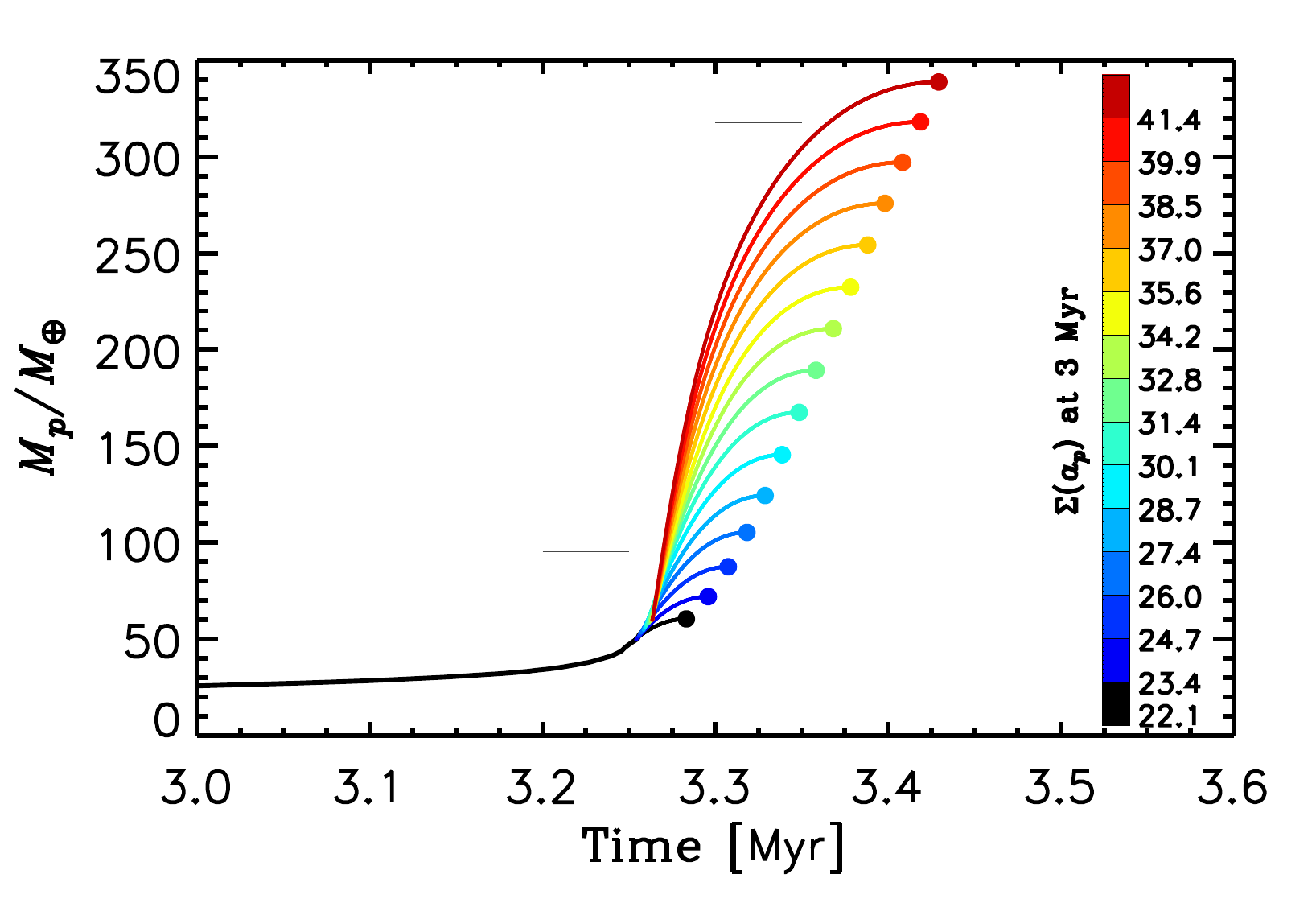}%
    \includegraphics[]{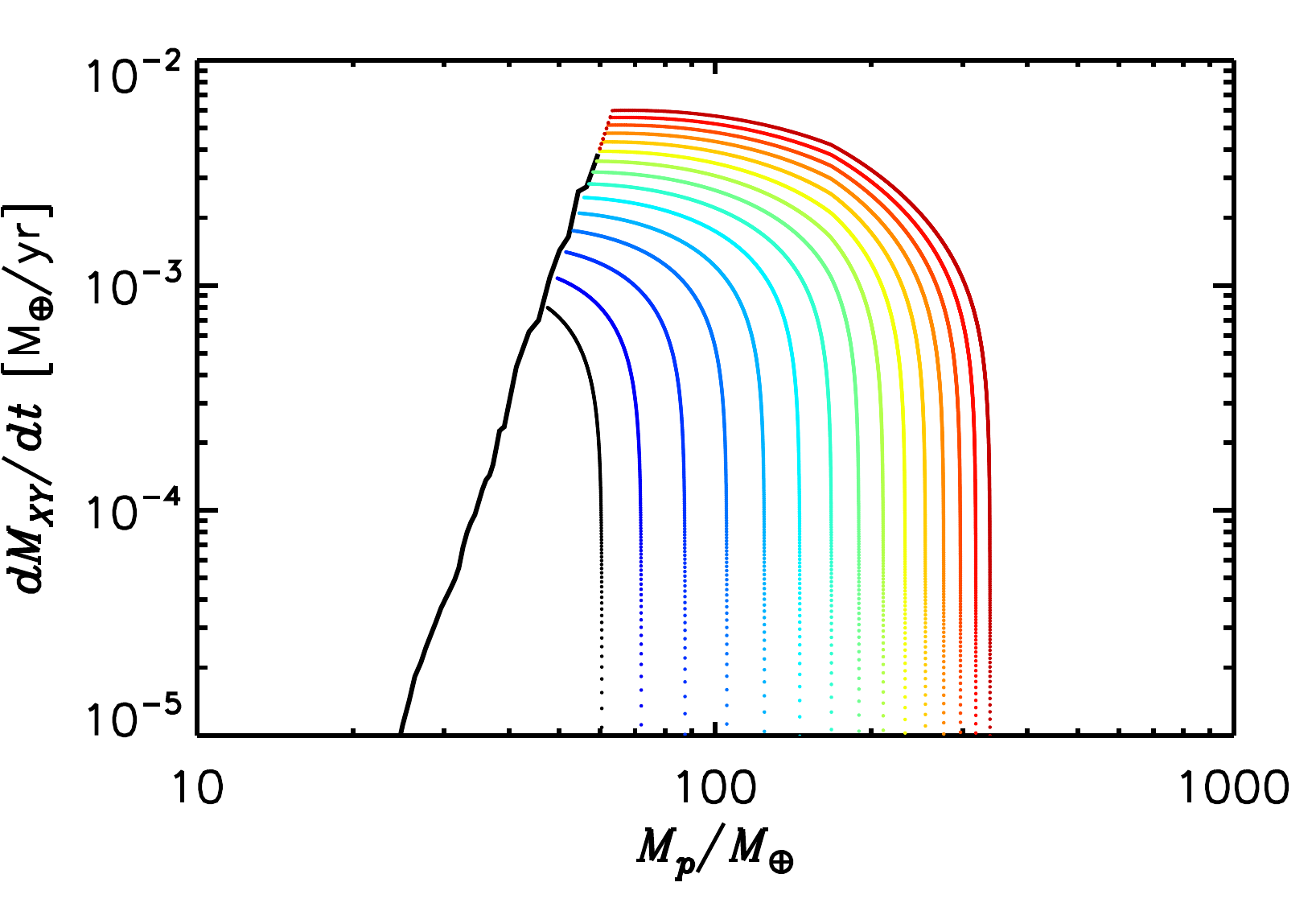}}
    \resizebox{1.0\linewidth}{!}{%
    \includegraphics[]{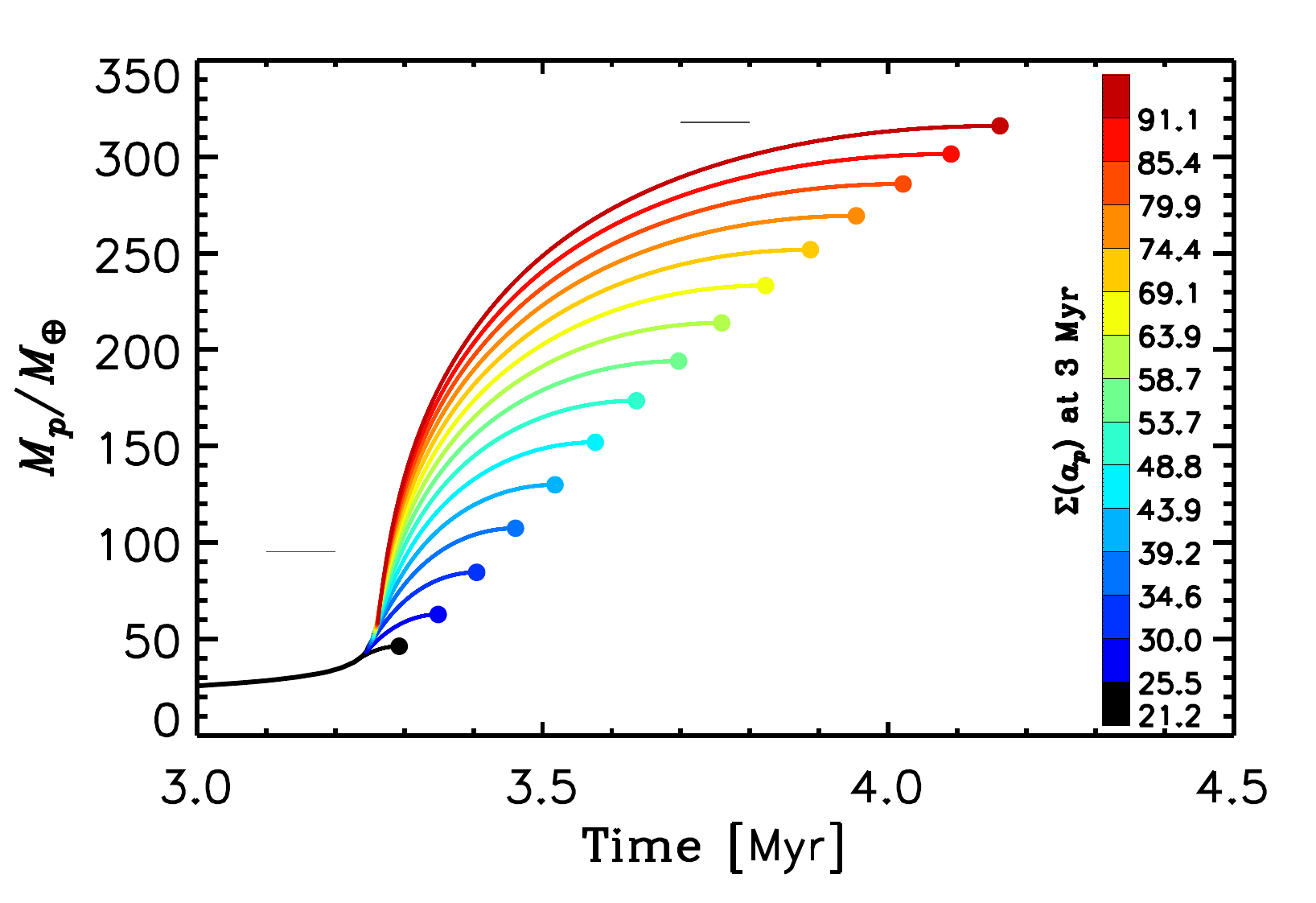}%
    \includegraphics[]{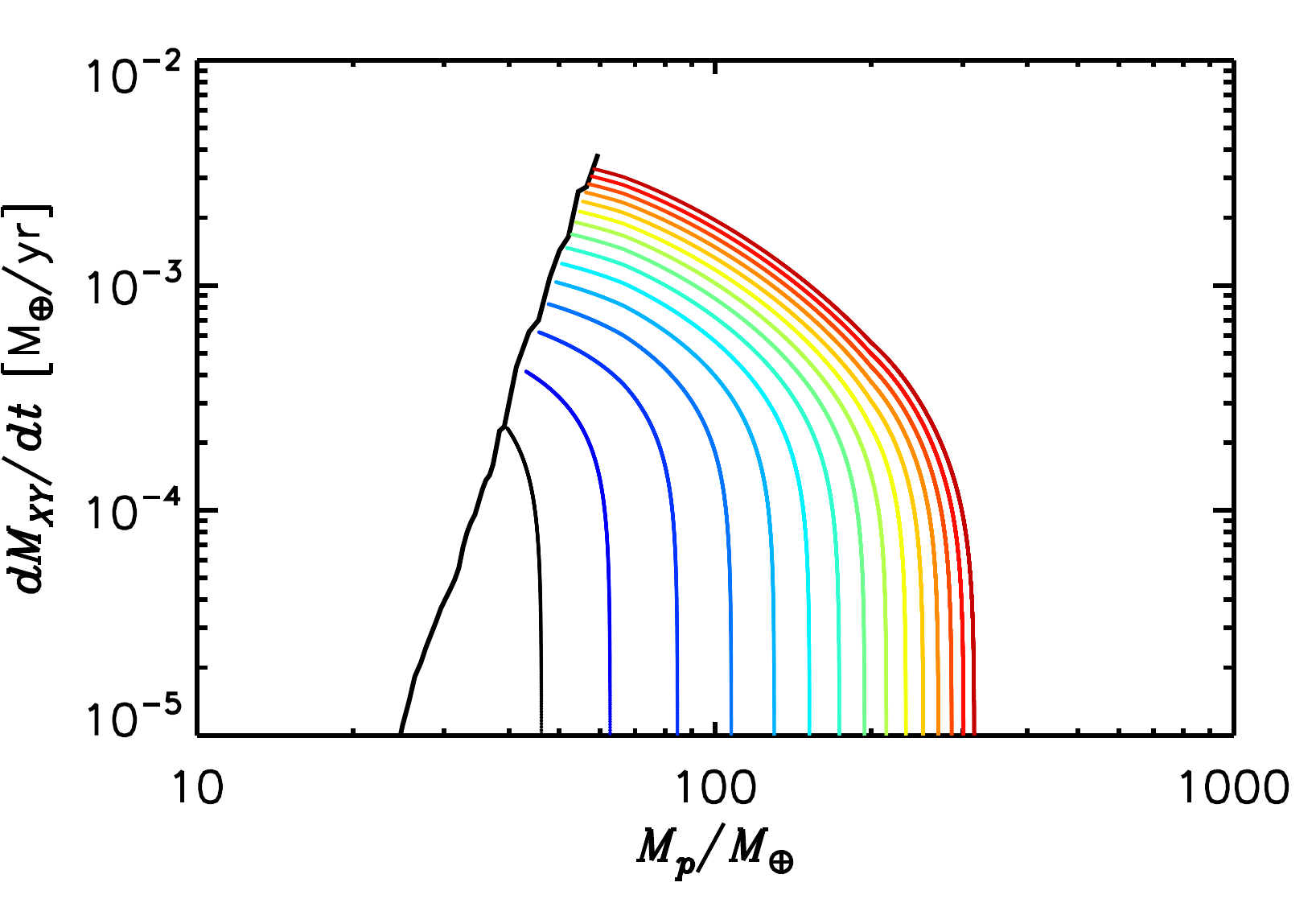}}
    \caption{
    Left: Growth tracks of the planet undergoing disk-limited accretion during 
    Phase~4. The solid circles indicate the final mass when the nebula around
    the planet's orbit has dispersed. Different colors indicate different values
    of the disk mass, $\mathcal{M}_{D}$, at $t=0$. The curves are color-coded 
    according to the value of the gas density at the planet location, $\Sigma(\ap)$ 
    in $\mathrm{g\,cm^{-2}}$, after $3\,\Myr$.
    The lower and upper horizontal segments to the left of the tracks indicate,
    respectively, the masses of Saturn and Jupiter.
    Right: Gas accretion rate versus planet mass for the same models as in the 
    left panels. The turbulence viscosity parameter of the nebula is $\alpha=4\times 10^{-3}$
    in the top panels and $\alpha=4\times 10^{-4}$ in the bottom panels.
    Thick black curves represent results from the structure calculations, in which
    accretion is limited by the envelope's thermal pressure. 
    }
    \label{fig:fmass}
\end{figure*}

In fact, the disk models presented below predict, at that time of the 
evolution ($t\approx 3.3\,\Myr$) and for models forming Jupiter-mass 
planets, disk-limited accretion rates of a few to several times 
$10^{-3}\,\Mearth\,\mathrm{yr}^{-1}$, 
depending on the level of gas turbulence viscosity (see \cifig{fig:fmass}). 
These accretion rates are comparable with the values dictated by envelope 
contraction in the structure calculation when $\Mp \approx 60\,\Mearth$, 
as indicated in \cifig{fig:acc}.
At that point of the evolution, $\Mc\approx 16\,\Mearth$ and the heavy-element 
accretion rate is $\approx 10^{-4}\,\Mearth\,\mathrm{yr}^{-1}$.

In order to simulate the evolution in Phase~4 until the planet achieves 
its final mass, we consider the results from the structure calculation discussed 
above up to a time between $3.2$ and $3.3\,\Myr$ (see \cifig{fig:mzmxy}), 
when $\Mc\approx 16\,\Mearth$ and the total mass of the planet is 
$\Mp\approx 60\,\Mearth$.

For increasing planet masses not satisfying the inequality in \cieq{eq:gap_con},
disk-limited accretion is first of ``Bondi-type'', $\dMe\propto M^{3}_{p}$,
and then of ``Hill-type'', $\dMe\propto M_{p}$ \citep{gennaro2008}. 
Typically, however, these regimes can provide accretion rates larger, or much
larger, than cooling-limited accretion does at those planet masses. 
Once the condition~(\ref{eq:gap_con}) is satisfied, $\dMe$ is a relatively weak
function of planet mass at first, and then it rapidly declines as $\Mp$ 
increases. 

The mass growth of the planet in Phase~4 is calculated by integrating 
the disk-limited accretion rates computed by \citet{lissauer2009} and 
\citet{bodenheimer2013}. The integration starts at the point when gas accretion 
dictated by contraction exceeds disk-limited accretion, according to the local 
disk conditions. The calculation proceeds as follows. 
It is assumed that, during its evolution, the planet orbits at a fixed distance 
($a=5.2\,\AU$) from the Sun. 
The nebula gas has a viscosity parameter, $\alpha$ \citep{S&S1973}, constant 
both in space and time. 
Given an initial ($t=0$) gas surface density, $\Sigma \propto \ap/a$ (with an 
exponential cut-off beyond $\approx 40\,\AU$), where $a$ is again the heliocentric 
distance, gas is removed from the disk by accretion on the planet $d\Me/dt$, 
accretion on the Sun $\dot{\mathcal{M}}_{\nu}$, and photo-evaporation 
(i.e., disk wind) $\dot{\mathcal{M}}_{w}$. 

The accretion through the disk and onto the Sun, driven by viscous diffusion, 
is modeled by assuming steady-state conditions \citep{pringle1981},
\begin{equation}
    \frac{d\mathcal{M}_{\nu}}{dt}=3\pi\nu\Sigma,
    \label{eq:dac}
\end{equation}
in which  $\partial(\nu\Sigma)/\partial a\approx 0$.
Here $\nu$ represents the kinematic viscosity and is given by 
$\nu=\alpha H^{2} \Omega$, where $H$ is the pressure scale-height of the nebula 
and $\Omega=\sqrt{G \Msun/a^{3}}$ the local Keplerian frequency. 
The disk's thermal structure is such that $H\propto a^{9/7}$ \citep[e.g.,][]{chiang2010},
which makes $\nu$ nearly proportional to the heliocentric distance. 
The photo-evaporation rate is assumed to be constant, both in time and space, 
with $\dot{\mathcal{M}}_{w}\approx 10^{-8}\,\Msun\,\mathrm{yr}^{-1}$ 
\citep[e.g.,][]{gorti2016,picogna2019}. The gas density at the orbital distance
of the planet is calculated from the total disk mass $\mathcal{M}_{D}$, 
as the disk depletes, by preserving the initial slope $d\ln{\Sigma}/d\ln{a}$ 
of the distribution:
\begin{equation}
    -\frac{d\mathcal{M}_{D}}{dt}=\frac{d\mathcal{M}_{\nu}}{dt}+\dot{\mathcal{M}}_{w}+\frac{d\Me}{dt}
    \label{eq:dotMD}
\end{equation}
and $\Sigma(\ap,t)\propto \mathcal{M}_{D}(t)/(\pi a^{2}_{p})$. 
Preservation of $d\ln{\Sigma}/d\ln{a}$ with time stems from the fact that, 
in a steady-state disk, $\Sigma$ is nearly proportional to $1/\nu$. 
The mass $\Mp$ is integrated until $\mathcal{M}_{D}=0$.
In Phase~4 it is also assumed that $\dMp=\dMe$, a reasonable approximation 
since the gas accretion rate can be a factor of $10$, or more, larger than
the accretion rate of heavy elements around the onset of disk-limited accretion,
and the ratio increases as $\Mp$ grows (assuming that accretion of solids
is only caused by expansion of the feeding zone, see \cieq{eq:dMZdM}).

The disk models also assume that the initial reservoir of solids from which
the planet forms is the same as in \cieq{eq:sigma0}, independently of the
initial value adopted for the local gas surface density. 
This is not a global attribute of the disk (the solid component is not 
modeled), but rather a local property, spanning a few \AU\ in radius over 
the planet formation region. 
Globally (i.e., over tens of \AU) the initial total mass of solids 
may still be related to the initial total mass of gas.

Neglecting gas removal by planetary accretion, the computed lifetime of 
such a nebula model would be $\approx 3.5\,\Myr$ for $\alpha=4\times 10^{-3}$ 
and an initial disk mass of $0.1\,\Msun$. For $\alpha=4\times 10^{-4}$, 
the lifetime would be $\approx 8\,\Myr$, as mostly determined by the factor 
of $10$ difference in $\alpha$ 
\citep[the disk dispersal timescale is $\propto \alpha^{-1/3}$ for a constant 
$\dot{\mathcal{M}}_{w}$, e.g.,][]{gorti2015}. 
At the latter viscosity level, an initial nebula whose mass was $0.05\,\Msun$ 
would last $\approx 4.4\,\Myr$.

The evolution of the planet mass is found by solving the differential equation
\begin{equation}
    \frac{d\Mp}{dt}=\Psi(\Mp,\ap,H,\alpha)\Sigma(\ap,t),
    \label{eq:dotMp}
\end{equation}
where $\Psi$ is a parametric function of $\alpha$ derived from three-dimensional 
hydrodynamic simulations \citep{lissauer2009,bodenheimer2013}.
The local value of the gas surface density $\Sigma(\ap,t)$ depends on 
$\mathcal{M}_{D}(t)$, which is determined by integrating \cieq{eq:dotMD}. 
\cieq{eq:dotMp} is solved numerically by means of an Adams-Bashforth-Moulton 
method of variable order with adaptive step-size and error control.
The initial time at which the integration starts is taken from the structure 
calculation as the time at which the right-hand side of \cieq{eq:dotMp}
drops below the rate of accretion dictated by contraction (see \cifig{fig:acc}).
If a disk model predicts gas accretion rates on the planet greater than
that in \cifig{fig:acc} when $\Mp=60\,\Mearth$, we apply an extrapolation 
of $d\Me/dt$ obtained from the structure calculation data. 
More specifically, we use a linear regression in $\log{\dMe}$--$\log{\Mp}$
space, in the range $\Mp=25$--$60\,\Mearth$.
The extrapolated $d\Me/dt$ is applied until it exceeds the disk-limited 
accretion rate, i.e., the right-hand side of \cieq{eq:dotMp}.
Note that, as long as the planet is in contact with the nebula, i.e., 
during Phases~1 through 3, the gas disk model (of the type considered here)
is not expected to affect its evolution much.

Results from the mass integration under the assumption of disk-limited accretion 
are plotted in \cifig{fig:fmass}. The left-top panel shows the growth tracks 
of the planet in disks whose initial mass varies between $0.089$ and 
$0.102\,\Msun$ 
and $\alpha=4\times 10^{-3}$.  The left-bottom panel shows models in which 
the planet grows in less viscous disks, $\alpha=4\times 10^{-4}$, whose initial 
mass varies between $0.036$ and $0.048\,\Msun$. 
The gas surface density after $3\,\Myr$, at the radial location of the planet, 
$\Sigma(\ap)$, 
is indicated in the left panels in units of $\mathrm{g\,cm^{-2}}$. Solid circles
mark track ends, when $\Sigma(\ap)=0$. The right panels of the figure show
the gas accretion rate as a function of the planet mass for the same cases as
in the left panels.
In the Figure, the transition between Phases~3 and 4 occurs
at the beginning (leftmost positions) of the colored curves, 
and is dependent on the disk model.

In the models illustrated in the top panels, corresponding to more viscous disks,
the initial density ($t=0$) at the planet orbit ranges from 
$\approx 630$ to $\approx 730\,\mathrm{g\,cm^{-2}}$, 
yielding an initial
gas-to-solids mass ratio comparable to that reported by \citet{pollack1994}.
At $t=3\,\Myr$, $\Sigma(\ap)$ declines to values between $\approx 22$ and
$\approx 41\,\mathrm{g\,cm^{-2}}$. The final mass of the planet after gas
disperses ranges from $60$ to $339\,\Mearth$ and is achieved between
$t\approx 3.28$ and $\approx 3.43\,\Myr$. Final masses similar to Jupiter's
are achieved at $t\approx 3.42\,\Myr$, one million years after the beginning 
of Phase~3 and $\approx 150\,000$ years after the beginning of Phase~4.
The top-right panel of \cifig{fig:fmass} indicates that disk-limited accretion 
sets in at masses $\Mp\lesssim 60\,\Mearth$ when the initial gas surface density 
around the planet's orbit is $\Sigma\lesssim 700\,\mathrm{g\,cm^{-2}}$.

In the models corresponding to less viscous disks, illustrated in the bottom 
panels of \cifig{fig:fmass}, $\Sigma(\ap)$ at $t=0$ varies between $\approx 260$ 
and $\approx 340\,\mathrm{g\,cm^{-2}}$ whereas, after $3\,\Myr$, the gas surface
density at the planet's orbit drops to values between $\approx 21$ and
$\approx 91\,\mathrm{g\,cm^{-2}}$. 
In these examples, the solids necessary to form the planet would
initially be spread over a radial region about $2.5$ times as wide as 
in the $\alpha=4\times 10^{-3}$ disks if the local gas-to-solids mass ratio 
was the same, hence requiring some redistribution during (or prior to) 
planet assembly.
In the cases shown in the plot, the 
mass of the planet after gas dispersal ranges from $46$ to $316\,\Mearth$. 
Isolation starts between $t\approx 3.3$ and $\approx 4.2\,\Myr$. At the latter
time the planet would achieve a final mass similar to Jupiter's, $1.8\,\Myr$ 
after the beginning of Phase~3 and $\approx 930\,000$ years after the beginning 
of Phase~4.
Owing to the decreased ability of the disk to deliver gas to the planet, which
arises from the combination of a smaller accretion rate, $d\mathcal{M}_{\nu}/dt$, 
and deeper tidal gap at any given planet mass, disk-limited accretion sets in 
prior to $\Mp\approx 60\,\Mearth$ for the models considered here
(see bottom-right panel).

\begin{figure*}
    \centering
    \resizebox{1.0\linewidth}{!}{%
    \includegraphics[]{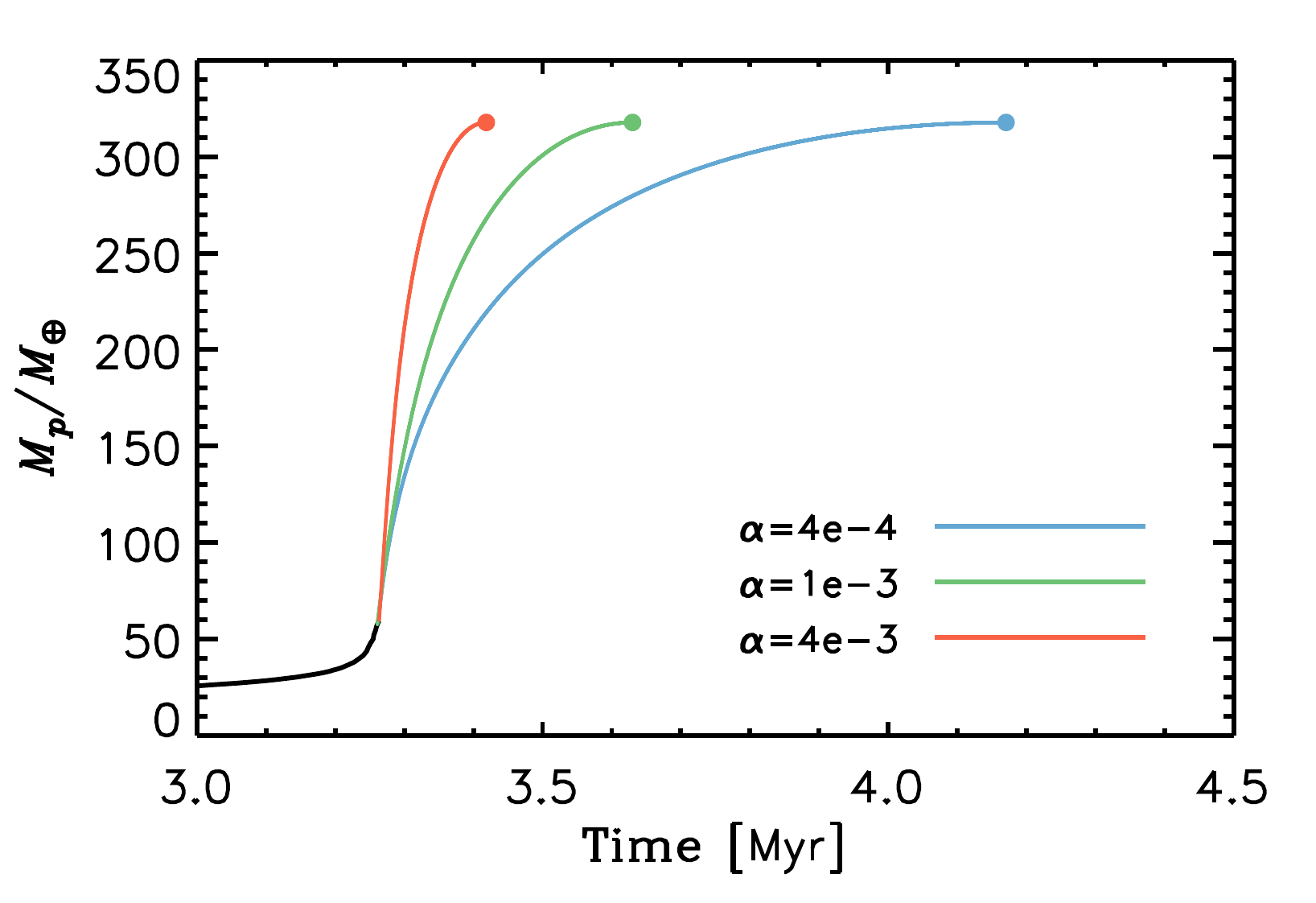}%
    \includegraphics[]{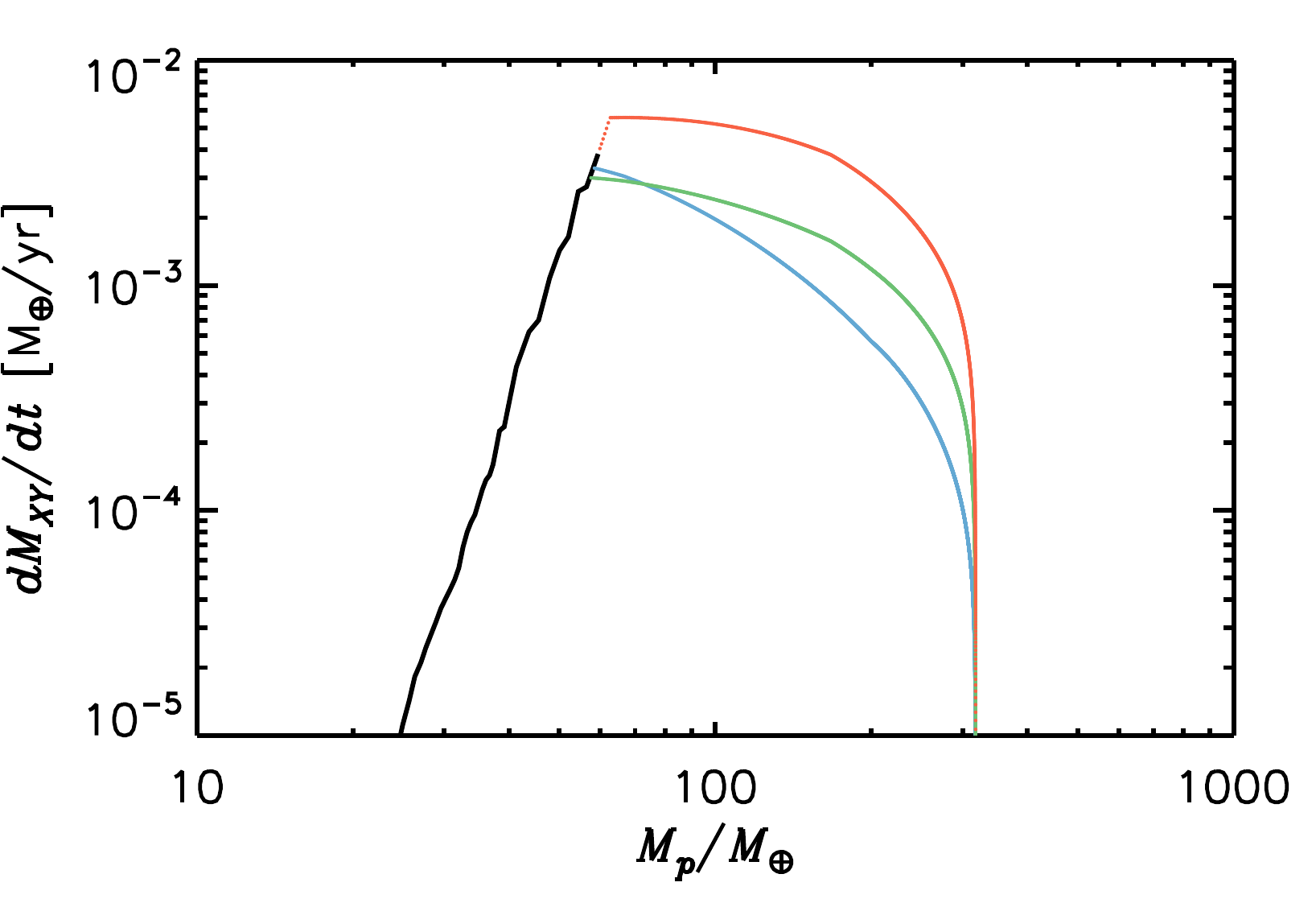}}
    \caption{
    Growth tracks and accretion rates as in \cifig{fig:fmass}, for three different
    values of the turbulence viscosity parameter $\alpha$, as indicated. 
    The initial disk mass is chosen as to result in a final planet mass 
    (after disk dispersal) equal to Jupiter's (see text for further details).
    }
    \label{fig:fmass2}
\end{figure*}
We apply the same modeling procedure to choose an initial ($t=0$) gas
surface density that would result in a final mass equal to Jupiter's, for 
three different values of the turbulence parameter $\alpha=4\times 10^{-4}$,
$10^{-3}$, and $4\times 10^{-3}$. Results are shown in \cifig{fig:fmass2}. 
In these models, the initial $\Sigma(\ap)$ ranges from $339$  to 
$722\,\mathrm{g\,cm^{-2}}$, for increasing values of $\alpha$, and formation
is complete between $t\approx 3.42$ and $4.17\,\Myr$. The initial disk 
(gaseous) mass is somewhat less than $0.05\,\Msun$ in the two less viscous
cases and $0.1\,\Msun$ in the highest viscosity case.
In all cases, the formation time is consistent with observational estimates 
of \emph{dust} lifetimes in circumstellar disks 
\citep[e.g.,][ and references therein]{gorti2016,ercolano2017}.

At the level of viscosity considered in the models displayed in \cifig{fig:fmass}
and \ref{fig:fmass2}, the accretion rate of gas declines with time when disk-limited
accretion sets in because, at those planet masses, disk-limited accretion is already 
past its maximum, $\partial\Psi/\partial\Mp<0$ in \cieq{eq:dotMp} 
\citep[see e.g.,][]{lissauer2009,bodenheimer2013},
and also because $\Sigma$ declines with time. 
For a given surface density, disk-limited rates start to decline for $\Mp$ 
satisfying \cieq{eq:gap_con} (within a factor of order unity), which occurs at several 
tens of Earth's masses ($\alpha=4\times 10^{-3}$) or less ($\alpha=4\times 10^{-4}$).

The right panel of \cifig{fig:fmass2} indicates that disk-limited accretion 
starts when $\Mp\approx 58\,\Mearth$ for the two cases with lowest viscosity, 
and that it starts earlier when $\alpha=0.001$. This implies that disk-limited 
accretion is (somewhat) lower at that planet mass when $\alpha=0.001$.
But since $\Psi$ in \cieq{eq:dotMp} is smaller for $\alpha=4\times 10^{-4}$
than it is for $\alpha=0.001$, the difference is caused by the lower surface 
density $\Sigma(\ap)$ at intermediate viscosity at the epoch when disk-limited 
accretion begins.

\begin{figure}
    \centering
    \resizebox{1.0\linewidth}{!}{%
    \includegraphics[]{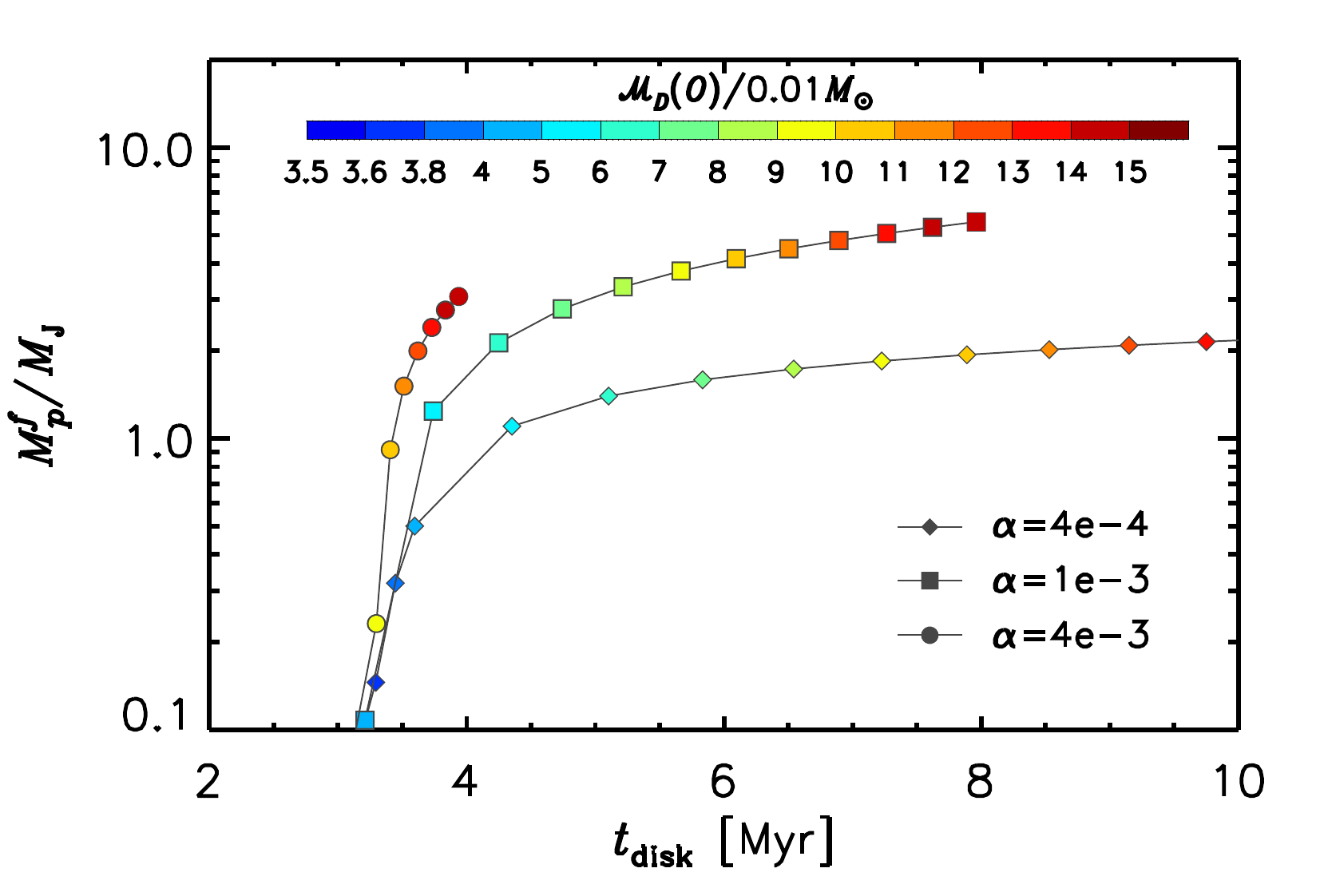}}
    \caption{
    Final mass of a planet, $M^{f}_{p}$, versus the disk's gas lifetime,
    $t_{\mathrm{disk}}$. 
    The vertical axis is in units of Jupiter's mass, $M_{\mathrm{J}}$.
    Different symbols indicate
    different values of the turbulence parameter $\alpha$ in the disk, 
    as reported in the legend. 
    Each symbol is color-coded according to the initial value of the 
    disk mass $\mathcal{M}_{D}(0)$, as indicated in the color bar in units of
    $0.01\,\Msun$.
    }
    \label{fig:mf}
\end{figure}
By applying the same turbulence viscosity as in the models displayed in
\cifig{fig:fmass2}, we integrate the disk-limited accretion rates of the right 
panel for a number of initial disk masses, $\mathcal{M}_{D}(0)$. 
As mentioned above, the gaseous nebula is supposed to extend over 
a radial distance of roughly $50\,\AU$.
Results are plotted in \cifig{fig:mf} for the final mass of the planet,
$M^{f}_{p}$, as a function of $t_{\mathrm{disk}}$, the nebula lifetime. 
Different symbols indicate 
the three different levels of turbulence viscosity (see legend) and colors render 
$\mathcal{M}_{D}(0)$. Within a timescale $\lesssim 5\,\Myr$, giant planets more 
massive than several times Jupiter's mass (\Mjup) can more easily form when gas 
viscosity is high. Low viscosity ($\alpha\ll 0.001$) disks are generally incapable 
of forming gas giants exceeding $\approx 2\,\Mjup$, despite the fact that they can
live much longer (for equal values of $\mathcal{M}_{D}(0)$). 
As a reference, for the largest
value of the disk mass used and $\alpha=4\times 10^{-4}$, the planet's final mass 
is $M^{f}_{p}\approx 2.25\,\Mjup$ and it would still be $<2.5\,\Mjup$ even if
$\mathcal{M}_{D}(0)$ was $0.2\,\Msun$.
It should be noted that, even though observations seem to suggest 
that average timescales for dust removal from disks are in the range
$3$--$5\,\Myr$, gas dispersal timescales could be longer 
\citep[e.g.,][]{gorti2016,ercolano2017}.

\subsubsection{Planetary Structure During Phase~4}
\label{sec:EDLA}

The structure and evolution of the planet during the phase of disk-limited
accretion is calculated by applying the gas accretion rates plotted in the
right panel of \cifig{fig:fmass2}, for the two cases with $\alpha=4\times 10^{-3}$
and $4\times 10^{-4}$, referred to as models \aqt\ and \aqq, respectively.
The structure calculation still assumes hydrostatic equilibrium, and the 
accreting gas is assumed to impact the planet's surface at a velocity close 
to free-fall, $v_{\mathrm{ff}}$. The density at $\Rp$ is derived from
the gas accretion rate and the free-fall velocity, 
$\propto \dot{M}_{XY}/(R^{2}_{p} v_{\mathrm{ff}})$.
The temperature at the surface is determined by an approximate solution 
of the radiation diffusion equation, applying the total luminosity of 
the planet (produced by contraction and accretion).
Quantitative details can be found in \citet{bodenheimer2000b}.

In this phase, the accretion of solids assumes continued 
supply of planetesimals from the edges of the gap in the solids' distribution
(see \cifig{fig:numsig}), according to \cieq{eq:dMZdM}. The accretion 
rate of heavy elements can therefore be written in terms of the gas accretion 
rate, as
\begin{equation}
    \frac{d\Mc}{dt}=%
    \left(\frac{C}{M^{2/3}_{p}-C}\right) \frac{d\Me}{dt}.
    \label{eq:MZdotDLA}
\end{equation}
The constant is chosen by using the ratio $\dMc/\dMe$ prior to the onset of 
disk-limited accretion, at the largest core mass for which the swarm of solids 
is evolved ($\Mc\approx 16\,\Mearth$).

The size distribution of accreted planetesimals does not change during this 
phase and accretion rates as a function of planetesimal radius are obtained by 
re-scaling the solids' mass in each size bin by $\dMc$ in \cieq{eq:MZdotDLA}. 
The relative contributions versus $R$ are thus constant during the evolution, 
and most of the heavy elements continue to be delivered by planetesimals 
larger than $10\,\mathrm{km}$ in radius (see \cifig{fig:accvsr}).
Because of ongoing accretion of solids, envelope-planetesimal interaction
calculations, including the calculation of grain opacity, are also performed 
during this phase (see \cisec{sec:sc}).

\begin{figure*}
    \centering
    \resizebox{1.0\linewidth}{!}{%
    \includegraphics[]{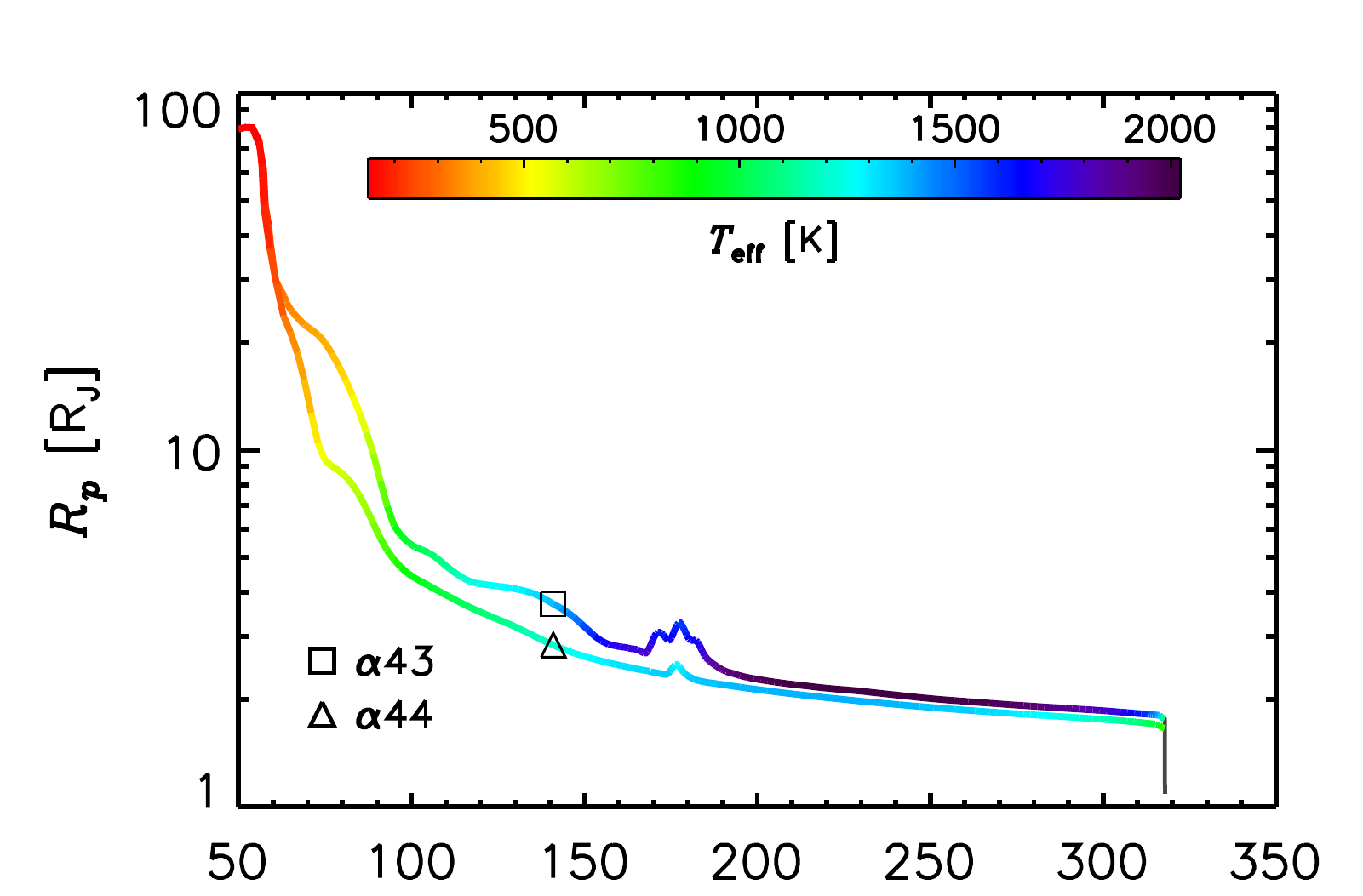}%
    \includegraphics[]{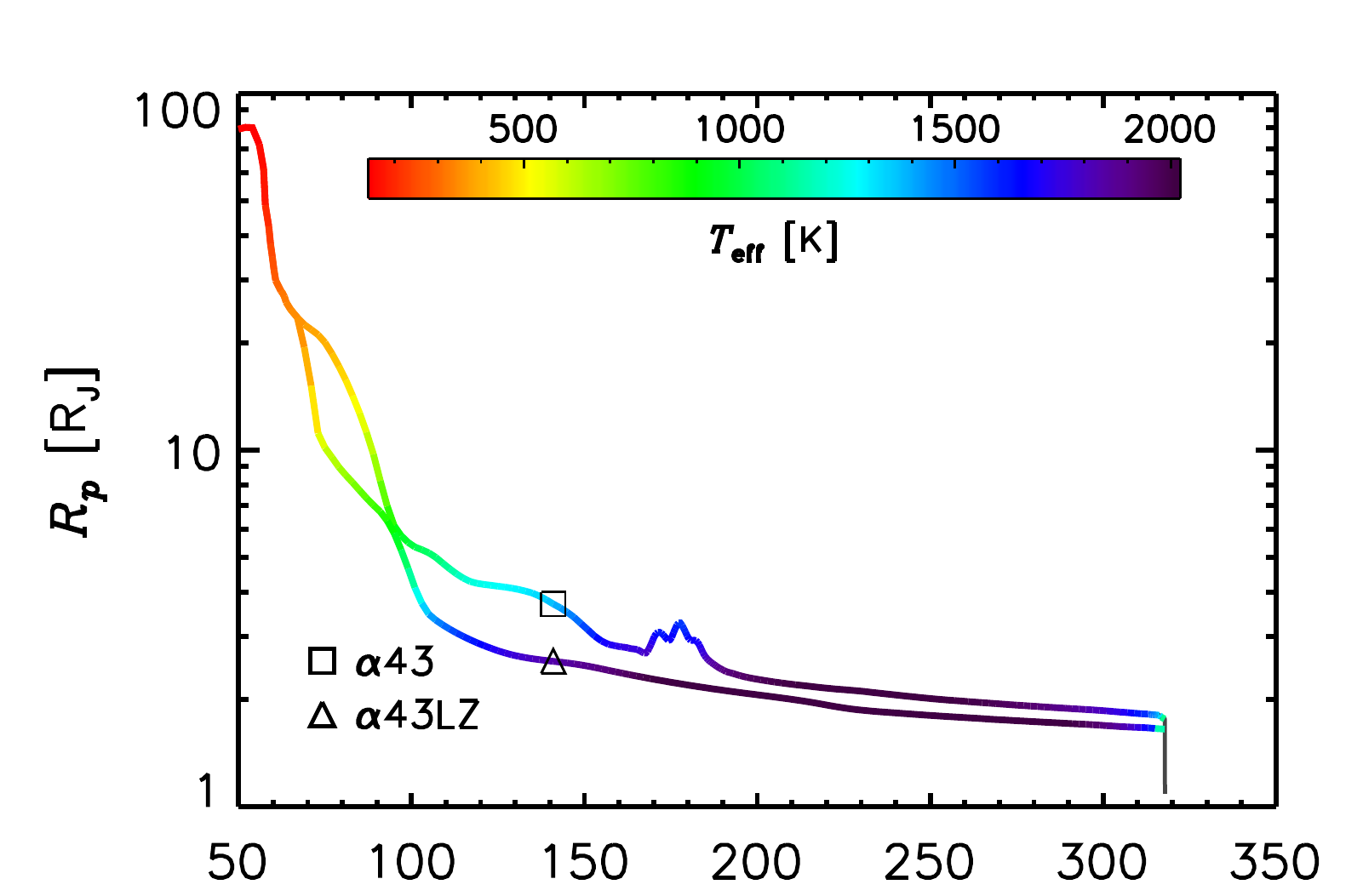}}
    \resizebox{1.0\linewidth}{!}{%
    \includegraphics[]{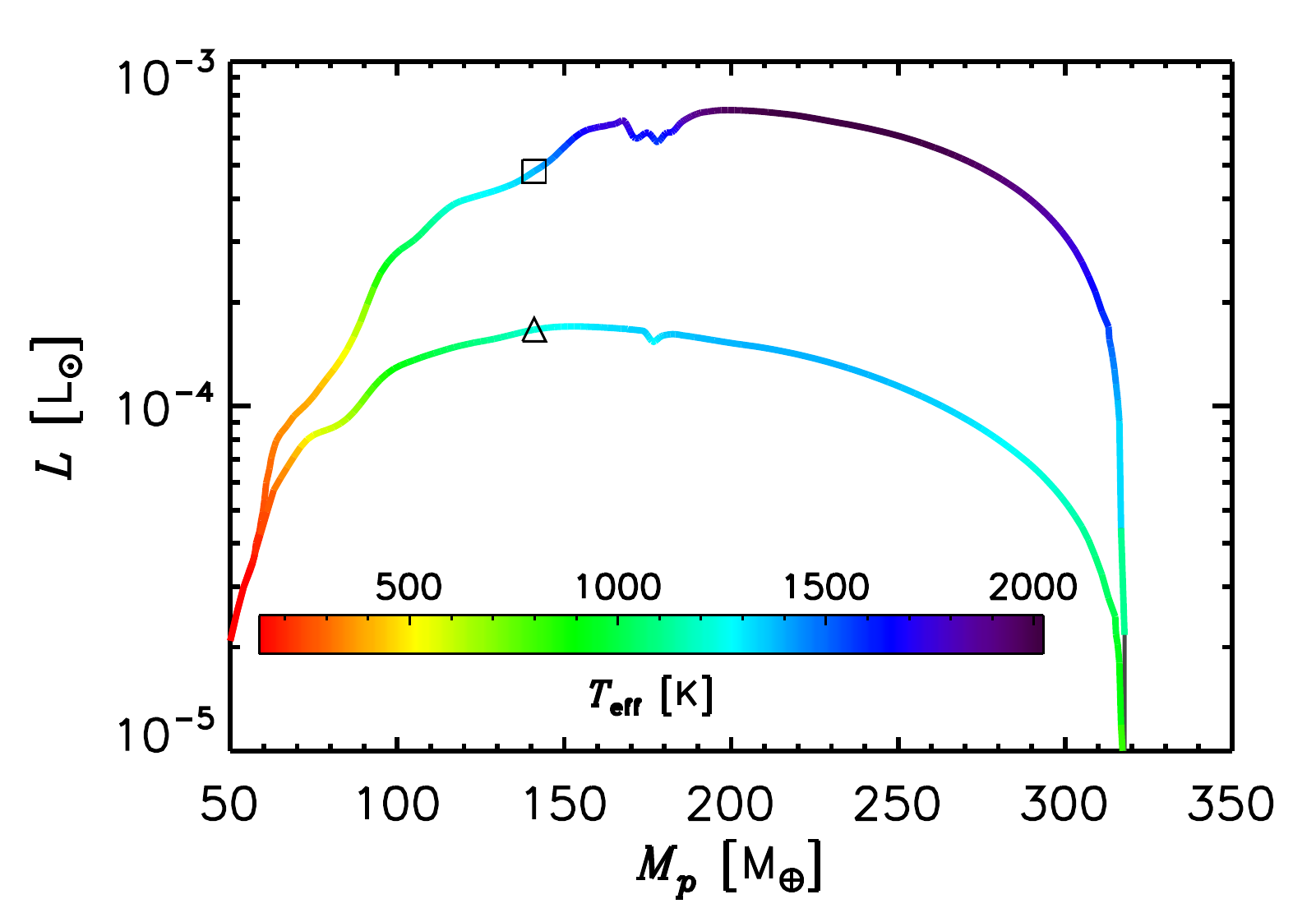}%
    \includegraphics[]{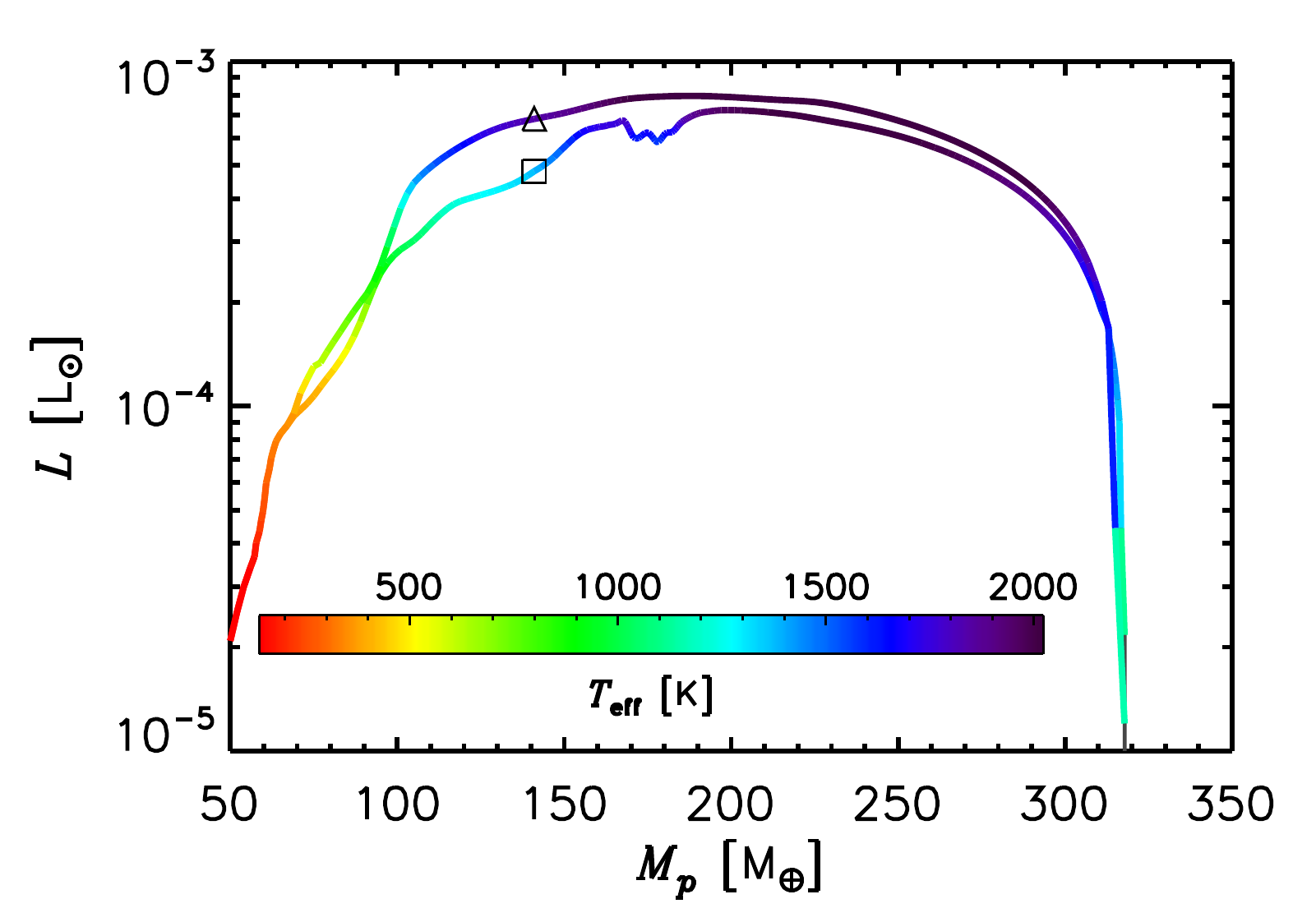}}
    \caption{
    Evolution of the planet during disk-limited accretion, according to 
    disk models described
    in \cisec{sec:DLA} and presented in \cifig{fig:fmass2}. Model \aqt\ and
    \aqq\ refer, respectively, to the nebula models with turbulence parameter
    $\alpha=4\times 10^{-3}$ and $4\times 10^{-4}$ (see legends in the top
    panels). Model \aqtlz\ refers to a variation of model \aqt, with lower heavy
    element content (see text).
    The planet radius (top) and luminosity (bottom) are plotted as a function 
    of the planet mass. The color bar renders the effective temperature of 
    the planet $T_{\mathrm{eff}}$, \cieq{eq:Teff}. 
    The transition between Phase~3 and 4 slightly depends on 
    the applied disk model (see \cifig{fig:fmass2}).
    The dark-gray (vertical)
    portion of the curves represents the isolation phase
    (see \cisec{sec:IP}).
    }
    \label{fig:p3}
\end{figure*}
The total mass of heavy elements at the end of Phase~4 (when $\Mp=1\,\Mjup$) 
is $\Mc\approx 20\,\Mearth$, similar in the two cases because the ratio 
$\dMc/\dMe$ at the end of Phase~3 (i.e., constant $C$ in \cieq{eq:MZdotDLA}) 
is comparable (note that the beginning of Phase~4 is somewhat different 
in the two models, see \cifig{fig:fmass2}).
The planet would collect this amount of solids if, according to \cieq{eq:MZiso},
it cleared planetesimals from a region of radial width $\approx 2.3\,\Rhill$ 
on either side of its orbit. 
Direct calculations performed by \citet{gennaro2015} indicate that, 
for a Jupiter-mass planet, $b\approx 2.5$ on average for 
$R\gtrsim 1\,\mathrm{km}$ planetesimals (which provide the bulk of 
the heavy-element mass in these models).

A variant of model \aqt\ is also considered, in which the supply of
solids is assumed to end around the epoch when disk-limited accretion begins
(model \aqtlz). In this model the total mass of heavy elements at the end
of formation is $\Mc=16.1\,\Mearth$. Once accretion of solids stops, there 
is no energy delivery by incoming planetesimals. Thereafter, it is assumed 
that dust grains in the outer envelope are only supplied by the gas accretion 
flow and the opacity calculation follows the approach of \citet{gennaro2016}.

Maximum rates of gas accretion are $\dMe\approx 3.3\times 10^{-3}$ and 
$\approx 5.6\times 10^{-3}\,\Mearth\,\mathrm{yr}^{-1}$, respectively 
in the less and more viscous nebula. Correspondingly, maximum accretion 
rates of heavy elements are $\dMc\approx 1.2\times 10^{-4}$ and 
$\approx 1.7\times 10^{-4}\,\Mearth\,\mathrm{yr}^{-1}$.
During this phase, as the planet mass grows by a factor of $\approx 5.5$,
the heavy-element mass increases by $25$\% at most. Apart from
compositional changes in the interior, most of the impact of solids' 
accretion during this time is due to the dust contribution to the opacity 
of the outer envelope's layers, which can affect contraction.

Results for the evolution of the planet radius and luminosity during 
the disk-limited accretion phase are presented in \cifig{fig:p3}.
Dark-gray (vertical)
portions of the curves represent evolution during the isolation 
phase, described in \cisec{sec:IP}.
As soon as the accretion rate of gas cannot keep up with the contraction
rate of the envelope, the planet starts to shrink. In doubling its mass,
from $\Mp=50$ to $100\,\Mearth$, the radius reduces by a factor of $\approx 16$, 
a contraction in volume of over $4000$ times (i.e., by more than $99.97$\%). 
And by the end of formation, the planet's volume shrinks another $96$\%. 
The radius $\Rp$ at the end of Phase~4 is comparable in the three models, 
ranging from $1.65\,\Rjup$ (model \aqtlz) to $1.76\,\Rjup$ (model \aqt).

As expected, the luminosity is lowest in model \aqq\ during most of 
the disk-limited accretion phase (see bottom panels), because of 
the relatively slow rate at which energy is supplied by gas accretion
(this part of formation in model \aqq\ lasts several times as long
as in the other models, see left panel of \cifig{fig:fmass2}). 
For $\Mp>100\,\Mearth$, 
the luminosity of model \aqt\ is between $2$ and $6$ times as large
as that of model \aqq, when compared at the same planet mass, and
about $2.6$ as large at the end of Phase~4.

Model \aqtlz\ is more luminous than model \aqt\ (but by $\lesssim 50$\%) 
during most of the disk-limited accretion phase, likely due to its 
less opaque outer envelope (therefore the planet also contracts 
somewhat faster initially, see top-right panel). However, the cooling 
rate slows down as $\Mp$ grows (note that energy delivered by gas 
accretion versus planet mass is similar in the two models) and, 
by the end of Phase~4,
$L$ drops below the luminosity of model \aqt\ (see \cifig{fig:p3},
bottom-right panel).
According to these models,
at the end of formation the young Jupiter is between several $10^{3}$
and $10^{4}$ times as luminous as it is today.

The curves in \cifig{fig:p3} are color-coded in terms of the effective
temperature of the planet, defined by
\begin{equation}
    L=4\pi \sigma_{\mathrm{SB}} R_{p}^{2} T^{4}_{\mathrm{eff}}, 
    \label{eq:Teff}
\end{equation}
where $\sigma_{\mathrm{SB}}$ is the Stefan-Boltzmann constant.
The effective temperature peaks at $\approx 2000$ and $\approx2200\,\K$,
respectively, in models \aqt\ and \aqtlz. In the less viscous nebula
model, $T_{\mathrm{eff}}$ has a maximum at $1400\,\K$. However, 
the planet in model \aqq\ remains ``warm'' ($T_{\mathrm{eff}}>1000\,\K$) 
for about $7\times 10^{5}$ years, a long period of time compared to only
$1.4\times 10^{5}$ years in models \aqt\ and \aqtlz. In all three cases,
this state lasts from when $\Mp\approx 100\,\Mearth$ until formation is 
nearly complete.
At the end of Phase~4, $T_{\mathrm{eff}}$ ranges between $\approx 750$ 
(model \aqq) and $\approx 940\,\K$ (model \aqt).

\subsection{Isolation Phase}
\label{sec:IP}

\begin{figure}
    \centering
    \resizebox{1.0\linewidth}{!}{%
    \includegraphics[]{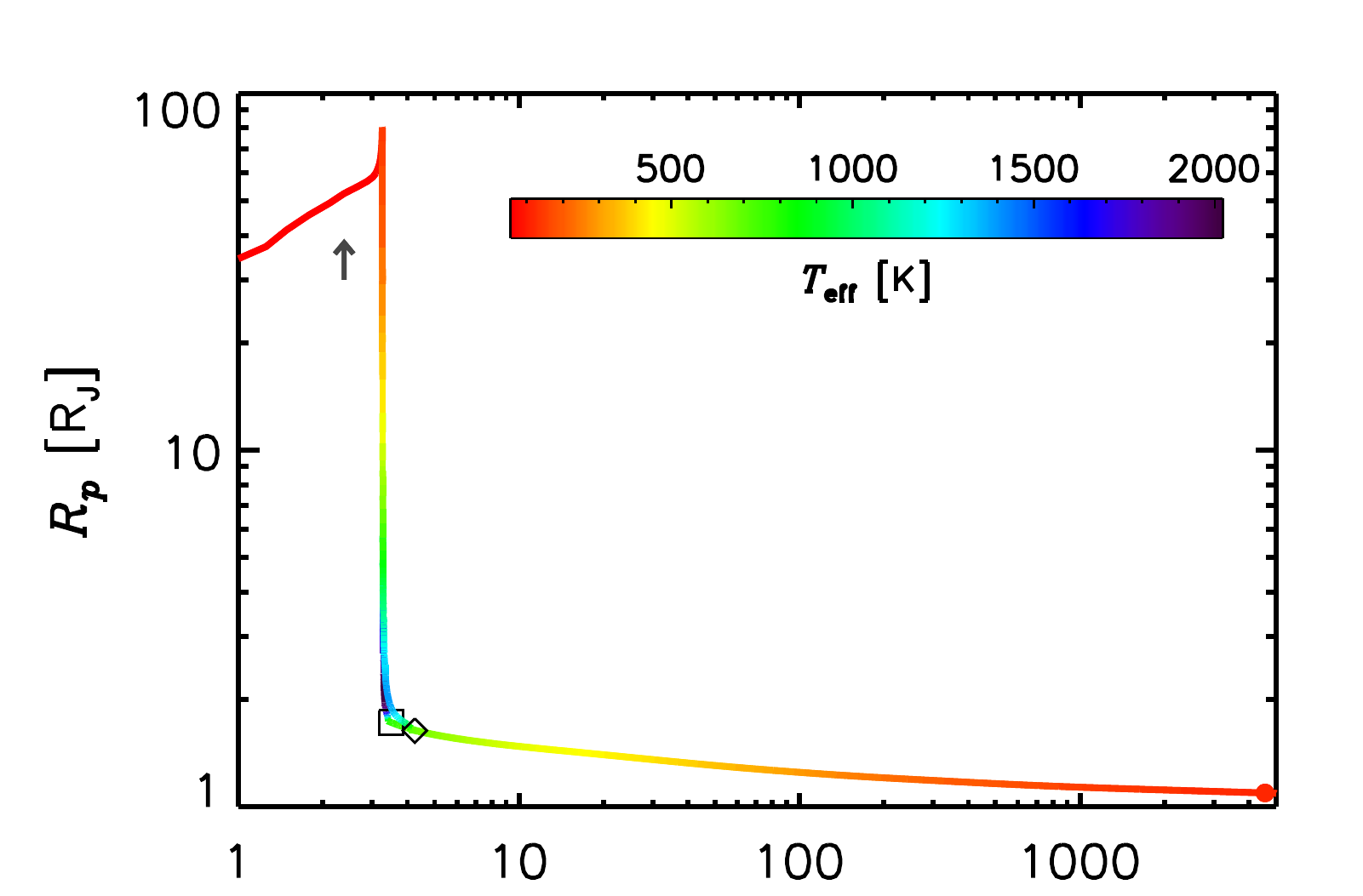}}
    \resizebox{1.0\linewidth}{!}{%
    \includegraphics[]{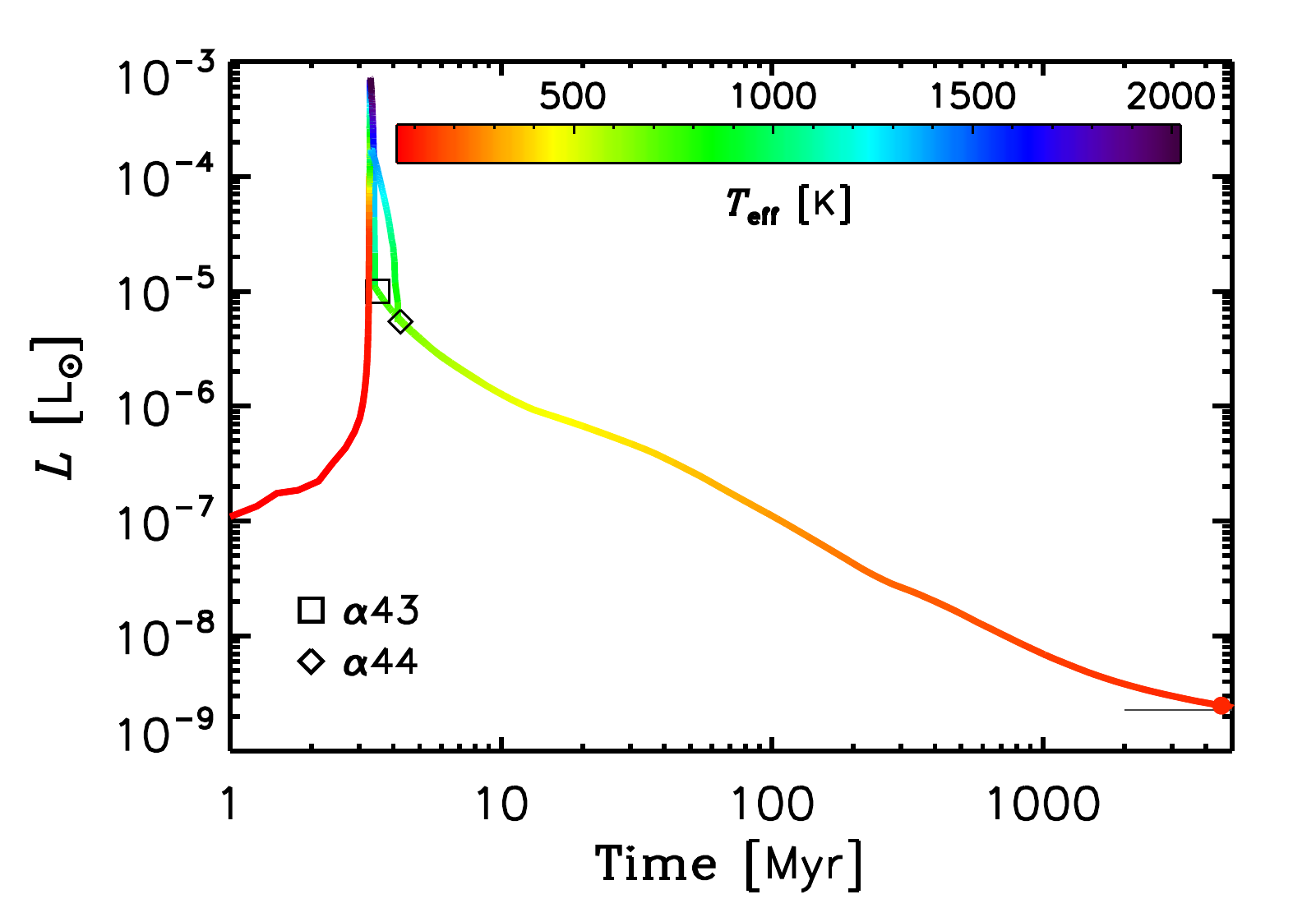}}
    \caption{
    The same quantities as in \cifig{fig:p3} are plotted for models 
    \aqt\ and \aqq\ as a function of time, starting from Phase~2 and 
    continuing to the present epoch.
    Phase~2 ends at $t\approx 2.38\,\Myr$ in both cases (see arrow);
    Phase~3 ends at $t\approx 3.3\,\Myr$, with a slight 
    dependence on the disk model (see \cifig{fig:fmass2}).
    The square and diamond symbols indicate the end of Phase~4 in the two
    models (see legend in the bottom panel). 
    The solid circles mark the age of the solar system, $t=4567\,\Myr$.
    The color bar renders the planet's effective temperature (see \cieq{eq:Teff}).
    During the isolation phase the two models are hardly distinguishable.
    The horizontal segment in the bottom panel indicates Jupiter's current
    luminosity (see text).
    }
    \label{fig:iso}
\end{figure}

The evolution during the isolation phase is calculated until 
$t\approx 5000\,\Myr$, somewhat longer than the reference age 
of the solar system, $4567\,\Myr$ \citep{connelly2012}.
Standard photospheric boundary conditions \citep[e.g.,][]{cox2006,kippenhahn2013} 
are applied in this phase, as the planet cools and contracts, 
and the envelope is assumed to be free of dust.

During isolation, the energy absorbed from the Sun can provide 
a significant contribution to the thermal state of the upper 
envelope, and hence it may alter the contraction of Jupiter. 
Solar irradiation is taken into account following the approach 
outlined in \citet{gennaro2016}. 
The energy flux impinging on the planet and absorbed by 
the envelope is parameterized in terms of an ``equilibrium'' 
temperature $T_{\mathrm{eq}}$, such that
\begin{equation}
    4\sigma_{\mathrm{SB}} T^{4}_{\mathrm{eq}}=
    \frac{\Lsun}{4\pi a^{2}}(1-\mathcal{A}),
    \label{eq:T4eq}
\end{equation}
where $\mathcal{A}$ represents the planet's albedo. 
The total luminosity of the planet is then written as
\begin{equation}
    L=L_{\mathrm{int}}+
      4\pi \sigma_{\mathrm{SB}} R_{p}^{2} T^{4}_{\mathrm{eq}}.
    \label{eq:Ltot}
\end{equation}
Loosely speaking, the internal luminosity $L_{\mathrm{int}}$ 
is the power generated by internal energy sources, which the 
planet would have if it was not irradiated by the Sun. 
Strictly speaking, the irradiation flux applied to the top 
part of the envelope can affect the temperature structure in
the underlying layers and, therefore, $L_{\mathrm{int}}$ can 
also depend on $T_{\mathrm{eq}}$, although the effect is 
expected to be small in the case of Jupiter.
According to \cieq{eq:Teff} and \cieq{eq:Ltot}, $T_{\mathrm{eff}}$ 
would converge to $T_{\mathrm{eq}}$ as the planet cools.

We use a present-day value for Jupiter's albedo $\mathcal{A}=0.343$,
based on Pioneer and Voyager~1 observations \citep{hanel2003}, 
and for the Sun's luminosity ($\Lsun$), 
which correspond to a constant $T_{\mathrm{eq}}\simeq 110\,\K$. 
This approximation somewhat overestimates the second term on 
the right-hand side of \cieq{eq:Ltot} since the Sun's luminosity has risen 
by about $30$\% over the past $4000\,\Myr$ \citep{ribas2010,spada2013}.
The consequences are quantified in Appendix~\ref{sec:CSC}, 
where results are presented of a version of model \aqt\ that uses 
a time-dependent solar irradiation.
Results from using an alternative value for the albedo are also 
discussed in the Appendix.

\cifig{fig:iso} shows the evolution of Jupiter's radius and luminosity
during most of the planet's history, including most of Phase~2
as well as Phases~3, 4, and the isolation phase
(see caption of \cifig{fig:iso} for further details).
Only models \aqt\ and \aqq\ are plotted (see legend in the
bottom panel). Their behavior is very similar during the isolation
phase, and very similar to that of model \aqtlz. The curves are again
rendered according to $T_{\mathrm{eff}}$. 
Square and diamond symbols mark the end of Phase~4 in the two models,
whereas the solid circles on the right-hand ends of the curves mark the age 
of the solar system. 

Although the three models begin the isolation phase with luminosities 
differing by a factor of $\approx 1.8$ ($\Rp$ differs by much less, 
about $7\%$), $L$ is basically the same by $t=30\,\Myr$.
Thereafter, the models can be hardly distinguished.
At the age of the solar system, model variations in radius and 
luminosity are below $1$\%, with $\Rp=1.09\,\Rjup$, 
$L=2.47\times 10^{-9}\,\Lsun$, about $10$\% in excess of Jupiter's 
average emitted power 
\citep[$2.25\times 10^{-9}\,\Lsun$, as measured from Cassini data by][]{lil2012}, 
and $T_{\mathrm{eff}}=123\,\K$, close to the values of $124$ or $125\,\K$, 
derived respectively from Voyager~1 \citep{hanel2003} and Cassini 
measurements \citep{lil2012}.

\section{Discussion and Conclusions}
\label{sec:dac}

The formation of a giant planet by core-nucleated accretion requires the growth
of a condensed core, several times the Earth's mass, to which gas can bind.
This work assumes that the initial reservoir of solids (tens of meters to
hundreds of kilometers in size), from which the core grows, is locally available.
This assumption implies a high surface density of solids around 
the formation region. We do not speculate about the origins of these solids.
However, if accessible, a sufficiently large concentration of planetesimals 
can indeed provide an efficient mechanism for planetary growth 
\citep[see also][]{voelkel2020}.

The calculations of \citetalias{gennaro2014} showed that a planet's
tenuous atmosphere can substantially enhance the accretion of solids 
during the early stages of the planet's growth (Phase~1; see \cifig{fig:mzmxy}), 
and an evolving swarm of planetesimals (which accounts for mutual interactions
and gas drag, among other effects) contributes to reduce the growth timescale 
of $\Mc$ until it achieves a few Earth masses (compare models in the left panel 
of \cifig{fig:acomp}).
However, once $\dMc$ drops below $\dMe$ and formation enters Phase~2, 
planet's growth slows down (see again \cifig{fig:acomp}). 
Compared to previous models \citep[e.g., $\sigma 10$ of][which
uses the same $\sigma^{0}_{Z}$ at $5.2\,\AU$ as assumed here]{naor2010} 
the core mass is less and the accretion rates of gas and solids are
lower, all of which contribute to a lower luminosity.
Importantly, the smaller $\dMe$ indicates that the cooling 
(i.e., contraction) timescale of the envelope, $\propto \Mc\Me/L$, 
is longer in the new model. 
As a result, compared to model $\sigma 10$, 
Phase~2 lasts much longer: $2\,\Myr$ versus $0.5\,\Myr$. 
It should be noted that the short growth timescale predicted by 
the old model may actually be difficult to reconcile with the formation 
of Jupiter, since it would require a relatively massive nebula (a few 
to several times $0.01\,\Msun$) dispersing in a little over $1\,\Myr$,
assuming formation begins early in the evolution of the nebula
\citep[see, e.g.,][]{najita2014}.
The present calculations obviate this problem.

Recently, \citet{alibert2018} presented a scenario for Jupiter's
formation based on the accretion of centimeter-size solids followed
by the accretion of planetesimals. In their calculations, the planet's
mass varies from $5\,\Mearth$ to $16\,\Mearth$ after $1\,\Myr$,
a range of values that include the results illustrated in \cifig{fig:mzmxy}.

During Phase~2, $\Me$ increases about linearly in time, at an average
rate of $\approx 4.8\times 10^{-6}\,\Mearth\,\mathrm{yr}^{-1}$, whereas
the heavy-element mass grows on average at 
$\approx 1.2\times 10^{-6}\,\Mearth\,\mathrm{yr}^{-1}$. 
This result is also similar to the findings of \citet{alibert2018}, 
who concluded that, between $\approx 1$ and $\approx 3\,\Myr$, 
the planet accreted heavy elements at a rate of order
$10^{-6}\,\Mearth\,\mathrm{yr}^{-1}$.
The longest phase of formation is also the faintest (apart from 
the initial stages of Phase~1, when $\Mc\ll 1\,\Mearth$), see 
\cifig{fig:lum}, with luminosities hovering just 
above $10^{-7}\,\Lsun$, and not exceeding a few times this value. 
The energy radiated per unit time is similar to that provided by solids' 
accretion, \cieq{eq:Lacc}.

The relatively slow contraction of the envelope affects also the beginning
of Phase~3 ($\Me=\Mc$), as the planet takes over $0.8\,\Myr$ to double its
mass, from $20$ to $40\,\Mearth$ (including heavy elements), at an average
gas accretion rate of $2\times 10^{-5}\,\Mearth\,\mathrm{yr}^{-1}$.
This outcome is against the general credence that, past the cross-over 
mass, the growth timescale of a planet is much shorter than the remainder 
of the nebula lifetime. 
However, it helps to reconcile the formation of planets like Saturn, for
which gas starvation by gap formation in a protosolar nebula is generally
ineffective \citep[][and references therin]{lissauer2009}.
The outcome is also in accord with the results of \citet{alibert2018},
whose models show that Jupiter attains $\Mp\approx 50\,\Mearth$ at around
$3\,\Myr$.

The onset of rapid gas accretion ($\dMe\gg 10^{-3}\,\Mearth\,\mathrm{yr}^{-1}$)
occurs somewhat after $t\approx 3.2\,\Myr$, when $\Mp$ is nearly 
$60\,\Mearth$. 
A circumstellar disk of this age around a solar-mass star may 
sustain an accretion rate (through the disk and on the star) of order
$10^{-8}\,\Msun\,\mathrm{yr}^{-1}$ \citep{hartmann2016}, and only 
a fraction is made available to the planet because of disk-planet 
tidal interactions.
From this time onward, the evolution of the planet is tied to 
that of the surrounding nebula. Simplified accretion disk models, 
connected to Phases~1, 2, and 3 of the planet's evolution, indicate 
that the formation of Jupiter can occur within $3.4$--$4.2\,\Myr$ 
(see \cifig{fig:fmass2}), in a disk whose initial mass is between 
$0.05$ and $0.1\,\Msun$ (and typical disk's photo-evaporation rates).
Depending on the level of kinematic viscosity and disk mass, gas 
giants can be generated ranging from sub-Saturn to several Jupiter
masses, within a few to several $\Myr$ (see \cifig{fig:fmass} and 
\cifig{fig:mf}).
For a given disk model, the single largest factor that determines the 
planet's final mass is likely the duration of Phase~2, which is fairly
independent of nebula conditions (provided the disk is not too old 
when the phase begins).

As the planet evolves along Phase~4, undergoing disk-limited gas 
accretion, it rapidly contracts as $\Mp$ increases. The planet's luminosity
rises to values between $10^{-4}$ and $10^{-3}\,\Lsun$.
Since the planet's mass is $\gtrsim 100\,\Mearth$, and hence its orbit
would lie in a tidally-produced gap (in the gaseous disk), 
these may represent the most favorable conditions to observe a giant planet
in formation. The window of opportunity may be short, $\sim 10^{5}$ years,
for more luminous planets, but relatively long, $\sim 10^{6}$ years,
for fainter planets.
In these models, Jupiter's effective temperature drops below $1000\,\K$ 
by the end of formation, and the longer the formation time the lower 
is $T_{\mathrm{eff}}$.
It should be emphasized that the power emitted by the planet during the 
disk-limited accretion phase is primarily that arising from gravitational 
energy released by the infalling gas, at a surface shock, under 
the assumption that all energy released there is radiated. 
The intrinsic luminosity arising from the planet's interior is likely 
considerably lower 
\citep[see, e.g.,][]{marley2007,mordasini2012a,mordasini2017}.

The planet acquires most of its heavy-element content, 
$\Mc\approx 16\,\Mearth$, prior to the onset of disk-limited accretion, 
during which time we model the detailed evolution of the planetesimal swarm 
and the interactions of the solids with the planet. 
Of note is that \citet{alibert2018} also found that up to $20\,\Mearth$
worth of heavy elements can be delivered prior to Phase~4.
Thereafter, we assume 
that the planet accretes from the edge of the feeding zone, as this region 
expands into an undepleted reservoir of solids. Only about $4\,\Mearth$ 
worth of heavy elements are accreted at later stages, a mass determined 
by the gas-to-solids accretion rates at the beginning of Phase~4
and in agreement with the assessment of \citet{shibata2019}.
Although Jupiter 
could in principle keep accreting planetesimals well after formation, 
a mechanism is required to deliver solids within a few Hill radii from 
the planet's orbit. 
The heavy-element content of the young Jupiter would be consistent with 
the predictions of \citet{saumon2004} and \citet{militzer2008}, based 
on interior models of current Jupiter matching gravity data, but 
it would fall short of other predictions 
\citep[e.g.,][]{nettelmann2012,debras2019}.

The evolution in isolation of observables such as radius and luminosity 
(or $T_{\mathrm{eff}}$) soon converges in the three models \aqt,
\aqtlz, and \aqq, and is basically the same after about $30\,\Myr$.
The planet's radius is mostly influenced by the envelope's thermal
state, its composition, and equation of state. This is likely the reason 
that all models, including those presented in Appendix~\ref{sec:CSC}, 
result in similar radii at the age of the solar system, exceeding 
by $6$\% to $9$\% Jupiter's present (volumetric mean) radius.
The lack of higher abundances of heavy elements in the envelope 
contributes to the larger value of $\Rp$ predicted by these models.

The luminosity of the planet (about $10$\% in excess of the observed 
value, see \cifig{fig:iso}) can be affected by the amount of power 
absorbed from solar radiation. 
The impact of albedo $\mathcal{A}$, and quite likely its history, 
is non-trivial.
Estimates of $\mathcal{A}$ were revised upward by using Cassini
measurements \citep[see][]{lil2018} and, as reported in the Appendix, 
the higher albedo can lower luminosities in the models by $15$\% at 
$t=4567\,\Myr$. 
Likewise, a time-varying solar luminosity can impact Jupiter's power 
output by several percent, at the age of the solar system, as also 
discussed in the Appendix.

At the current epoch, in Jupiter models \aqt\ and \aqq, the radius of 
the heavy-element core is $R_{Z}\approx 20\,800\,\mathrm{km}$,
about $27$\% of the planet's radius. This result does not agree with 
the static (non-evolved) models of the planet that are needed to explain
the gravity measurements acquired by the Juno spacecraft. 
As an example, in a model by \citet{debras2019} the heavy-element mass 
fraction $X_{Z}$ decreases with radius, and is considerably more dilute 
than in ours, with $X_{Z}$ decreasing to $10$\% at $60$\% of Jupiter's radius. 
In \citet{wahl2017}, the heavy-element dilute core has a constant $X_{Z}$ 
with radius, extending out to about $0.5\,\Rjup$.
These interior structures are not consistent with the results of models 
presented herein, which calculate the full formation and evolution history 
of the planet.
Even if heavy-element deposition in the envelope and the resulting 
compositional gradients were taken into account, significant discrepancies 
would likely persist \citep{lissauer2020}.

\appendix
\section{Model Comparisons and Refinements}
\label{sec:CSC}

\begin{figure}
    \centering
    \resizebox{1.0\linewidth}{!}{%
    \includegraphics[]{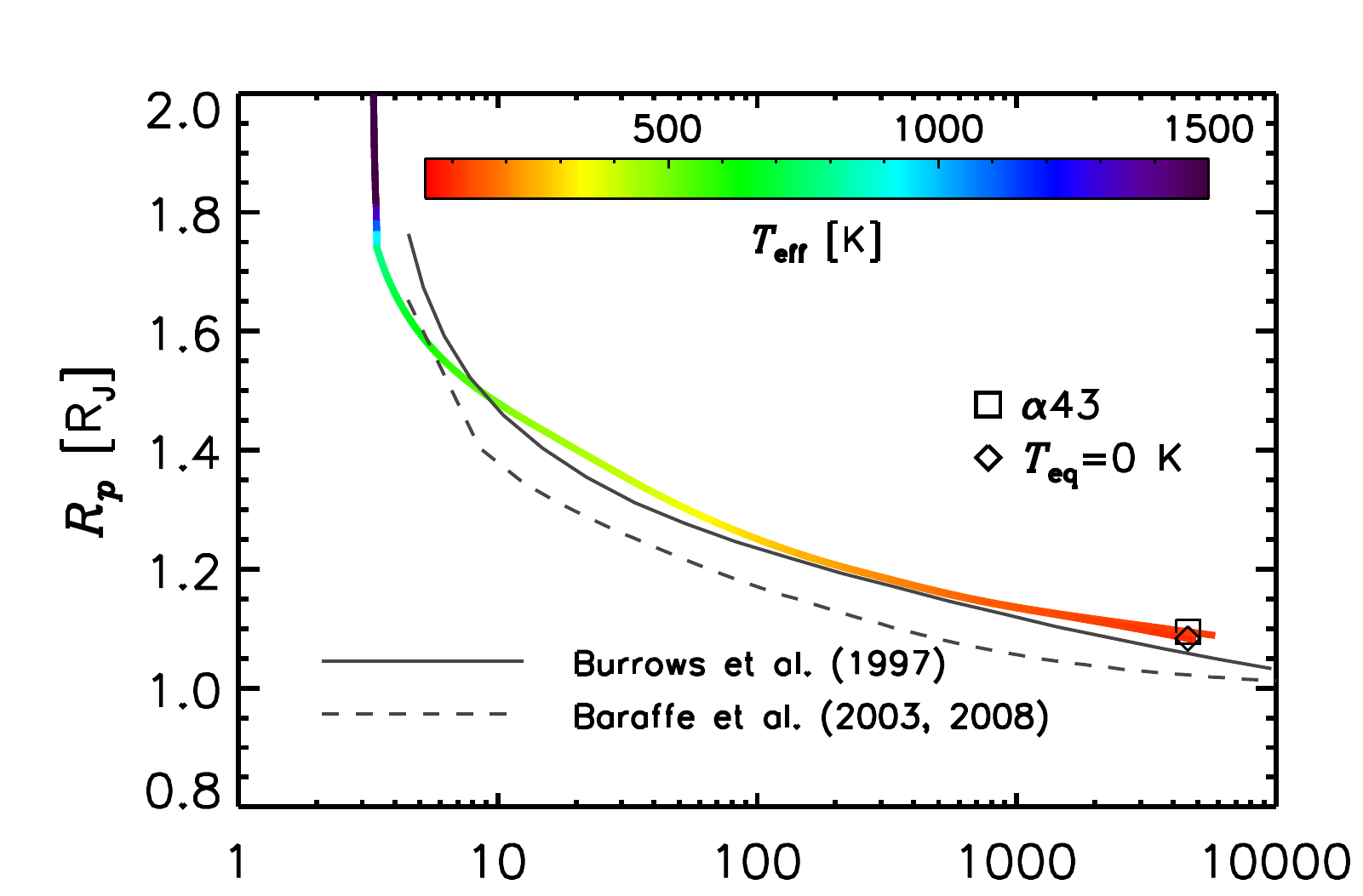}}
    \resizebox{1.0\linewidth}{!}{%
    \includegraphics[]{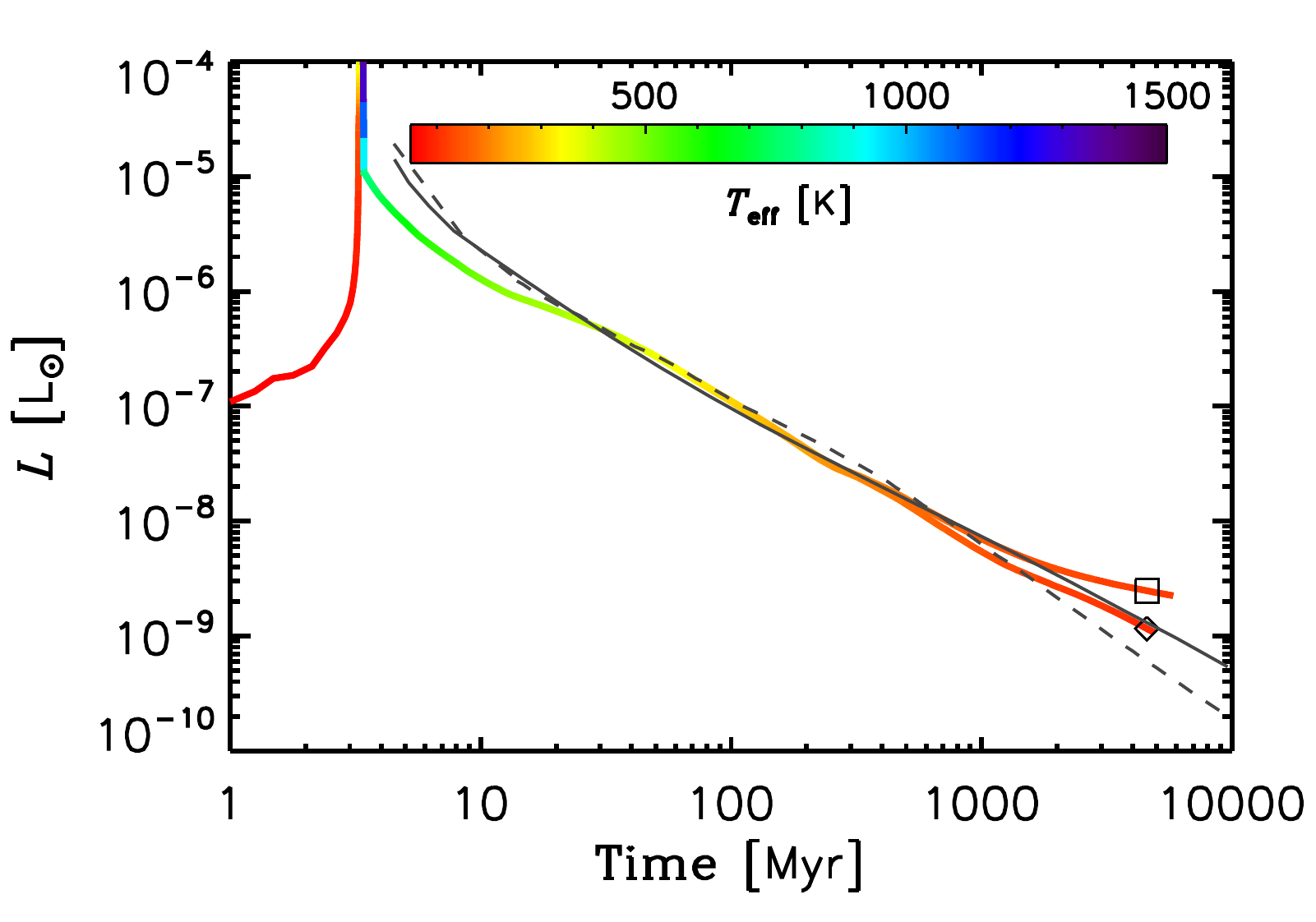}}
    \caption{
    Comparison of the radius (top) and luminosity evolution (bottom)
    of model \aqt\ and of an analogous model that 
    neglects solar irradiation ($T_{\mathrm{eq}}=0\,\K$, see legend
    in the top panel)
    against Jupiter's post-formation models of \citet{burrows1997} 
    and \citet{baraffe2003,baraffe2008}, as indicated, which do not 
    include solar 
    irradiation. Color bars render the planet's effective temperature
    of the present calculations.
    }
    \label{fig:bb}
\end{figure}
\cifig{fig:bb} shows the results for the radius (top) and luminosity
evolution (bottom) from model \aqt\ and from a version of the same model
that does not include solar irradiation (i.e., $T_{\mathrm{eq}}=0\,\K$ 
in \cieq{eq:Ltot} during the evolution in isolation).
The curves are marked, respectively, by a square and a diamond.
The figure also shows results from the post-formation calculations of 
\citet{burrows1997} and \citet{baraffe2003,baraffe2008} 
(see legend in the top panel). 
These latter models assume non-gray atmospheres, a more sophisticated 
treatment of radiation transfer compared to the frequency-independent
approximation \citep[e.g.,][]{gray1992,m&m} adopted in the calculations
presented herein. For purposes of comparison, the evolutionary tracks 
of \citeauthor{burrows1997} and \citeauthor{baraffe2003} are shifted
in time so that they start at the end of Phase~4
of our formation calculation. 
The case with $T_{\mathrm{eq}}=0\,\K$ is illustrated because the non-gray 
atmosphere calculations neglect solar irradiation during evolution. 
As can be seen in the figure by comparing the irradiated and 
non-irradiated model, the effects are mostly negligible on $\Rp$, 
whereas absorption of solar radiation affects the planet's power 
output after several $100\,\Myr$ (the difference is less than $10$\%
at $t=500\,\Myr$).
At the age of the solar system, marked by the symbols, maximum 
differences in radius between gray and non-gray atmosphere calculations 
are within $7$\%. 
The post-formation luminosity tracks are similar in all models, up to
$t\approx 1000\,\Myr$. They begin to diverge afterward, with the gray
atmosphere (non-irradiated) calculation more closely following 
\citeauthor{burrows1997}'s model.

\begin{figure}
    \centering
    \resizebox{1.0\linewidth}{!}{%
    \includegraphics[]{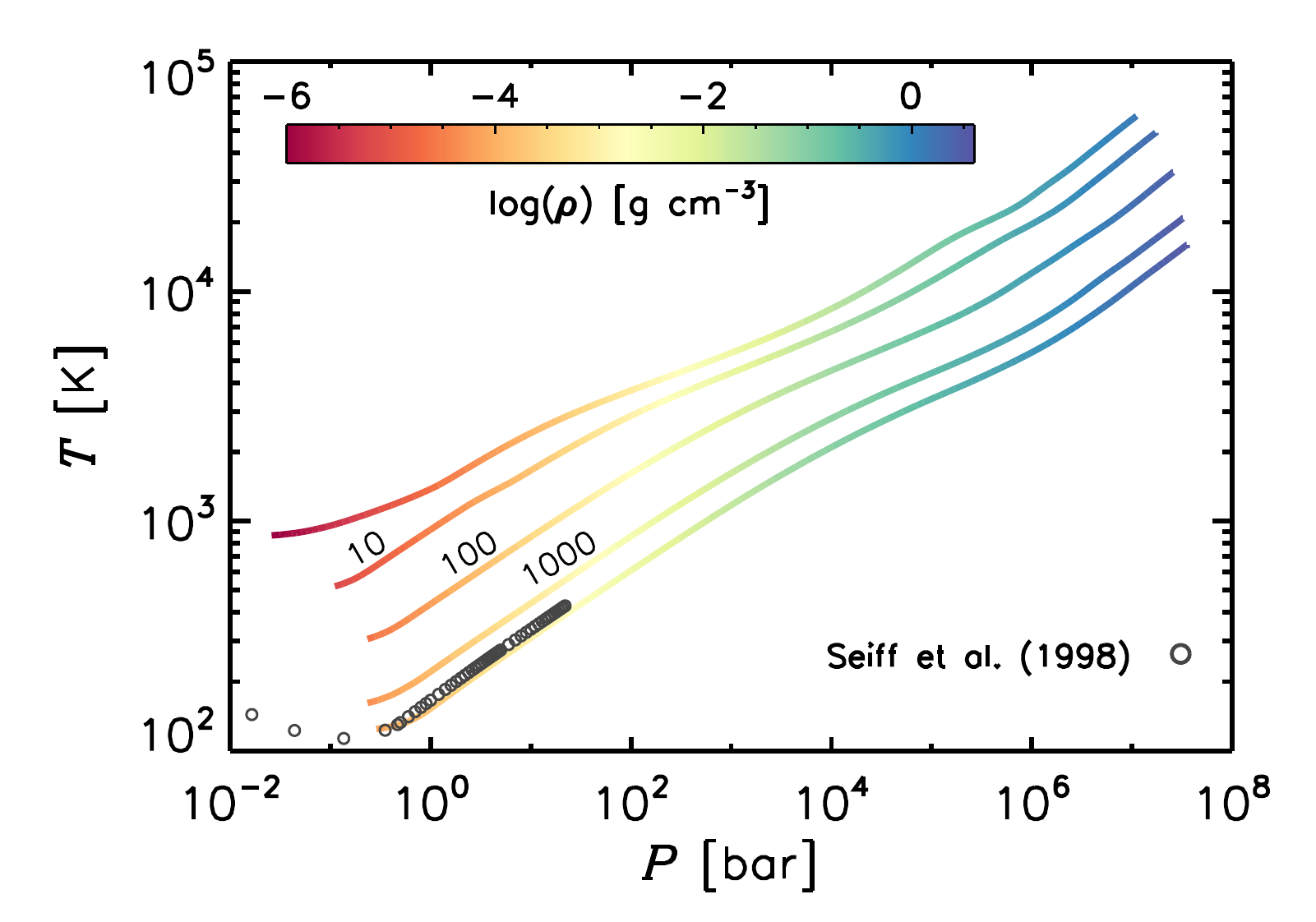}}
    \caption{
    Temperature versus pressure in the planet's envelope at various 
    post-formation epochs, in a variant of model \aqt\ with a
    compressible core (see text for details on the model).
    The uppermost curve refers to a time soon after formation ends
    ($t\approx 3.5\,\Myr$). Numbers on the three middle curves mark 
    the epochs in $\Myr$. The lowest curve refers to $t=4.6\,\Myr$.
    The open circles represent results from the analysis of \citet{seiff1998},
    based on the data acquired by the Galileo entry probe.
    The curves are color-coded in terms of the gas density at each layer.
    }
    \label{fig:seiff}
\end{figure}
Temperatures and pressures inside the planet are plotted in \cifig{fig:seiff}
at a few epochs after formation, with the top curve indicating a stage
right after formation ends (see figure's caption for details). 
The next three curves beneath represent epochs at $10$, $100$, and $1000\,\Myr$, 
respectively, and the last one corresponds to the current age of the planet. 
The colors render the envelope density in logarithmic scale.
Over-plotted to the curves are temperature and pressure data from an analysis 
by \citet{seiff1998}, which is based on the measurements acquired in 1995 
by the Galileo entry probe, during its hour-long descent in Jupiter's 
atmosphere. The temperature difference at $1\,\mathrm{bar}$ is about $8$\%.

These are results from a version of model \aqt\ that includes a compressible
core, as discussed further below. At the core boundary $R_{Z}$, the values of
density, pressure, and temperature at the current age are $\approx 3.6\,\rhou$, 
$3.5\times 10^{7}\,\mathrm{bar}$ ($3500\,\mathrm{GPa}$), and 
$1.6\times 10^{4}\,\K$, respectively.
For the in\-com\-pressible-core version of the model (\aqt), in which $R_{Z}$ 
is larger by a factor of about $1.6$ at $t\approx 4600\,\Myr$, these 
values are instead $\approx 2.7\,\rhou$, $2\times 10^{7}\,\mathrm{bar}$ 
($2000\,\mathrm{GPa}$), and $1.3\times 10^{4}\,\K$, respectively.

\begin{figure}
    \centering
    \resizebox{1.0\linewidth}{!}{%
    \includegraphics[]{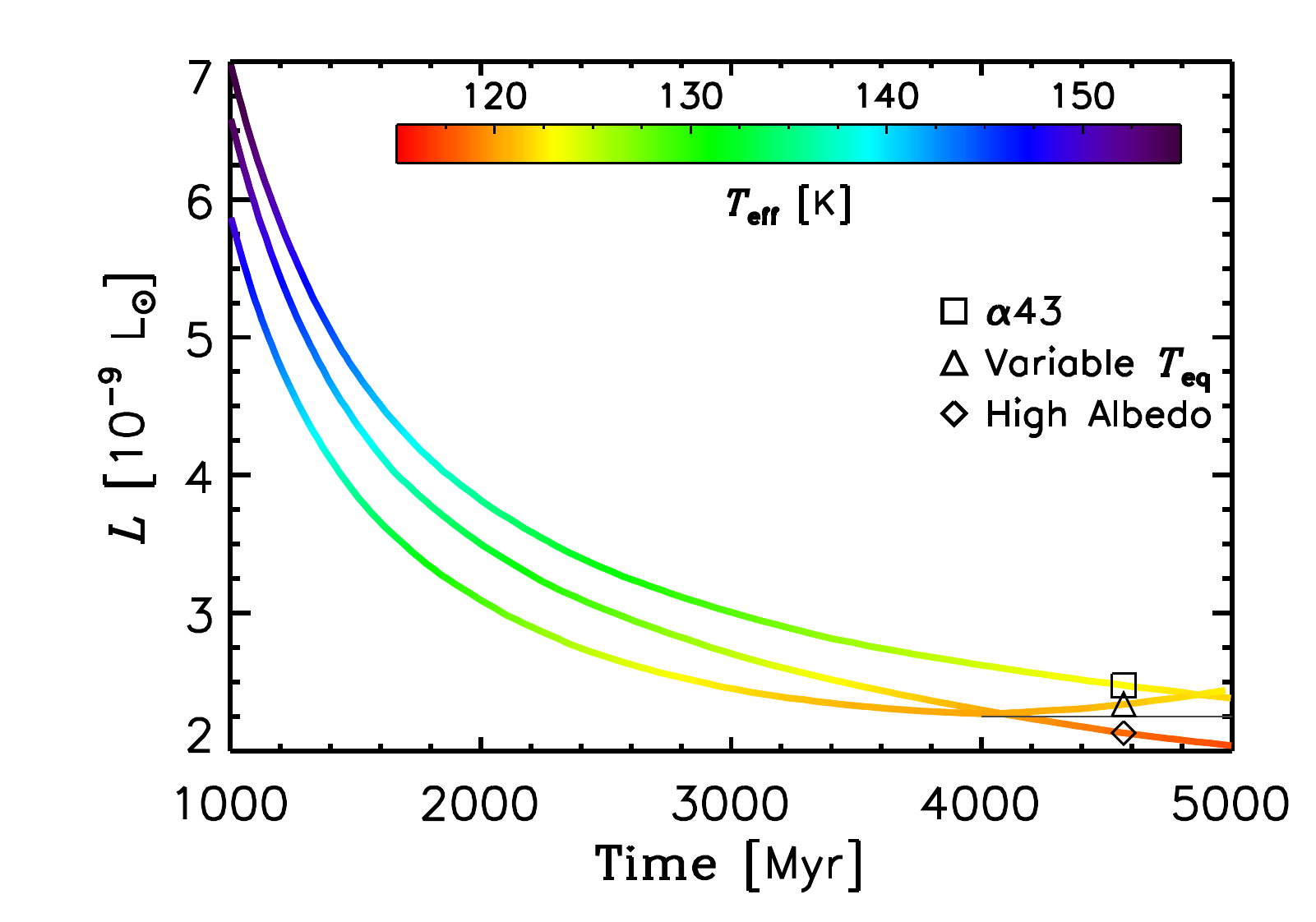}}
    \caption{
    Jupiter's luminosity during isolation for variations 
    of model \aqt\ (curve marked by an open square).
    The diamond symbol indicates a model in which a higher, constant 
    albedo $\mathcal{A}=0.503$ is applied to account for absorption 
    from solar irradiation.
    The triangle indicates a model in which $\mathcal{A}=0.343$ 
    (as in \aqt) but the solar luminosity varies according to 
    a stellar model of \citet{spada2013}.
    The symbols also mark the solar system's age, $t=4567\,\Myr$.
    The color bar renders the planet's effective temperature
    and the horizontal segment indicates Jupiter's current luminosity.
    }
    \label{fig:ap}
\end{figure}
As shown in the bottom panel of \cifig{fig:bb}, solar irradiation
can affect Jupiter's luminosity over most of its history. The amount
of external radiation absorbed by the planet depends on its albedo, 
$\mathcal{A}$.
Not only are variations of $\mathcal{A}$ over time unknown, but 
also its current value has differing estimates.
Early Pioneer and Voyager~1 observations resulted in $\mathcal{A}=0.343$ 
\citep{hanel2003}, a value that has been long applied in studies of 
Jupiter's post-formation evolution \citep[e.g,][]{fortney2003,mankovich2016}.
However, an analysis of Cassini multi-instrument observations \citep{lil2018}
provided a higher estimate of Jupiter's albedo, $\mathcal{A}=0.503$,
which would reduce the equilibrium temperature to 
$T_{\mathrm{eq}}\simeq 102.7\,\K$.
A calculation of model \aqt, applying this lower equilibrium temperature, 
results in a planet radius only marginally different from that illustrated 
in \cifig{fig:iso} (at $t=4567\,\Myr$), whereas the effect on the luminosity 
is larger, as shown in \cifig{fig:ap} (see curve marked by a diamond), 
bringing $L$ within $5$\% of the observed value.

The Sun's luminosity has changed over the age of the solar system. 
In order to gauge the effect of the varying solar power output on
Jupiter's luminosity, a calculation of model \aqt\ is performed by
using a time-dependent solar luminosity from a solar-type stellar 
model of \citet{spada2013}. Results are illustrated in \cifig{fig:ap}. 
The luminosity, marked by a triangle in the figure, is lower than 
in model \aqt\ over most of the planet's history
but it becomes larger after $4800\,\Myr$. 
The planet radius in the three models in the figure is largely 
unaffected and, at the age of the solar system, is the same within 
$1$\%.

\begin{figure}
    \centering
    \resizebox{1.0\linewidth}{!}{%
    \includegraphics[]{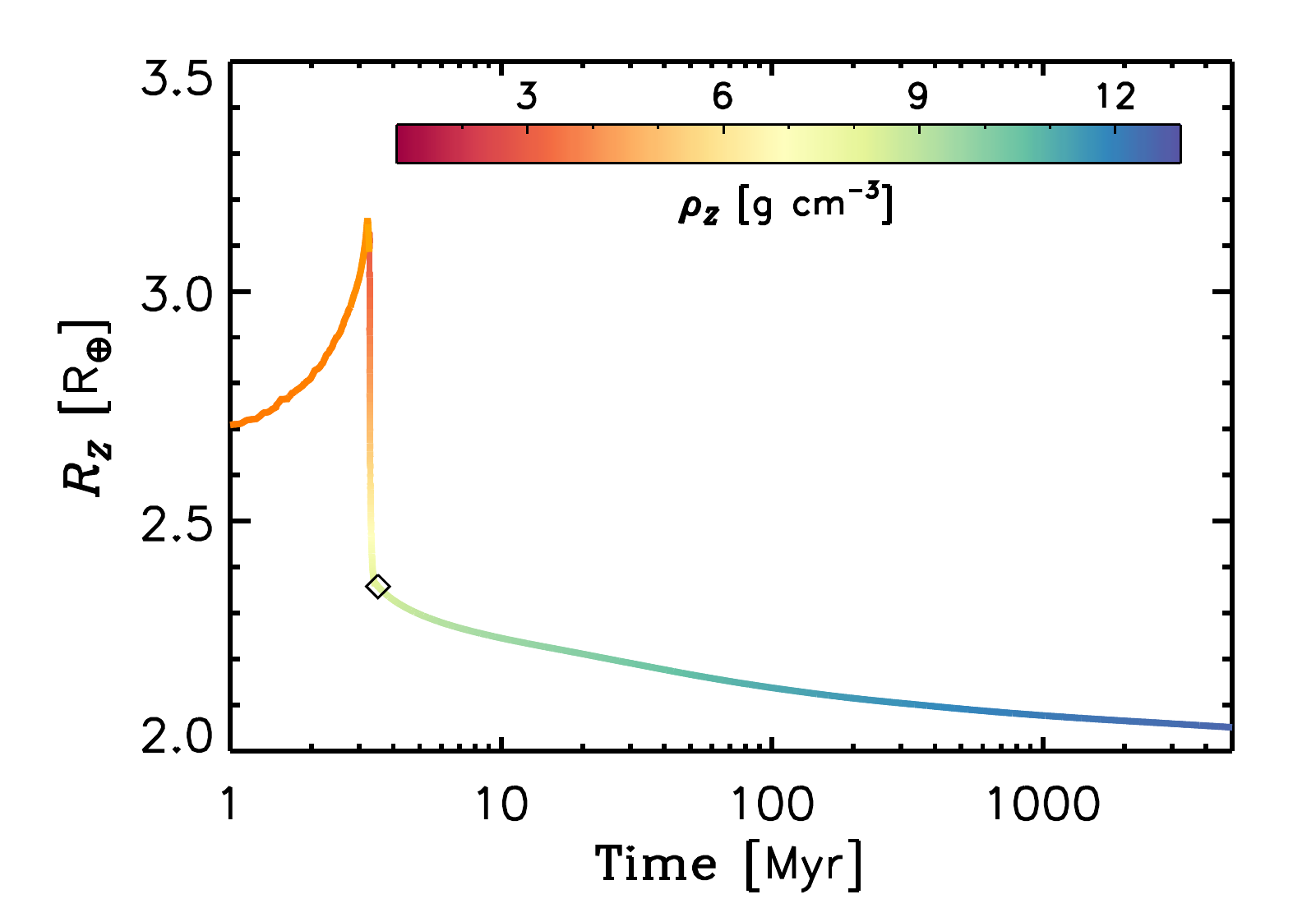}}
    \caption{
    Radius of the ice and silicate core versus time in a version of 
    model \aqt\ in which compression of the heavy elements is taken 
    into account as described in the text. 
    The curve is color-coded according to the core's mean density 
    (mass-to-volume ratio). The heavy-element composition is the same 
    as that of the accreted solids, H$_{2}$O and silicates in a ratio 
    $58/42$ by mass. The diamond symbol indicates the end of Phase~4 
    and beginning of isolation.
    }
    \label{fig:RZ}
\end{figure}
As mentioned in \cisec{sec:sc}, the calculations assume that accreted solids
sediment to the center, eventually accumulating on an incompressible core of
heavy elements.
But self-gravity and pressure exerted by the overlying gaseous envelope 
can compress the core.
To relax this approximation, structure models of the core are calculated 
according to the procedure described in \citet{gennaro2016}. The core 
is assumed to be in hydrostatic equilibrium and to have a fully adiabatic 
interior, a good approximation given the relatively high interior 
temperatures (we are mostly concerned with late Phase~2 and beyond). 
The composition is that of accreted solids, ice and silicates in a mass 
ratio of $58/42$.
Given $\Mc$, and temperature and pressure at the base of the envelope,
the core radius $R_{Z}$ is provided by interpolation of values from a 
dense table based on these structure models. 

\cifig{fig:RZ} illustrates results from a calculation of a version of model 
\aqt\ that accounts for a compressible core. The radius $R_{Z}$ is plotted
as a function of time and the diamond symbol on the curve marks the end of 
formation. The curve is color-coded according to the core's mean density, 
$\rho_{Z}$. 
The radius $R_{Z}$ increases, as $\Mc$ grows, up to a maximum of a little 
over $3.1\,\Rearth$, when $\Mc\approx 15\,\Mearth$. This happens in Phase~3, 
before rapid contraction begins.
From this point, and until the end of formation, the pressure exerted 
at the base of the envelope increases by a factor of $25$, from $\approx 44$ 
to $1110\,\mathrm{GPa}$. Although the temperature at $R_{Z}$ rises from
$\approx 2\times 10^{4}$ to $\approx 5.8\times 10^{4}\,\K$ and $\Mc$ 
grows, the core shrinks by about $25$\%, to $2.37\,\Rearth$.
At the end of formation, differences in planet radius are marginal
(a few percent), whereas the luminosity is lower than in model \aqt, 
by about $35$\%.
Contraction of the core continues during the isolation phase, as 
the envelope cools and the pressure exerted on it increases. 
At $t=4567\,\Myr$, $R_{Z}\approx 2.05\,\Rearth$ and both 
$L=2.44\times 10^{-9}\,\Lsun$ and $\Rp=1.06\,\Rjup$
are within a few percent of the values obtained from model \aqt.

\section*{Acknowledgements}

We wish to thank two anonymous reviewers whose comments 
helped improve this manuscript.
Primary support for this research was provided by NASA's 
Emerging Worlds program 
(proposals 80HQTR19T0071, 18-EW18\_\-2-0060, and
15-EW15\_\-2-0007).
Additional support was provided by NASA's Exoplanets Research
Program (proposal 18-XRP18\_\-2-0106) and through NASA grant
NNX14AG92G.
Computational resources supporting this work were provided 
by the NASA High-End Computing (HEC) Program through the 
NASA Advanced Supercomputing (NAS) Division at 
Ames Research Center.

\printcredits



\end{document}